\documentclass[aps,pra,reprint,superscriptaddress,amsmath,amssymb]{revtex4-2}

\usepackage{CJK}
\usepackage{color}
\usepackage{graphicx}
\usepackage[colorlinks]{hyperref}
\usepackage{orcidlink}
\usepackage{bm}
\renewcommand{\d}{\mathrm{d}}
\renewcommand{\Im}{\mathrm{Im}}
\renewcommand{\Re}{\mathrm{Re}}

\begin{document}

\begin{CJK*}{UTF8}{gbsn}

\title{Direct measurement of the quantum geometric tensor in pseudo-Hermitian systems}

\author{Ze-Hao Huang~\orcidlink{0000-0002-1180-7673}}
\email{zhhuang98@smail.nju.edu.cn}
\affiliation{National Laboratory of Solid State Microstructures and School of Physics, Nanjing University, Nanjing 210093, China}
\affiliation{Collaborative Innovation Center of Advanced Microstructures, Nanjing University, Nanjing 210093, China}

\author{Hai-Tao Ding}
\email{htding.9@nus.edu.sg}
\affiliation{Centre for Quantum Technologies, National University of Singapore, 3 Science Drive 2, Singapore 117543}
\affiliation{MajuLab, CNRS-UNS-NUS-NTU International Joint Research Unit, Singapore UMI 3654, Singapore}

\author{Li-Jun Lang~\orcidlink{0000-0001-6038-8340}}
\email{ljlang@scnu.edu.cn}
\affiliation {Guangdong Basic Research Center of Excellence for Structure and Fundamental Interactions of Matter, Guangdong Provincial Key Laboratory of Quantum Engineering and Quantum Materials, School of Physics, South China Normal University, Guangzhou 510006, China}
\affiliation {Guangdong-Hong Kong Joint Laboratory of Quantum Matter, Frontier Research Institute for Physics, South China Normal University, Guangzhou 510006, China}

\date{\today}

\begin{abstract}

The quantum geometric tensor (QGT) fundamentally encodes the geometry and topology of quantum states in both Hermitian and non-Hermitian regimes.
While adiabatic perturbation theory links its real part (quantum metric) and imaginary part (Berry curvature) to energy fluctuations and generalized forces, respectively, in Hermitian systems, direct measurement of the QGT, which is defined using both left and right eigenstates of a non-Hermitian Hamiltonian, remains challenging.
Here we develop two quantum simulation schemes to directly extract all components of the QGT in pseudo-Hermitian systems with real spectra.
Each scheme independently determines the complete QGT using generalized expectation values of either the energy fluctuation operator or the generalized force operator with respect to two time-evolved states prepared through distinct nonadiabatic evolutions, thereby establishing two self-contained measurement protocols.
We illustrate the validity of these schemes on two $q$-deformed two-band models: one with nontrivial topology and the other with a nonvanishing off-diagonal quantum metric.
Numerical simulations demonstrate that, for suitably chosen nonadiabatic ramp velocities, both schemes achieve high-fidelity agreement with theoretical predictions for measuring the QGT in both models and successfully capture the topological phase transition of the first model using Chern numbers calculated from Berry curvatures.
For larger velocities, the generalized force scheme yields greater accuracy for the real part of the QGT, while the energy fluctuation scheme better captures its imaginary part.
This work establishes a framework for extending dynamical measurement schemes from Hermitian to pseudo-Hermitian systems with real spectra.

\end{abstract}

\maketitle

\end{CJK*}

\section{Introduction} \label{sec: introduction}

The quantum geometric tensor (QGT)~\cite{Torma2022,Torma2023,SLZhu2008} is a fundamental quantity in quantum mechanics, characterizing the geometric and topological properties of quantum systems.
Its symmetric real part, known as the quantum metric tensor (QMT)~\cite{Provost1980}, quantifies the quantum distance and Fisher information~\cite{Braunstein1994} in parameter space and governs phenomena including flat-band superfluidity~\cite{Peotta2015,Julku2016,LLiang2017}, orbital magnetic susceptibility~\cite{YGao2015,Piechon2016}, and the intrinsic nonlinear Hall effect~\cite{ZZDu2021,AYGao2023,NZWang2023}.
The antisymmetric imaginary part, corresponding to the Berry curvature~\cite{DXiao2010}, determines the geometric phase~\cite{Simon1983,Berry1984} and topological invariants such as the Chern number~\cite{Thouless1982}, which are central to topological quantum matter~\cite{Hasan2010,XLQi2011,Armitage2018,SLZhu2006SHE,DWZhang2018}.
Both components are also key to characterizing quantum phase transitions~\cite{Carollo2005,SLZhu2006,Venuti2007,SJGu2008}.
Alongside theoretical progress in generalizing the QGT from its original Abelian form for pure states in nondegenerate systems to the non-Abelian form in degenerate systems~\cite{YQMa2010,HTDing2024} and to mixed states~\cite{XYHou2024,ZZhou2024,QYWang2025},
experimental advances have enabled its measurement across diverse platforms, including superconducting qubits~\cite{Schroer2014,Roushan2014,XSTan2018,XSTan2019,XSTan2019b,XSTan2021,TQChen2024}, ultracold atoms~\cite{Tran2017,Ozawa2018,Asteria2019,QXLv2021,HTDing2022,CRYi2023,SLZhu2007}, nitrogen-vacancy (NV) centers~\cite{MYu2020,MChen2022,XYZhang2023}, exciton-photon polaritons~\cite{Gianfrate2020}, and, most recently, solid-state systems~\cite{Kim2025}.

Unique phenomena revealed in non-Hermitian systems beyond Hermitian ones~\cite{Ghatak2019,Ashida2020,Bergholtz2021} have motivated the generalization of the QGT to non-Hermitian systems, thereby yielding two alternative definitions:
the right-right (RR) formalism, defined using only right eigenstates~\cite{Solnyshkov2021,Alon2024,Cuerda2024b,JFRen2024},
and the left-right (LR) formalism, which involves both left and right eigenstates~\cite{Brody2013,DJZhang2019,GYSun2022,CYC2024,Tzeng2021,YTTu2023,ANFan2020,PHe2021,YQZhu2021,PHe2023}.
Both formalisms are important for understanding wave-packet dynamics~\cite{YMRobinHu2025,Silberstein2020,JHWang2022}.
While the Berry curvature in both formalisms characterizes the same topological phases through identical Chern numbers obtained from its integration over the Brillouin zone~\cite{HTShen2018}, the QMT exhibits distinct roles:
specifically, the RR formalism identifies localization transitions~\cite{JFRen2024}, whereas the LR formalism detects quantum phase transitions~\cite{DJZhang2019,GYSun2022,CYC2024}, exceptional points~\cite{Tzeng2021,YTTu2023}, and underpins several other phenomena~\cite{FQin2025,KChen2025,PHe2021}.
Experimentally, despite successful demonstrations of the RR formalism~\cite{QLiao2021,Cuerda2024,YMRobinHu2024}, the QGT in the LR formalism has not yet been measured.
We therefore focus exclusively on the LR formalism hereafter.

Pseudo-Hermitian systems, a special class of non-Hermitian systems with spectra consisting of either real eigenvalues or complex-conjugate pairs~\cite{Mostafazadeh2002,*Mostafazadeh2002b,*Mostafazadeh2002c,Mostafazadeh2003}, have attracted interest due to their dynamical similarities to Hermitian systems~\cite{Pinske2019} and novel features such as enhanced quantum sensing~\cite{YMChu2020}.
Measuring the QGT in pseudo-Hermitian systems with real spectra is still an important experimental challenge.
While Ref.~\cite{YQZhu2021} proposed an indirect approach using Rabi frequencies, practical schemes for direct measurement of the QGT are essential to advance experimental studies.

We develop two independent schemes for the direct measurement of the QGT in pseudo-Hermitian systems with real spectra.
These schemes combine adiabatic perturbation theory for describing time-dependent quantum dynamics~\cite{Rigolin2008,DeGrandi2010b,Nenciu1992,QZhang2019} with quantum circuits capable of measuring generalized expectation values~\cite{ZHHuang2023,Wagner2024,Chiribella2024}, defined as expectationlike quantities of an operator with respect to two states, in a form analogous to weak values~\cite{Dressel2014,Hofmann2012,YLWen2023}.
The first scheme extracts both the real and imaginary parts of the QGT by measuring the generalized expectation value of the energy fluctuation operator $\Delta^2 H \equiv (H-\langle\!\langle H \rangle\!\rangle)^2$ within a unified experimental framework, where $\langle\!\langle H \rangle\!\rangle$ denotes the generalized expectation value of the system Hamiltonian.
The second scheme utilizes the generalized force operator $f_\mu \equiv -\partial_\mu H$ and requires two distinct generalized expectation value measurements to separately determine the real and imaginary components of the QGT.
We verify both schemes on two pseudo-Hermitian $q$-deformed two-band models: one exhibiting nontrivial topology and the other featuring a nonvanishing off-diagonal quantum metric.
Numerical simulations demonstrate high-fidelity QGT measurement for both schemes.
Notably, for the first model, the Chern number phase diagram can be accurately reconstructed using either scheme, whereas conventional expectation value approaches allow only the generalized force method~\cite{Schroer2014}.
As the velocity in nonadiabatic evolutions increases, the energy fluctuation scheme exhibits greater deviations in the real part of the QGT, while the generalized force scheme shows larger inaccuracies in the imaginary part.

The paper is structured as follows.
In Sec.~\ref{sec: QGT}, we introduce the QGT in the LR formalism for non-Hermitian systems, which becomes a Hermitian quantity in pseudo-Hermitian systems with real spectra.
In Sec.~\ref{sec: schemes}, we propose two distinct schemes for directly measuring the complete QGT in such systems.
In Sec.~\ref{sec: examples}, we provide numerical verification of both schemes using two illustrative examples.
In Sec.~\ref{sec: exp}, we discuss experimental implementation.
Concluding remarks are given in Sec.~\ref{sec: conclusion}.

\section{Quantum Geometric Tensor in Pseudo-Hermitian Systems} \label{sec: QGT}

We consider a generic $N$-band non-Hermitian Hamiltonian $H(\boldsymbol{\lambda})$ parametrized by $\boldsymbol{\lambda} \equiv (\lambda_1, \lambda_2, \dots)$.
The time-independent Schr\"odinger equations are
\begin{align}
    H(\boldsymbol{\lambda})|\phi^R_n(\boldsymbol{\lambda})\rangle &= E_n(\boldsymbol{\lambda}) |\phi^R_n(\boldsymbol{\lambda})\rangle, \\
    H^\dag(\boldsymbol{\lambda})|\phi^L_n(\boldsymbol{\lambda})\rangle &= E_n^*(\boldsymbol{\lambda}) |\phi^L_n(\boldsymbol{\lambda})\rangle,
\end{align}
where $|\phi^R_n(\boldsymbol{\lambda})\rangle$ and $|\phi^L_n(\boldsymbol{\lambda})\rangle$ denote the right and left eigenstates, respectively, and $E_n(\boldsymbol{\lambda})$ are the discrete, nondegenerate eigenvalues, with $n$ labeling the bands.
The right and left eigenstates are generally nonorthogonal, i.e., $\langle \phi^{R,L}_n(\boldsymbol{\lambda}) | \phi^{R,L}_m(\boldsymbol{\lambda}) \rangle \neq 0$, but together they form a complete biorthogonal basis, i.e., $\langle \phi_m^L (\boldsymbol{\lambda}) | \phi_n^R (\boldsymbol{\lambda}) \rangle = \delta_{mn}$ and $\sum_n |\phi_n^R (\boldsymbol{\lambda}) \rangle \langle \phi_n^L (\boldsymbol{\lambda})| = 1$~\cite{Brody2014}.
The LR formalism of the QGT for the $n$th band is defined as~\cite{DJZhang2019,YQZhu2021,CYC2024}
\begin{equation} \label{eq: biQGT}
    Q^n_{\mu\nu} := \langle \partial_\mu \phi_n^L | \partial_\nu \phi_n^R \rangle - \langle \partial_\mu \phi_n^L | \phi_n^R \rangle \langle \phi_n^L | \partial_\nu \phi_n^R \rangle,
\end{equation}
where $\partial_\mu \equiv \partial/\partial \lambda_\mu$, and $|\phi_n^{R,L}\rangle \equiv |\phi_n^{R,L}(\boldsymbol{\lambda})\rangle$ for brevity.

In general non-Hermitian systems, the QGT is not Hermitian, i.e., $Q^n_{\mu\nu} \neq [Q^n_{\nu\mu}]^*$, which leads to ambiguities in defining the quantum metric and Berry curvature~\cite{CYC2024}.
However, for pseudo-Hermitian systems with real spectra, which satisfy the pseudo-Hermiticity condition $H^\dagger = \eta H \eta^{-1}$ for some Hermitian, invertible operator $\eta$, the relation $|\phi^L_n(\boldsymbol{\lambda})\rangle = \eta|\phi^R_n(\boldsymbol{\lambda})\rangle$ restores the Hermiticity of the QGT.
This ensures unique definitions for the quantum metric and Berry curvature, given by $g^n_{\mu\nu} = \frac{1}{2} [ Q^n_{\mu\nu} + (Q^n_{\nu\mu})^* ]$ and $F^n_{\mu\nu} = i [ Q^n_{\mu\nu} - (Q^n_{\nu\mu})^* ]$, which coincide with those in Hermitian systems.

\section{Direct Measurement Schemes} \label{sec: schemes}

In this section, we propose two self-contained schemes for directly measuring the QGT component $Q_{\mu\nu}^0(\boldsymbol{\lambda}^0)$ for the lowest band at a target parameter $\boldsymbol{\lambda}^0 \equiv (\lambda_1^0, \lambda_2^0, \cdots)$ in pseudo-Hermitian systems with real spectra.
Both schemes are based on preparing pairs of time-evolved states via nonadiabatic evolutions, described by adiabatic perturbation theory (see Appendix~\ref{sec: adiabatic perturbation})~\cite{Rigolin2008,DeGrandi2010b,Nenciu1992,QZhang2019}, and subsequently measuring generalized expectation values of specific operators between these states.
The generalized expectation value extends the conventional expectation value to any nonorthogonal pair of quantum states, and is defined for an operator $A$ and two states $|\psi_1\rangle$, $|\psi_2\rangle$ as
\begin{equation} \label{eq: generalized expectation value}
    \langle\!\langle A \rangle\!\rangle \equiv \frac{\langle \psi_1 | A | \psi_2 \rangle}{\langle \psi_1 | \psi_2 \rangle},
\end{equation}
which can be efficiently measured using quantum circuits (see Appendix~\ref{sec: quantum circuit})~\cite{ZHHuang2023,Wagner2024,Chiribella2024}.

\textit{State preparation.}
We first describe how to prepare the time-evolved states $|\psi_\mu(\boldsymbol{\lambda}^0, t)\rangle$, $|\psi_\mu^+(\boldsymbol{\lambda}^0, t)\rangle$, and $|\psi_\mu^-(\boldsymbol{\lambda}^0, t)\rangle$.
To prepare $|\psi_\mu(\boldsymbol{\lambda}^0, t)\rangle$, one initializes the system in the right ground state $|\phi^R_0(\boldsymbol{\lambda}^0 - \Delta\lambda\, \mathbf{e}_\mu)\rangle$ of $H(\boldsymbol{\lambda}^0 - \Delta\lambda\, \mathbf{e}_\mu)$, where $\Delta\lambda$ is a preset constant and $\mathbf{e}_\mu$ is the unit vector along the $\mu$th parameter direction.
Then, one evolves this state under the time-dependent Hamiltonian $H[\boldsymbol{\lambda}(t)]$ with a quadratic ramp along the $\mathbf{e}_\mu$ direction~\cite{Gritsev2012}:
\begin{equation} \label{eq: lambda evolution}
    \boldsymbol{\lambda}(t) = \boldsymbol{\lambda}^0 - \Delta\lambda\, \mathbf{e}_\mu + \frac{v^2 t^2}{4 \Delta\lambda}\, \mathbf{e}_\mu, \quad t \in [0, t_f]
\end{equation}
where $v$ is the final ramp velocity and $t_f = 2\Delta\lambda / v$ is the total evolution time.
This quadratic ramp profile ensures zero initial velocity [$\dot{\boldsymbol{\lambda}}(0) = 0$], which suppresses oscillations from the initial state~\cite{Gritsev2012}.
The ramp velocity $v$ should be chosen sufficiently small to ensure the nonadiabatic response remains in the linear regime (see Appendix~\ref{sec: adiabatic perturbation} for details).

Similarly, to prepare $|\psi_\mu^+(\boldsymbol{\lambda}^0, t)\rangle$ and $|\psi_\mu^-(\boldsymbol{\lambda}^0, t)\rangle$, one initializes the system in the left ground state $|\phi^L_0(\boldsymbol{\lambda}^0 - \Delta\lambda\, \mathbf{e}_\mu)\rangle$ and evolves it under $H^\dagger[\boldsymbol{\lambda}(t)]$ and $-H^\dagger[\boldsymbol{\lambda}(t)]$, respectively, using the same ramp profile as in Eq.~\eqref{eq: lambda evolution}.

\textit{Measurement schemes.}
We now explain how to select appropriate pairs of these time-evolved states to construct generalized expectation values of either the energy fluctuation operator $\Delta^2 H \equiv (H-\langle\!\langle H \rangle\!\rangle)^2$ or the generalized force operator $f_\mu \equiv -\partial_\mu H$ for QGT measurement.
The explicit connections between these generalized expectation values and the QGT components are derived in detail in Appendix~\ref{sec: derivation}.

\textit{Scheme 1: Using energy fluctuation operator $\Delta^2 H$.}
The first scheme enables direct measurement of both the real and imaginary parts of the QGT by measuring the generalized expectation value of the energy fluctuation operator $\Delta^2 H (\boldsymbol{\lambda}^0) \equiv [H (\boldsymbol{\lambda}^0) - \langle\!\langle H (\boldsymbol{\lambda}^0) \rangle\!\rangle]^2$ between the time-evolved states $|\psi_\mu^+(\boldsymbol{\lambda}^0, t_f)\rangle$ and $|\psi_\nu(\boldsymbol{\lambda}^0, t_f)\rangle$:
\begin{equation} \label{eq: energy fluctuation}
    \frac{\langle \psi_\mu^+(\boldsymbol{\lambda}^0, t_f) | \Delta^2 H (\boldsymbol{\lambda}^0) | \psi_\nu(\boldsymbol{\lambda}^0, t_f) \rangle}{\langle \psi_\mu^+(\boldsymbol{\lambda}^0, t_f) | \psi_\nu(\boldsymbol{\lambda}^0, t_f) \rangle} \approx v^2 Q_{\mu\nu}^0(\boldsymbol{\lambda}^0),
\end{equation}
where $\langle\!\langle H (\boldsymbol{\lambda}^0) \rangle\!\rangle$ is the generalized expectation value of the Hamiltonian, defined as $\frac{\langle \psi_\mu^+(\boldsymbol{\lambda}^0,t_f) | H (\boldsymbol{\lambda}^0) | \psi_\nu(\boldsymbol{\lambda}^0,t_f) \rangle}{\langle \psi_\mu^+(\boldsymbol{\lambda}^0,t_f) | \psi_\nu(\boldsymbol{\lambda}^0,t_f) \rangle}$.
By dividing the measured value by $v^2$, one obtains the desired QGT component $Q_{\mu\nu}^0(\boldsymbol{\lambda}^0)$, including both its real and imaginary parts.

\textit{Scheme 2: Using generalized force operator $f_\mu$.}
The second scheme separately determines the real and imaginary parts of the QGT using the generalized force operator $f_\mu(\boldsymbol{\lambda}^0) \equiv -\partial_\mu H(\boldsymbol{\lambda}^0)$.
The imaginary part of the QGT is obtained from the generalized expectation value with $|\psi_\nu^+(\boldsymbol{\lambda}^0, t_f)\rangle$ and $|\psi_\nu(\boldsymbol{\lambda}^0, t_f)\rangle$:
\begin{align}
    &\frac{\langle \psi_\nu^+(\boldsymbol{\lambda}^0, t_f) | f_\mu(\boldsymbol{\lambda}^0) | \psi_\nu(\boldsymbol{\lambda}^0, t_f) \rangle}{\langle \psi_\nu^+(\boldsymbol{\lambda}^0, t_f) | \psi_\nu(\boldsymbol{\lambda}^0, t_f) \rangle} \notag \\
    &\approx \langle\!\langle f_\mu (\boldsymbol{\lambda}^0)\rangle\!\rangle - 2 v\, \Im\left[Q_{\mu\nu}^0(\boldsymbol{\lambda}^0)\right] \label{eq: generalized force im} \\
    &= \langle\!\langle f_\mu (\boldsymbol{\lambda}^0)\rangle\!\rangle + v\, F_{\mu\nu}^0(\boldsymbol{\lambda}^0),
\end{align}
where the constant
\begin{equation} \label{eq: constant}
    \langle\!\langle f_\mu (\boldsymbol{\lambda}^0)\rangle\!\rangle = \frac{\langle \phi^L_0(\boldsymbol{\lambda}^0) | f_\mu(\boldsymbol{\lambda}^0) | \phi^R_0(\boldsymbol{\lambda}^0) \rangle}{\langle \phi^L_0(\boldsymbol{\lambda}^0) | \phi^R_0(\boldsymbol{\lambda}^0) \rangle}
\end{equation}
is the generalized expectation value of $f_\mu$ with respect to the instantaneous left and right ground states at $\boldsymbol{\lambda}^0$.
After subtracting this constant, the coefficient of $v$ yields the Berry curvature $F_{\mu\nu}^0(\boldsymbol{\lambda}^0) = -2\,\Im[Q_{\mu\nu}^0(\boldsymbol{\lambda}^0)]$.

To obtain the real part of the QGT, we measure the generalized expectation value with $|\psi_\nu^-(\boldsymbol{\lambda}^0, t_f)\rangle$ and $|\psi_\nu(\boldsymbol{\lambda}^0, t_f)\rangle$:
\begin{align}
    & \frac{\langle \psi_\nu^-(\boldsymbol{\lambda}^0, t_f) | f_\mu(\boldsymbol{\lambda}^0) | \psi_\nu(\boldsymbol{\lambda}^0, t_f) \rangle}{\langle \psi_\nu^-(\boldsymbol{\lambda}^0, t_f) | \psi_\nu(\boldsymbol{\lambda}^0, t_f) \rangle} \notag \\
    &\approx \langle\!\langle f_\mu (\boldsymbol{\lambda}^0)\rangle\!\rangle + i 2v\, \Re\left[Q_{\mu\nu}^0(\boldsymbol{\lambda}^0)\right] \label{eq: generalized force re} \\
    &= \langle\!\langle f_\mu (\boldsymbol{\lambda}^0)\rangle\!\rangle + i 2v\, g_{\mu\nu}^0(\boldsymbol{\lambda}^0).
\end{align}
The imaginary part of this result, divided by $2v$ after subtracting the constant, yields the quantum metric $g_{\mu\nu}^0(\boldsymbol{\lambda}^0) = \Re[Q_{\mu\nu}^0(\boldsymbol{\lambda}^0)]$.

\section{Examples} \label{sec: examples}

Pseudo-Hermitian systems with real spectra can be realized through two complementary approaches: (i) in real space, by introducing nonreciprocal hopping in lattice models~\cite{YLong2022}; and (ii) in momentum space, by replacing conventional generator matrices with their $q$-deformed counterparts~\cite{YQZhu2021}, which naturally emerge from the Fourier transformation of lattice models featuring nonreciprocal hopping and staggered onsite potentials (see Appendix~\ref{sec: mapping} for details).
In this section, we employ $q$-deformed matrices to construct two distinct two-band models in momentum space for benchmarking the proposed measurement schemes.
The general form of their Hamiltonian is
\begin{equation} \label{eq: H}
    H = \mathbf{d} \cdot \tilde{\bm{\sigma}} = d_x \tilde\sigma_x + d_y \tilde\sigma_y + d_z \tilde\sigma_z,
\end{equation}
where $\mathbf{d} = (d_x, d_y, d_z)$ is the Bloch vector, which varies between models, and $\tilde{\bm{\sigma}} = (\tilde\sigma_x, \tilde\sigma_y, \tilde\sigma_z)$ are the $q$-deformed Pauli matrices defined as
\begin{equation} \label{eq: q-deformed Pauli matrices}
    \tilde\sigma_x = \begin{bmatrix} 0 & a \\ b & 0 \end{bmatrix}, \quad
    \tilde\sigma_y = \begin{bmatrix} 0 & -i a \\ i b & 0 \end{bmatrix}, \quad
    \tilde\sigma_z = \begin{bmatrix} q^{-1} & 0 \\ 0 & -q \end{bmatrix},
\end{equation}
with $a = \sqrt{(1 + q^2)/2}$, $b = \sqrt{(1 + q^{-2})/2}$, and $q > 0$.
These systems satisfy the pseudo-Hermiticity condition $H^\dag = \eta H \eta^{-1}$ with $\eta = \mathrm{diag}(q^{-1/2}, q^{1/2})$, possess real eigenvalues $E_\pm = d d_z \pm \sqrt{ab (d_x^2 + d_y^2) + c^2 d_z^2}$ with $c = (1 + q^2)/(2q)$ and $d = (1 - q^2)/(2q)$, and reduce to Hermitian systems when $q = 1$.
Throughout the benchmark tests, we fix $q = 3$.

\subsection{Model I with nontrivial topology} \label{subsec: model1}

The first model is defined by the Bloch vector $\mathbf{d}(\theta, \phi) = \frac{1}{2} (\Omega_1 \sin \theta \cos \phi,\, \Omega_1 \sin \theta \sin \phi,\, \Delta_1 \cos \theta + \Delta_2)$, which can be mapped to a Haldane-type honeycomb lattice model with nonreciprocal hopping (see Appendix~\ref{sec: mapping}).
This model exhibits nontrivial Chern numbers $C = \frac{1}{2\pi} \int_S F_{\theta\phi}\,\d S = \int_0^{\pi} F_{\theta\phi}\,\d \theta$, where the Berry curvature is $F_{\theta\phi} = 2\,\Im[Q_{\phi\theta}]$.
Notably, the off-diagonal quantum metric vanishes throughout the parameter space, i.e., $\Re[Q_{\theta\phi}] = \Re[Q_{\phi\theta}] = 0$.

To measure the QGT at the target parameters $(\theta^0, \phi^0)$, we set the preset constants $\Delta \theta = \Delta \phi = \pi/2$.
The time-evolved states $|\psi_\theta[(\theta^0, \phi^0),t_f]\rangle$ and $|\psi_\phi[(\theta^0, \phi^0),t_f]\rangle$ are generated from the initial right eigenstates $|\phi^R_0(\theta^0-\frac{\pi}{2}, \phi^0)\rangle$ and $|\phi^R_0(\theta^0, \phi^0-\frac{\pi}{2})\rangle$, respectively, by time evolution under $H[\theta(t),\phi(t)]$ with distinct time-dependent parameters:
\begin{align}
    \theta\text{ direction:} &\quad \theta(t) = \theta^0 - \frac{\pi}{2} + \frac{v^2 t^2}{2\pi},\quad \phi(t) = \phi^0, \\
    \phi\text{ direction:} &\quad \theta(t) = \theta^0,\quad \phi(t) = \phi^0 - \frac{\pi}{2} + \frac{v^2 t^2}{2\pi}.
\end{align}
Similarly, the time-evolved states $|\psi_\theta^\pm[(\theta^0, \phi^0),t_f]\rangle$ and $|\psi_\phi^\pm[(\theta^0, \phi^0),t_f]\rangle$ are obtained by evolving the initial left eigenstates $|\phi^L_0(\theta^0-\frac{\pi}{2}, \phi^0)\rangle$ and $|\phi^L_0(\theta^0, \phi^0-\frac{\pi}{2})\rangle$ under $\pm H^\dag[\theta(t),\phi(t)]$.
After preparing these states, the QGT components can be measured using the generalized expectation values of either the energy fluctuation operator [Eq.~\eqref{eq: energy fluctuation}] or the generalized force operators [Eqs.~\eqref{eq: generalized force im} and \eqref{eq: generalized force re}] with the corresponding state pairs.
For this model, the generalized force operators are
\begin{align} \label{eq: generalized force operators spin}
    f_\theta &= - \textstyle \frac{1}{2} (\Omega_1 \cos \theta \cos \phi,\, \Omega_1 \cos \theta \sin \phi,\,-\Delta_1 \sin \theta) \cdot \tilde{\bm{\sigma}}, \notag \\
    f_\phi &= - \textstyle \frac{1}{2} (-\Omega_1 \sin \theta \sin \phi,\, \Omega_1 \sin \theta \cos \phi,\, 0) \cdot \tilde{\bm{\sigma}}.
\end{align}

\begin{figure}
    \includegraphics[width=\columnwidth]{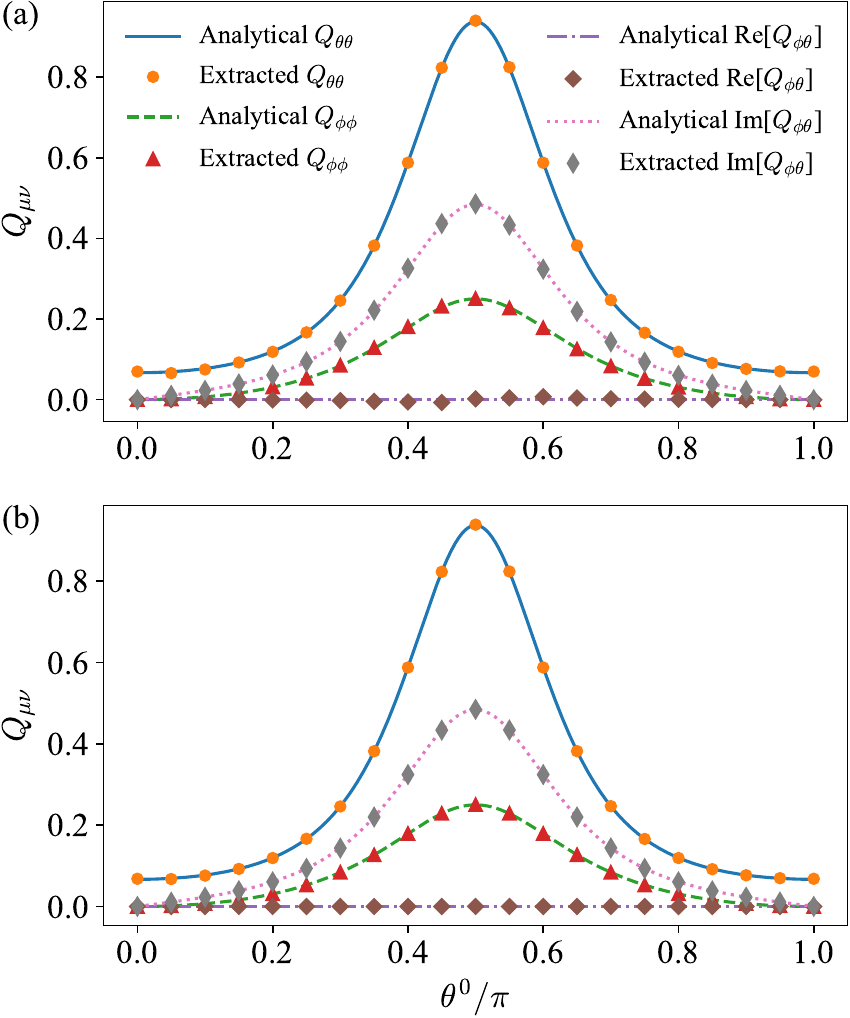}
    \caption{QGT components for model I, extracted using (a) the energy fluctuation scheme and (b) the generalized force scheme.
    Analytical results are shown as lines, while numerical data are represented by markers.
    Specifically, $Q_{\theta\theta}$ is shown with solid lines and circles, $Q_{\phi\phi}$ with dashed lines and triangles, $\Re[Q_{\phi\theta}]$ with dashed-dotted lines and diamonds, and $\Im[Q_{\phi\theta}]$ with dotted lines and thin diamonds.
    The parameters are set to $\Omega_1/2\pi = 10\,\mathrm{MHz}$, $\Delta_1/2\pi = 15\,\mathrm{MHz}$, $\Delta_2 = 0$, and $\phi^0 = 0$.
    }
    \label{fig: pH_spin_QGT}
\end{figure}

As a demonstration, we extract the QGT components $Q_{\theta\theta}$, $Q_{\phi\phi}$, $\Re[Q_{\phi\theta}]$, and $\Im[Q_{\phi\theta}]$ for model I at $\Omega_1/2\pi = 10\,\mathrm{MHz}$, $\Delta_1/2\pi = 15\,\mathrm{MHz}$, $\Delta_2 = 0$, and $\phi^0 = 0$, with $\theta^0$ sampled at $N = 21$ evenly spaced points over $[0, \pi]$.
The final ramp velocity and total evolution time are set to $v = 1\,\mu\mathrm{s}^{-1}$ and $t_f = \pi\,\mu\mathrm{s}$, respectively.
Figure~\ref{fig: pH_spin_QGT} shows excellent agreement between numerically extracted data from both schemes and analytical predictions.

\begin{figure}
    \includegraphics[width=\columnwidth]{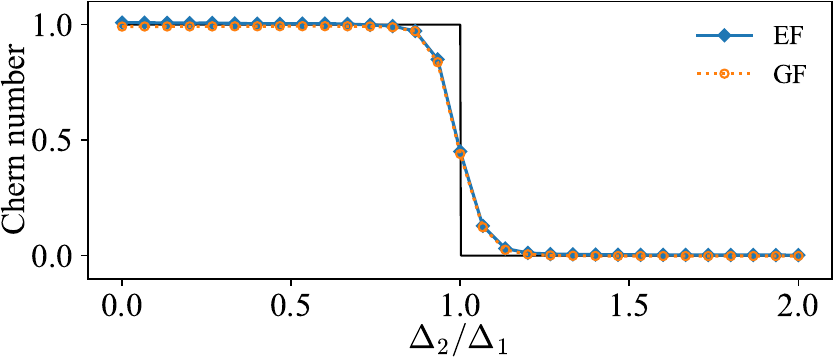}
    \caption{Chern number phase diagram for model I.
    The Chern numbers are calculated by integrating Berry curvatures extracted using the energy fluctuation scheme (diamonds) and the generalized force scheme (circles), and are compared with analytical results (solid line).
    The parameters are fixed at $\Omega_1/2\pi = 10\,\mathrm{MHz}$ and $\Delta_1/2\pi = 15\,\mathrm{MHz}$, while $\Delta_2/2\pi$ is varied from $0$ to $30\,\mathrm{MHz}$, corresponding to $\Delta_2/\Delta_1$ ranging from $0$ to $2$.
    }
    \label{fig: pH_spin_Chern}
\end{figure}

For fixed $\Omega_1/2\pi = 10\,\mathrm{MHz}$ and $\Delta_1/2\pi = 15\,\mathrm{MHz}$, a topological transition occurs at $\Delta_2/\Delta_1 = 1$.
Figure~\ref{fig: pH_spin_Chern} compares Chern numbers calculated from Berry curvatures extracted via both schemes across $\Delta_2/2\pi \in [0,30]\,\mathrm{MHz}$ ($\Delta_2/\Delta_1 \in [0,2]$).
For each $\Delta_2$, $F_{\theta\phi}$ is integrated over $\theta \in [0, \pi]$ using $N = 21$ discrete points. 
Both schemes agree with theory except near the critical point $\Delta_2/\Delta_1=1$, where numerical accuracy improves with an increased number of sampling points.

\subsection{Model II with a nonvanishing off-diagonal quantum metric}

The Bloch vector for the second model is $\mathbf{d}(x, y) = \frac{B}{2} [\sin(x+y) \cos(xy),\, \sin(x+y) \sin(xy),\, \cos(x+y)]$~\cite{XYZhang2023}, which is specifically designed to yield a nonvanishing off-diagonal quantum metric component, i.e., $\Re[Q_{xy}] = \Re[Q_{yx}] \neq 0$.

We adopt the same scheme parameters as in model I ($\Delta x = \Delta y = \pi/2$, $v = 1\,\mu\mathrm{s}^{-1}$, $t_f = \pi\,\mu\mathrm{s}$), with parameter ramps defined as
\begin{align}
    x\text{ direction:} &\quad x(t) = x^0 - \frac{\pi}{2} + \frac{v^2 t^2}{2\pi},\quad y(t) = y^0, \\
    y\text{ direction:} &\quad x(t) = x^0,\quad y(t) = y^0 - \frac{\pi}{2} + \frac{v^2 t^2}{2\pi}.
\end{align}
The time-evolved states $|\psi_{x,y}[(x^0, y^0),t_f]\rangle$ are prepared following the same procedure as in Sec.~\ref{subsec: model1}, and the corresponding states $|\psi_{x,y}^+[(x^0, y^0),t_f]\rangle$ and $|\psi_{x,y}^-[(x^0, y^0),t_f]\rangle$ are generated analogously.
The generalized force operators for this model are
\begin{align} \label{eq: generalized force operators xy}
    f_x &= - \textstyle \frac{B}{2} [\cos(x+y) \cos(xy) - y \sin(x+y) \sin(xy), \notag\\
    &\qquad \qquad \cos(x+y) \sin(xy) + y \sin(x+y) \cos(xy), \notag\\
    &\qquad \qquad -\sin(x+y)] \cdot \tilde{\bm{\sigma}}, \notag \\
    f_y &= - \textstyle \frac{B}{2} [\cos(x+y) \cos(xy) - x \sin(x+y) \sin(xy), \notag\\
    &\qquad \qquad \cos(x+y) \sin(xy) + x \sin(x+y) \cos(xy), \notag\\
    &\qquad \qquad -\sin(x+y)] \cdot \tilde{\bm{\sigma}}.
\end{align}

\begin{figure}
    \includegraphics[width=\columnwidth]{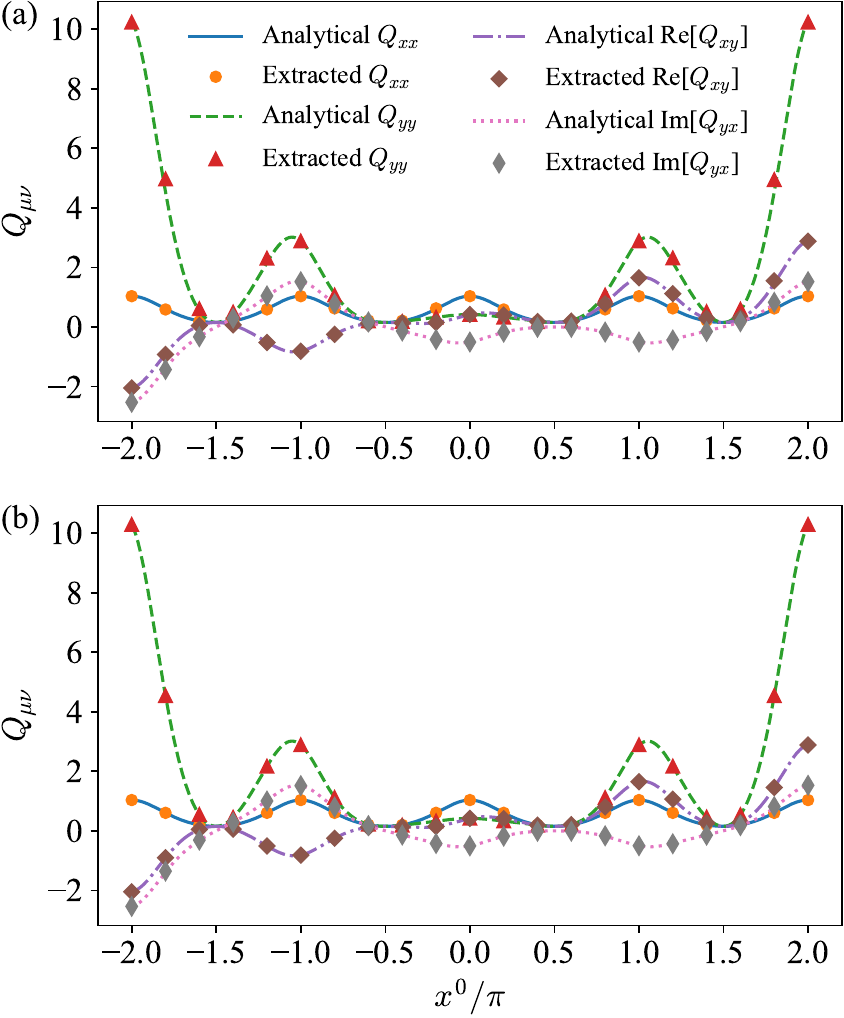}
    \caption{QGT components for model II, extracted using (a) the energy fluctuation scheme and (b) the generalized force scheme.
    Analytical results are shown as lines, while numerical data are represented by markers, following the same style conventions as in Fig.~\ref{fig: pH_spin_QGT}, with subscripts adjusted for the present model.
    The parameters are set to $B/2\pi = 10\,\mathrm{MHz}$ and $y^0 = \pi/2$.
    }
    \label{fig: pH_xy_QGT}
\end{figure}

For $B/2\pi = 10\,\mathrm{MHz}$ and $y^0 = \pi/2$, with $x^0$ sampled at $N = 21$ evenly spaced points over $[-2\pi, 2\pi]$, Fig.~\ref{fig: pH_xy_QGT} demonstrates high-fidelity extraction of $Q_{xx}$, $Q_{yy}$, $\Re[Q_{yx}]$, and $\Im[Q_{yx}]$ using both measurement schemes.

\section{Experimental Implementation} \label{sec: exp}

The proposed schemes for measuring the QGT components in pseudo-Hermitian systems with real spectra are compatible with current quantum devices, such as nitrogen-vacancy (NV) centers~\cite{YWu2019,WQLiu2020,YWu2024,YWu2025} and superconducting qubits~\cite{Dogra2021}.
The implementation proceeds in three main steps. 
Nonunitary evolutions in \textit{step~1} and \textit{step~2} can be realized by embedding them into larger unitary dynamics using the dilation method~\cite{Gunther2008,YWu2019,WQLiu2020,YWu2024,YWu2025}.

\textit{Step 1: Ground-state preparation.}
Initialize the qubit system in a computational basis state, such as $[0,1]^T$ for a single qubit, which corresponds to the ground state of $H(\boldsymbol{\lambda})$ at a reference parameter point [e.g., $(\theta,\phi) = (0,0)$ for model I or $(x,y) = (2 m \pi, 2 n \pi)$ with $m, n \in \mathbb{Z}$ for model II].
Then, adiabatically vary the parameters $\boldsymbol{\lambda}$ to the target point offset by $\Delta \lambda$ along the $\mu$th direction, i.e., $\boldsymbol{\lambda}^0 - \Delta \lambda \mathbf{e}_\mu$, to prepare the right and left ground states $|\phi^R_0(\boldsymbol{\lambda}^0 - \Delta \lambda \mathbf{e}_\mu)\rangle$ and $|\phi^L_0(\boldsymbol{\lambda}^0 - \Delta \lambda \mathbf{e}_\mu)\rangle$.
The variation velocity should satisfy the adiabatic condition $|v| \ll \min_{\boldsymbol{\lambda}} |\Delta E| / |\langle \phi_1^L | \partial_\mu H | \phi_0^R \rangle|$, where $\Delta E$ is the energy gap between the ground and first excited states.

\textit{Step 2: State evolution.}
Evolve the prepared right ground state $|\phi^R_0(\boldsymbol{\lambda}^0 - \Delta \lambda \mathbf{e}_\mu)\rangle$ under the time-dependent Hamiltonian $H[\boldsymbol{\lambda}(t)]$ to obtain $|\psi_\mu(\boldsymbol{\lambda}^0,t_f)\rangle$.
Similarly, evolve the left ground state $|\phi^L_0(\boldsymbol{\lambda}^0 - \Delta \lambda \mathbf{e}_\mu)\rangle$ under $\pm H^\dagger[\boldsymbol{\lambda}(t)]$ to obtain $|\psi_\mu^\pm(\boldsymbol{\lambda}^0,t_f)\rangle$.
In all cases, the parameter trajectory $\boldsymbol{\lambda}(t)$ follows Eq.~\eqref{eq: lambda evolution}.

\textit{Step 3: Generalized expectation value measurement.}
Measure the generalized expectation value, including both real and imaginary parts, using the quantum circuit described in Refs.~\cite{ZHHuang2023,Wagner2024,Chiribella2024} (see Appendix~\ref{sec: quantum circuit} for details).
For the energy fluctuation scheme, use $|\psi_1\rangle = |\psi_\mu^+(\boldsymbol{\lambda}^0,t_f)\rangle$ and $|\psi_2\rangle = |\psi_\nu(\boldsymbol{\lambda}^0,t_f)\rangle$ as the input states, with $A = \Delta^2 H \equiv (H-\langle\!\langle H \rangle\!\rangle)^2$ as the operator.
For the generalized force scheme, set $A = f_\mu \equiv -\partial_\mu H$ and measure the Berry curvature using $|\psi_1\rangle = |\psi_\nu^+(\boldsymbol{\lambda}^0,t_f)\rangle$, $|\psi_2\rangle = |\psi_\nu(\boldsymbol{\lambda}^0,t_f)\rangle$, and the quantum metric using $|\psi_1\rangle = |\psi_\nu^-(\boldsymbol{\lambda}^0,t_f)\rangle$, $|\psi_2\rangle = |\psi_\nu(\boldsymbol{\lambda}^0,t_f)\rangle$.
The constant in Eq.~\eqref{eq: constant} is obtained from the ground-state generalized expectation value at $\boldsymbol{\lambda}^0$ using $|\psi_1\rangle = |\phi^L_0(\boldsymbol{\lambda}^0)\rangle$ and $|\psi_2\rangle = |\phi^R_0(\boldsymbol{\lambda}^0)\rangle$, which can be prepared by adiabatically ramping $\boldsymbol{\lambda}$ from the reference parameter point to $\boldsymbol{\lambda}^0$.

\textit{Error analysis.}
Each of the three steps introduces distinct sources of error that affect the accuracy of the measured QGT components.
Decoherence errors primarily arise from the long adiabatic evolution time in \textit{step~1} and are inversely proportional to the minimum instantaneous energy gap $\min_{\boldsymbol{\lambda}} |\Delta E|$ encountered along the adiabatic path.
In \textit{step~2}, systematic errors in the measured QGT components occur when the ramp velocity $v$ exceeds the linear response regime, resulting in deviations from theoretical values. The sensitivity to $v$ differs between the two measurement schemes.
\textit{Step~3} introduces errors associated with circuit depth and gate fidelities, mainly from the implementation of the controlled-\textsc{swap} gate in the quantum circuit.

For the two models discussed in Sec.~\ref{sec: examples}, the total adiabatic ground-state preparation time is approximately $7.5\,\mu\mathrm{s}$ (see Appendix~\ref{sec: error}), which is well within the coherence times of current superconducting quantum processors~\cite{FTJin2025}. 
The nonadiabatic state evolution time is $t_f = \pi\,\mu\mathrm{s}$, also below current coherence times, and can be further reduced by decreasing $\Delta\lambda$.
The controlled-\textsc{swap} gate required for the quantum circuit reduces to a Fredkin gate, which can be decomposed into at least five two-qubit gates and several single-qubit gates~\cite{HFChau1995,Smolin1996,NKYu2015}. 
With current two-qubit gate fidelities exceeding 99\%~\cite{Google2025,IBM2025}, the overall circuit fidelity can remain above 95\%, rendering gate errors negligible in \textit{step~3}.

\begin{figure}
    \includegraphics[width=\columnwidth]{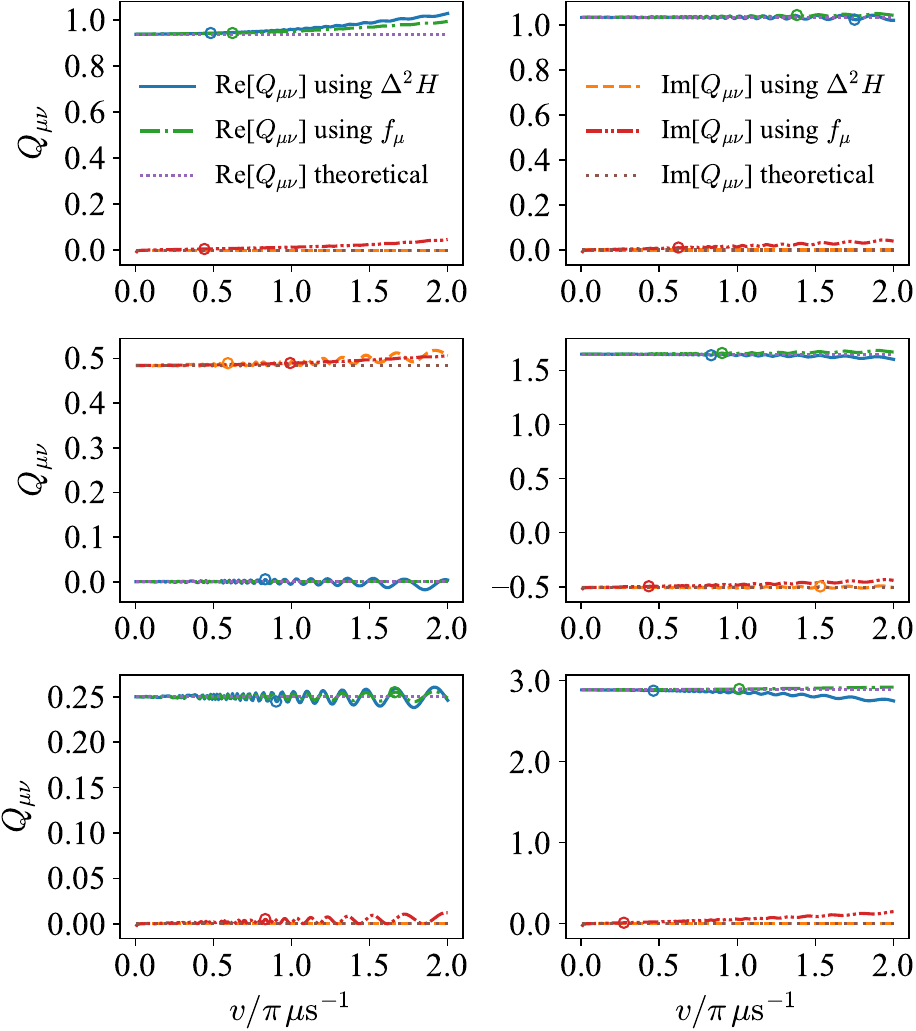}
    \caption{Dependence of measured QGT components on ramp velocity $v$ for both measurement schemes.
        Left panels: $Q_{\theta\theta}$, $Q_{\phi\theta}$, and $Q_{\phi\phi}$ (top to bottom) for model I at $(\theta^0, \phi^0) = (\pi/2, 0)$ with $\Omega_1/2\pi = 10\,\mathrm{MHz}$, $\Delta_1/2\pi = 15\,\mathrm{MHz}$, and $\Delta_2 = 0$.
        Right panels: $Q_{xx}$, $Q_{yx}$, and $Q_{yy}$ (top to bottom) for model II at $(x^0, y^0) = (\pi, \pi/2)$ with $B/2\pi = 10\,\mathrm{MHz}$.
        Numerical results from the energy fluctuation scheme ($\Delta^2 H$) and the generalized force scheme ($f_\mu$) are distinguished by line styles as indicated in the legend.
        Horizontal dashed lines denote theoretical values.
        Circles indicate the critical ramp velocities at which the absolute deviation between numerically extracted and theoretical values first exceeds $0.005$ for model I or $0.01$ for model II (see Table~\ref{table: critical velocities} in Appendix~\ref{sec: error}).}
    \label{fig: pH_robust}
\end{figure}

Figure~\ref{fig: pH_robust} illustrates how the measured QGT components depend on the final ramp velocity $v$ for both measurement schemes, using $(\theta^0, \phi^0) = (\pi/2, 0)$ for model I and $(x^0, y^0) = (\pi, \pi/2)$ for model II as representative parameter points.
High-fidelity QGT extraction is achieved when $|v| \lesssim \min_{\boldsymbol{\lambda}} |\Delta E| / 40$, providing a practical criterion for remaining within the linear response regime.
As $v$ increases, the energy fluctuation scheme ($\Delta^2 H$) is more robust for the imaginary part of the QGT (except for $Q_{\phi\theta}$ in model I), while the generalized force scheme ($f_\mu$) provides greater accuracy for the real part (except for $Q_{xx}$ in model II).
These observations offer guidance for selecting ramp velocities and measurement schemes to optimize experimental accuracy.
Further details and quantitative analysis are provided in Appendix~\ref{sec: error}.

\section{Conclusion and Discussion} \label{sec: conclusion}

We establish two independent schemes for directly measuring all components of the QGT, defined using both left and right eigenstates, in pseudo-Hermitian systems with real spectra.
The first scheme employs the energy fluctuation operator $\Delta^2 H \equiv (H-\langle\!\langle H \rangle\!\rangle)^2$, while the second utilizes the generalized force operator $f_\mu \equiv -\partial_\mu H$, with each providing a self-contained pathway to determine the complete QGT.
Numerical verification on two $q$-deformed two-band models (Figs.~\ref{fig: pH_spin_QGT} and \ref{fig: pH_xy_QGT}) demonstrates high-fidelity agreement with analytical predictions for both schemes when ramp velocities are chosen within the linear response regime of adiabatic perturbation theory.
Furthermore, the Chern number phase diagram of the first model (Fig.~\ref{fig: pH_spin_Chern}) confirms that both schemes can reconstruct topological invariants, a task previously achievable only through conventional expectation values of generalized force operators in Hermitian systems.
When ramp velocities exceed the linear response regime (see Fig.~\ref{fig: pH_robust}), each scheme exhibits complementary strengths: the energy fluctuation approach better preserves the fidelity of the imaginary part of the QGT (Berry curvature), whereas the generalized force method more reliably captures the real part (quantum metric).

Beyond two-band systems, both measurement schemes are expected to extend to many-body systems with real spectra, such as the $\mathcal{PT}$-symmetric regime of the non-Hermitian transverse-field Ising model~\cite{GYSun2022}. Achieving this generalization requires the capability to prepare relevant eigenstates, implement nonadiabatic evolutions, and realize $n$-qubit controlled-\textsc{swap} gates in experiments.

From a quantum measurement perspective, the generalized force operator requires fewer measurements than the energy fluctuation operator due to vanishing partial derivatives for certain measurement bases, such as the $\sigma_z$-component-free $f_\phi$ operator in Eqs.~\eqref{eq: generalized force operators spin}.
Consequently, the generalized force scheme provides superior measurement efficiency that scales with system dimension. 
Notably, both schemes remain applicable to Hermitian systems, establishing a unified experimental framework for directly measuring the QGT across Hermitian and non-Hermitian regimes with nonadiabatic response~\cite{Kolodrubetz2017}.
We expect that this framework can be extended to directly measure other physical quantities in the LR formalism via dynamic processes.

\begin{acknowledgments}
We thank S.-L. Zhu, Y.-Q. Zhu, P. He, and Z. Ma for helpful discussions.
This work was supported by the National Key Research and Development Program of China (Grant No.~2022YFA1405304), the Guangdong Basic and Applied Basic Research Foundation (Grant No.~2024A1515010188), the Guangdong Provincial Quantum Science Strategic Initiative (Grant No.~GDZX2204003), and the Startup Fund of South China Normal University.
\end{acknowledgments}

\section*{Data availability}

The data that support the findings of this article are openly available~\cite{dataset}, embargo periods may apply.

\appendix

\section{Adiabatic Perturbation Theory for Pseudo-Hermitian Systems} \label{sec: adiabatic perturbation}

In this Appendix, we overview the adiabatic perturbation theory for pseudo-Hermitian systems with real spectra~\cite{Rigolin2008,DeGrandi2010b,Nenciu1992,QZhang2019}, which describes the time evolution of quantum states under slowly varying parameters.

Consider a quantum state $|\psi[\lambda(t)]\rangle$ evolving under a time-dependent pseudo-Hermitian Hamiltonian $H[\lambda(t)]$ with real spectra, where $\lambda(t)$ is a slowly varying parameter with nonzero velocity $v_{\lambda}(t) \equiv \dot{\lambda}(t) \neq 0$.
The time-dependent Schr\"odinger equation reads as
\begin{equation} \label{eq: time-dependent Schrodinger equation}
    i \partial_t |\psi[\lambda(t)]\rangle = H[\lambda(t)] |\psi[\lambda(t)]\rangle,
\end{equation}
where the state $|\psi[\lambda(t)]\rangle$ can be expanded in terms of the instantaneous right eigenstates $|\phi^R_n[\lambda(t)]\rangle$ of $H[\lambda(t)]$ as
\begin{equation} \label{eq: expansion of psi}
    | \psi [\lambda(t)] \rangle = \sum_n a_n(t) | \phi^R_n[\lambda(t)] \rangle.
\end{equation}
The time-dependent coefficients $a_n(t)$ characterize the system's nonadiabatic response and are determined as follows.

During the evolution governed by $H[\lambda(t)]$, both the dynamical phases $\Theta_n(t) = \int_{0}^t \d t' E_n[\lambda(t')]$ and the geometric phases $\Phi_n(t) = -i \int_{0}^t \d t' \langle \phi^L_n[\lambda(t')] | \partial_{t'} \phi^R_n[\lambda(t')] \rangle$ are real, due to the real spectra and the relation between the left and right eigenstates $| \phi^L_n [\lambda(t)] \rangle = \eta | \phi^R_n [\lambda(t)] \rangle$ implied by the pseudo-Hermiticity condition $H[\lambda(t)]^\dag = \eta H[\lambda(t)] \eta^{-1}$, where $\eta$ is a time-independent Hermitian invertible operator.
This enables both phases to be factored out as a gauge transformation, allowing the expansion coefficients to be expressed as
\begin{equation} \label{eq: gauge transformation}
    a_n(t) = \alpha_n(t) \exp[-i \Theta_n(t)] \exp[-i \Phi_n(t)].
\end{equation}
Substituting Eqs.~\eqref{eq: expansion of psi} and \eqref{eq: gauge transformation} into Eq.~\eqref{eq: time-dependent Schrodinger equation} yields the equation of motion for $\alpha_n(t)$:
\begin{align} \label{eq: EoM alpha_n}
    \partial_t \alpha_n(t) &= - \sum_{m \ne n} \alpha_m(t) \langle \phi^L_n[\lambda(t)] | \partial_t \phi^R_m[\lambda(t)] \rangle \notag \\
     &\qquad \qquad \times \exp[-i \Theta_{mn}(t)] \exp[-i \Phi_{mn}(t)],
\end{align}
where $\Theta_{mn}(t) = \Theta_m(t) - \Theta_n(t)$ and $\Phi_{mn}(t) = \Phi_m(t) - \Phi_n(t)$.

Suppose the system is initially prepared in the instantaneous right ground state $|\phi^R_0[\lambda(0)]\rangle$ at $t=0$, i.e., $\alpha_n(0) = \delta_{n0}$, and the parameter $\lambda(t)$ varies sufficiently slowly such that the system remains close to $|\phi^R_0[\lambda(t)]\rangle$ throughout the evolution, i.e., $\alpha_0(t) \approx 1$ and $|\alpha_{n \neq 0}(t)| \ll 1$.
Integrating Eq.~\eqref{eq: EoM alpha_n} from $0$ to $t$ yields
\begin{align} \label{eq: alpha_n t}
    \alpha_{n \neq 0}(t) &= - \int_{0}^t \mathrm{d} t' \langle \phi^L_n[\lambda(t')] | \partial_{t'} \phi^R_0[\lambda(t')] \rangle \notag \\ 
    &\qquad \times \exp[-i \Theta_{0n}(t')] \exp[-i \Phi_{0n}(t')].
\end{align}
Applying integration by parts and retaining only the leading-order term in $\partial_{t'}$, we obtain
\begin{align} \label{eq: alpha_n t simplified by parts}
    \alpha_{n \neq 0}(t) &= i \frac{\langle \phi^L_n[\lambda(t')] | \partial_{t'} \phi^R_0[\lambda(t')] \rangle}{E_n[\lambda(t')] - E_0[\lambda(t')]} \notag \\ 
    &\qquad \times \exp[-i \Theta_{0n}(t')] \exp[-i \Phi_{0n}(t')] \bigg|_0^t + \cdots,
\end{align}
where the ellipsis denotes higher-order corrections in $\partial_{t'}$.
If the initial energy gap is sufficiently large, or if the parameter evolution is designed such that the initial evolution is adiabatic, the contribution from the lower limit vanishes.
Therefore, by substituting Eq.~\eqref{eq: alpha_n t simplified by parts} into Eq.~\eqref{eq: gauge transformation} and neglecting the global phase factor, the expansion coefficients $a_n(t)$ simplify to
\begin{equation} \label{eq: a_n t final}
    a_0(t) \approx 1, \quad a_{n \neq 0}(t) = i \frac{\langle \phi^L_n[\lambda(t)] | \partial_{t} \phi^R_0[\lambda(t)] \rangle}{E_n[\lambda(t)] - E_0[\lambda(t)]} + \cdots,
\end{equation}
where the ellipsis again denotes higher-order corrections in $\partial_{t}$.

For a monotonically evolving parameter $\lambda(t)$ as in Eq.~\eqref{eq: lambda evolution} of the main text, the time derivative can be rewritten as $\partial_t = v_{\lambda}(t) \partial_{\lambda}$, where $v_{\lambda}(t) = \dot{\lambda}(t)$ is the instantaneous ramping velocity.
This substitution allows Eqs.~\eqref{eq: alpha_n t}--\eqref{eq: a_n t final} to be reformulated with $\lambda$ as the independent variable.
If the system is initially prepared in $|\phi^R_0[\lambda(0)]\rangle$, the time-evolved state at time $t$ is
\begin{equation}
    | \psi[\lambda(t)] \rangle = | \phi^R_0[\lambda(t)] \rangle + \sum_{n \neq 0} a_n(t) | \phi^R_n[\lambda(t)] \rangle,
\end{equation}
where the expansion coefficients, describing the linear response to the parameter ramp $\lambda(t)$, are
\begin{equation} \label{eq: a_n t final v}
    a_n(t) = i v_{\lambda}(t) \frac{\langle \phi^L_n[\lambda(t)] | \partial_{\lambda} \phi^R_0[\lambda(t)] \rangle}{E_n[\lambda(t)] - E_0[\lambda(t)]} + \cdots,
\end{equation}
with the ellipsis denoting higher-order corrections in $v_{\lambda}(t)$, which are negligible for sufficiently slow variation of $\lambda(t)$.

\section{Quantum Circuit for Measuring Generalized Expectation Values} \label{sec: quantum circuit}

In this Appendix, we briefly introduce the quantum circuit for measuring the real and imaginary parts of generalized expectation values, as proposed in Ref.~\cite{ZHHuang2023}.

Our goal is to measure the generalized expectation value $\langle\!\langle A \rangle\!\rangle = \frac{\langle \psi_1 | A | \psi_2 \rangle}{\langle \psi_1 | \psi_2 \rangle}$ for arbitrary states $|\psi_1\rangle$, $|\psi_2\rangle$, and operator $A$. Since any operator $A$ can be decomposed into Hermitian and anti-Hermitian parts, $A = \frac{A + A^\dag}{2} + i \frac{A - A^\dag}{2i}$, it suffices to consider the measurement of generalized expectation values $\langle\!\langle O \rangle\!\rangle = \frac{\langle \psi_1 | O | \psi_2 \rangle}{\langle \psi_1 | \psi_2 \rangle}$ for experimentally accessible Hermitian observables $O$.

\begin{figure}
    \includegraphics[width=0.95\columnwidth]{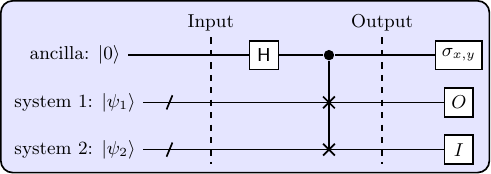}
    \caption{Quantum circuit for measuring the generalized expectation value $\langle\!\langle O \rangle\!\rangle = \frac{\langle \psi_1 | O | \psi_2 \rangle}{\langle \psi_1 | \psi_2 \rangle}$ via a controlled-\textsc{swap} operation.
    The ancilla qubit (top line) controls a \textsc{swap} operation between two $n$-qubit registers storing arbitrary states $|\psi_1\rangle$ and $|\psi_2\rangle$. 
    Measurement of $\sigma_{x(y)}$ on the ancilla yields the real (imaginary) part of $\langle \psi_1 | O | \psi_2 \rangle \langle \psi_2 | \psi_1 \rangle$, corresponding to the numerator of the generalized expectation value.
    The denominator $|\langle \psi_1 | \psi_2 \rangle|^2$ is obtained by setting $O$ as the identity operator $I$ and measuring $\sigma_x$ on the ancilla.
    }
    \label{fig: g_swap test}
\end{figure}

As illustrated in Fig.~\ref{fig: g_swap test}, the input state of the quantum circuit consists of three parts: an ancilla qubit initialized in $|0\rangle$, and two $n$-qubit registers prepared in arbitrary states $|\psi_1\rangle$ and $|\psi_2\rangle$,
\begin{equation} \label{eq: input state}
    |\Psi_\mathrm{input}\rangle = |0\rangle \otimes |\psi_1\rangle \otimes |\psi_2\rangle.
\end{equation}
A Hadamard gate $\mathsf{H}$ is applied to the ancilla, followed by a controlled-\textsc{swap} operation between the two registers, controlled by the ancilla. The output state is
\begin{equation} \label{eq: output state}
    |\Psi_\mathrm{output}\rangle = \frac{1}{\sqrt{2}} \left( |0\rangle \otimes |\psi_1\rangle \otimes |\psi_2\rangle + |1\rangle \otimes |\psi_2\rangle \otimes |\psi_1\rangle \right).
\end{equation}

The real (imaginary) part of the generalized expectation value $\langle\!\langle O \rangle\!\rangle = \frac{\langle \psi_1 | O | \psi_2 \rangle}{\langle \psi_1 | \psi_2 \rangle}$ is then obtained as the ratio of the expectation values of the operators $\sigma_{x(y)} \otimes O \otimes I$ and $\sigma_x \otimes I \otimes I$ on the output state $|\Psi_\mathrm{output}\rangle$:
\begin{equation} \label{eq: g_expectation}
    \Re\, (\Im) \left[ \langle\!\langle O \rangle\!\rangle \right] = \frac{\langle\Psi_\mathrm{output}|\sigma_{x(y)} \otimes O \otimes I |\Psi_\mathrm{output}\rangle}{\langle\Psi_\mathrm{output}|\sigma_x \otimes I \otimes I |\Psi_\mathrm{output}\rangle},
\end{equation}
where the Pauli operators $\sigma_{x,y}$ act on the ancilla qubit, and $I$ is the identity operator on the $n$-qubit register.
The numerator and denominator in Eq.~\eqref{eq: g_expectation} are evaluated as
\begin{align}
    &\langle\Psi_\mathrm{output}|\sigma_{x(y)} \otimes O \otimes I |\Psi_\mathrm{output}\rangle \notag \\
    & = \Re\, (\Im) \left[ \langle \psi_1 | O | \psi_2 \rangle \langle \psi_2 | \psi_1 \rangle \right], \\
    &\langle\Psi_\mathrm{output}|\sigma_x \otimes I \otimes I |\Psi_\mathrm{output}\rangle
    = |\langle \psi_1 | \psi_2 \rangle|^2,
\end{align}
where the expectation values are obtained by repeatedly preparing the input state [Eq.~\eqref{eq: input state}], applying the quantum gates to generate the output state [Eq.~\eqref{eq: output state}], and performing projective measurements corresponding to the relevant operators.

Therefore, this quantum circuit (Fig.~\ref{fig: g_swap test}) enables direct measurement of both the real and imaginary parts of generalized expectation values for arbitrary pairs of quantum states and Hermitian operators.

\section{Derivation of Eqs.~\eqref{eq: energy fluctuation}, \eqref{eq: generalized force im}, and \eqref{eq: generalized force re}} \label{sec: derivation}

In this Appendix, we provide detailed derivations for Eqs.~\eqref{eq: energy fluctuation}, \eqref{eq: generalized force im}, and \eqref{eq: generalized force re} in the main text.

Based on the adiabatic perturbation theory outlined in Appendix~\ref{sec: adiabatic perturbation}, the time-evolved states $|\psi_\mu(\boldsymbol{\lambda}^0, t_f)\rangle$ and $|\psi_\mu^\pm(\boldsymbol{\lambda}^0, t_f)\rangle$ can be written as
\begin{align}
    |\psi_\mu(\boldsymbol{\lambda}^0, t_f)\rangle &= |\phi^R_0(\boldsymbol{\lambda}^0)\rangle + \sum_{n \neq 0} a^R_{n,\mu}(\boldsymbol{\lambda}^0) |\phi^R_n(\boldsymbol{\lambda}^0)\rangle, \label{psi1 tar} \\
    |\psi_\mu^\pm(\boldsymbol{\lambda}^0, t_f)\rangle &= |\phi^L_0(\boldsymbol{\lambda}^0)\rangle \pm \sum_{n \neq 0} a^L_{n,\mu}(\boldsymbol{\lambda}^0) |\phi^L_n(\boldsymbol{\lambda}^0)\rangle, \label{psi2 3 tar}
\end{align}
where the expansion coefficients are
\begin{align}
    a^R_{n,\mu}(\boldsymbol{\lambda}^0) &\approx i v \frac{\langle \phi^L_n(\boldsymbol{\lambda}^0) | \partial_\mu \phi^R_0(\boldsymbol{\lambda}^0) \rangle}{E_n(\boldsymbol{\lambda}^0) - E_0(\boldsymbol{\lambda}^0)}, \label{anR} \\
    a^L_{n,\mu}(\boldsymbol{\lambda}^0) &\approx i v \frac{\langle \phi^R_n(\boldsymbol{\lambda}^0) | \partial_\mu \phi^L_0(\boldsymbol{\lambda}^0) \rangle}{E_n(\boldsymbol{\lambda}^0) - E_0(\boldsymbol{\lambda}^0)}. \label{anL}
\end{align}
Here, $\partial_\mu \equiv \partial / \partial \lambda_\mu$, $v \equiv v_\mu(t_f) = \dot{\lambda}_\mu(t_f)$, and higher-order corrections in $v$ are neglected.
The normalization factors for $|\psi_\mu(\boldsymbol{\lambda}^0, t_f)\rangle$ and $|\psi_\mu^\pm(\boldsymbol{\lambda}^0, t_f)\rangle$ are omitted here, as they appear in both the numerators and denominators of the generalized expectation values and thus cancel out in the final results.

We now evaluate the generalized expectation values of the energy fluctuation operator $\Delta^2 H(\boldsymbol{\lambda}^0)$ and the generalized force operator $f_\mu(\boldsymbol{\lambda}^0)$ by substituting Eqs.~\eqref{psi1 tar}--\eqref{anL} into the left-hand sides of Eqs.~\eqref{eq: energy fluctuation}, \eqref{eq: generalized force im}, and \eqref{eq: generalized force re}. For clarity, we adopt the biorthogonal relation $\langle \phi^L_m(\boldsymbol{\lambda}^0) | \phi^R_n(\boldsymbol{\lambda}^0) \rangle = \delta_{mn}$ and the completeness relation $\sum_n |\phi_n^R (\boldsymbol{\lambda}) \rangle \langle \phi_n^L (\boldsymbol{\lambda})| = 1$ in the following derivations.
With these, the QGT in the LR formalism [Eq.~\eqref{eq: biQGT}] simplifies to
\begin{equation}
    Q^n_{\mu\nu} = \sum_{m \neq n} \langle \partial_\mu \phi_n^L | \phi_m^R \rangle \langle \phi_m^L | \partial_\nu \phi_n^R \rangle. \label{eq: QGT simplified}
\end{equation}

The numerator of the generalized expectation value of the energy fluctuation operator $\Delta^2 H(\boldsymbol{\lambda}^0)$ on the left-hand side of Eq.~\eqref{eq: energy fluctuation} can be simplified as follows:
\begin{align}
    &\langle \psi_\mu^+(\boldsymbol{\lambda}^0,t_f) | \Delta^2 H(\boldsymbol{\lambda}^0) | \psi_\nu(\boldsymbol{\lambda}^0,t_f) \rangle \notag \\
    &\approx \langle \psi_\mu^+(\boldsymbol{\lambda}^0,t_f) | [H (\boldsymbol{\lambda}^0) -E_0 (\boldsymbol{\lambda}^0)]^2 | \psi_\nu(\boldsymbol{\lambda}^0,t_f) \rangle \notag \\
    &\approx v^2 Q_{\mu\nu}^0(\boldsymbol{\lambda}^0), \label{eq: eq6 numerator}
\end{align}
where $[H (\boldsymbol{\lambda}^0) -E_0 (\boldsymbol{\lambda}^0)]^2$ is referred to as the square of the absolute Hamiltonian, which differs from the energy fluctuation $\Delta^2 H (\boldsymbol{\lambda}^0) \equiv [H (\boldsymbol{\lambda}^0) - \langle\!\langle H (\boldsymbol{\lambda}^0) \rangle\!\rangle]^2$ only by higher-order negligible terms~\cite{XYZhang2023}. Specifically,
\begin{align}
    &\langle\!\langle [H(\boldsymbol{\lambda}^0)-E_0(\boldsymbol{\lambda}^0)]^2 \rangle\!\rangle - \langle\!\langle [H (\boldsymbol{\lambda}^0) - \langle\!\langle H (\boldsymbol{\lambda}^0) \rangle\!\rangle]^2 \rangle\!\rangle \notag \\
    &= [\langle\!\langle H(\boldsymbol{\lambda}^0) \rangle\!\rangle - E_0(\boldsymbol{\lambda}^0)]^2 \notag \\
    &= \left[\sum_{n \neq 0} a^{L*}_{n,\mu}(\boldsymbol{\lambda}^0) a^R_{n,\nu}(\boldsymbol{\lambda}^0) E_n(\boldsymbol{\lambda}^0)\right]^2 \sim O (v^4),
\end{align}
where $\langle\!\langle \cdot \rangle\!\rangle$ denotes $\frac{\langle \psi_\mu^+(\boldsymbol{\lambda}^0,t_f) | \cdot | \psi_\nu(\boldsymbol{\lambda}^0,t_f) \rangle}{\langle \psi_\mu^+(\boldsymbol{\lambda}^0,t_f) | \psi_\nu(\boldsymbol{\lambda}^0,t_f) \rangle}$.

Next, we calculate the numerators of the generalized expectation values for the generalized force operator $f_\mu(\boldsymbol{\lambda}^0) \equiv -\partial_\mu H(\boldsymbol{\lambda}^0)$, which appear on the left-hand sides of Eqs.~\eqref{eq: generalized force im} and \eqref{eq: generalized force re}:
\begin{align}
    &\langle \psi_\nu^\pm(\boldsymbol{\lambda}^0,t_f) | f_\mu(\boldsymbol{\lambda}^0) | \psi_\nu(\boldsymbol{\lambda}^0,t_f) \rangle \notag \\
    &\approx \langle \phi^L_0(\boldsymbol{\lambda}^0)| f_\mu(\boldsymbol{\lambda}^0) | \phi^R_0(\boldsymbol{\lambda}^0) \rangle + iv [Q_{\mu\nu}^0(\boldsymbol{\lambda}^0) \mp Q_{\nu\mu}^0(\boldsymbol{\lambda}^0)] \notag \\
    &= \langle \phi^L_0(\boldsymbol{\lambda}^0)| f_\mu(\boldsymbol{\lambda}^0) | \phi^R_0(\boldsymbol{\lambda}^0) \rangle + 2 v \left(-\Im,\, i\,\Re\right)[Q_{\mu\nu}^0(\boldsymbol{\lambda}^0)], \label{eq: eq8 numerator}
\end{align}
where we use the following relations for $m \neq n$:
\begin{align}
    \frac{\langle \phi^L_m(\boldsymbol{\lambda}^0)| f_\mu(\boldsymbol{\lambda}^0) |\phi^R_n(\boldsymbol{\lambda}^0)\rangle}{E_m(\boldsymbol{\lambda}^0) - E_n(\boldsymbol{\lambda}^0)} &= \langle \phi^L_m(\boldsymbol{\lambda}^0) | \partial_{\mu} \phi^R_n(\boldsymbol{\lambda}^0)\rangle, \notag\\
    &= - \langle \partial_{\mu} \phi^L_m(\boldsymbol{\lambda}^0) | \phi^R_n(\boldsymbol{\lambda}^0)\rangle.
\end{align}

The denominators of the generalized expectation values in Eqs.~\eqref{eq: energy fluctuation}, \eqref{eq: generalized force im}, and \eqref{eq: generalized force re} serve as normalization factors and, to leading order, reduce to $\langle \phi_0^L(\boldsymbol{\lambda}^0) | \phi_0^R(\boldsymbol{\lambda}^0) \rangle = 1$.
By combining Eqs.~\eqref{eq: eq6 numerator} and \eqref{eq: eq8 numerator} with this normalization factor, we recover the right-hand sides of Eqs.~\eqref{eq: energy fluctuation}, \eqref{eq: generalized force im}, and \eqref{eq: generalized force re} in the main text.

\section{Mapping Model I to a Haldane-type Honeycomb Lattice with Nonreciprocal Hopping} \label{sec: mapping}

In this Appendix, we present a Haldane-type honeycomb lattice model with nonreciprocal hopping as a concrete example of a pseudo-Hermitian system with real spectra in real space. We demonstrate how this model connects to model I in momentum space, establishing that both share identical topological properties and clarifying the physical origin of the $q$-deformed Pauli matrices. This mapping is analogous to the correspondence between a single-qubit Hamiltonian and the Haldane model~\cite{Haldane1988}, as discussed in Refs.~\cite{Schroer2014,Roushan2014}.

The Hamiltonian for the Haldane-type honeycomb lattice model is
\begin{equation} \label{eq: honeycomb}
    H = H_{\mathrm{NN}} + H_{\mathrm{NNN}} + H_{\mathrm{onsite}},
\end{equation}
where
\begin{align}
    H_{\mathrm{NN}} &= \sum_{\langle i,j \rangle} \left( a t_1\, c_i^\dag d_j + b t_1\, d_j^\dag c_i \right),\\
    H_{\mathrm{NNN}} &= \sum_{\langle\langle i,j \rangle\rangle} \left(q^{-1} t_2\, e^{i \nu_{ij} \phi} c_i^\dag c_j 
    + q t_2\, e^{i \nu_{ij} \phi} d_i^\dag d_j \right),\\
    H_{\mathrm{onsite}} &= \sum_{i} \left( q^{-1} M\, c_i^\dag c_i - q M\, d_i^\dag d_i \right).
\end{align}
Here, $c_i^\dag$ ($d_i^\dag$) and $c_i$ ($d_i$) are creation and annihilation operators for sublattices A (B) at site $i$; $a t_1$ and $b t_1$ are nonreciprocal nearest-neighbor (NN) hopping amplitudes; $q^{-1} t_2$ and $q t_2$ are next-nearest-neighbor (NNN) hopping amplitudes on sublattices A and B, respectively; $\nu_{ij}$ indicates the direction of the NNN hopping ($+1$ for counterclockwise, $-1$ for clockwise); $\phi$ is the NNN hopping phase; and $q^{-1} M$ and $-q M$ are the onsite potentials for sublattices A and B, respectively.
The parameters $t_1$, $t_2$, $M$, and $\phi$ are real, while $a = \sqrt{(1 + q^2)/2}$, $b = \sqrt{(1 + q^{-2})/2}$, and $q > 0$ are the $q$-deformed parameters as defined in Sec.~\ref{sec: examples}.
This model can be realized in experimental platforms supporting nonreciprocal coupling, such as electric circuits~\cite{YLong2022,RJLi2025}.

The effective Hamiltonians near the two inequivalent Dirac points $\mathbf{k}_\pm^0$ in the Brillouin zone are
\begin{equation}
    H_\pm = \mp v_F k^x_\pm\, \tilde\sigma_x + v_F k^y_\pm\, \tilde\sigma_y + m_\pm\, \tilde\sigma_z,
\end{equation}
where $v_F = \frac{3}{2} t_1$ is the Fermi velocity, $\mathbf{k}_\pm = \mathbf{k} - \mathbf{k}_\pm^0$ is the momentum measured from each Dirac point, and $m_\pm = M \pm 3\sqrt{3}\, t_2 \sin\phi$ are the effective mass terms.
The $q$-deformed Pauli matrices $\tilde\sigma_{x,y,z}$ naturally arise from the nonreciprocal hopping.
Analogous to the Haldane model, the system exhibits a nontrivial topological phase with nonzero Chern number when the mass terms at the two Dirac points have opposite signs, i.e., $|M| < |3\sqrt{3}\, t_2 \sin\phi|$.

To establish the connection to model I, we expand model I near the poles $\theta = 0$ and $\theta = \pi$, obtaining
\begin{equation}
    H_\pm^\mathrm{I} = \frac{1}{2} \Omega_1 \sin \theta \cos \phi\, \tilde\sigma_x + \frac{1}{2} \Omega_1 \sin \theta \sin \phi\, \tilde\sigma_y + m_\pm^\mathrm{I}\, \tilde\sigma_z,
\end{equation}
where the effective mass terms are $m_\pm^\mathrm{I} = (\Delta_2 \pm \Delta_1)/2$.
The criterion for a topologically nontrivial phase in model I is that the mass terms at the two poles have opposite signs, i.e., $|\Delta_2| < |\Delta_1|$, which matches the condition for the honeycomb lattice model.

By identifying $\Delta_2 \leftrightarrow 2 M$ and $\Delta_1 \leftrightarrow 6\sqrt{3}\, t_2 \sin\phi$, the poles of model I correspond to the two inequivalent Dirac points $\mathbf{k}_\pm^0$ of the honeycomb lattice model with nonreciprocal hopping.
Consequently, integrating the Berry curvature over the parameter space $(\theta, \phi)$ in model I is equivalent to integrating over the Brillouin zone $(k_x, k_y)$ in the lattice model, resulting in identical Chern numbers.
This establishes the topological equivalence between the two models.

\section{Detailed Error Analysis of Measurement Schemes for Models I and II} \label{sec: error}

In this Appendix, we provide a detailed analysis of the maximum total adiabatic evolution time ($7.5\,\mathrm{\mu s}$) and establish a practical criterion for the linear response regime of the ramp velocity $v$ in our measurement schemes. Specifically, we find that choosing the ramp velocity to satisfy $|v| \lesssim \min_{\boldsymbol{\lambda}} |\Delta E| / 40$ ensures high-fidelity extraction of QGT components for both models.

We first estimate the maximum total adiabatic evolution time required for ground-state preparation.
For representative parameters in model I ($\Omega_1/2\pi = 10\,\mathrm{MHz}$, $\Delta_1/2\pi = 15\,\mathrm{MHz}$, $\Delta_2 = 0$) and model II ($B/2\pi = 10\,\mathrm{MHz}$), the minimum instantaneous energy gap is $\min_{\boldsymbol{\lambda}} |\Delta E|/2\pi \sim 10\,\mathrm{MHz}$.
The largest distance traversed in parameter space during ground-state preparation is $3\pi/2$.
Choosing a variation velocity $|v| = 0.1 \times 2\pi\,\mathrm{\mu s}^{-1}$, which corresponds to $10^{-2} \times \min_{\boldsymbol{\lambda}} |\Delta E|$, yields a maximum total adiabatic evolution time of approximately $7.5\,\mathrm{\mu s}$.

Next, we establish a practical criterion for the linear response regime of the final ramp velocity $v$ in our measurement schemes.
Figure~\ref{fig: pH_robust} in the main text demonstrates that the measured QGT components remain highly accurate for ramp velocities below certain critical values.
We define the critical ramp velocity as the value at which the absolute deviation between the numerically extracted and theoretical QGT component first exceeds a specified threshold: $0.005$ for model I and $0.01$ for model II.
The critical velocities for each QGT component and measurement scheme, listed in Table~\ref{table: critical velocities}, are typically greater than $0.5 \pi\,\mathrm{\mu s}^{-1}$.
Given that $\min_{\boldsymbol{\lambda}} |\Delta E| \sim 10 \times 2\pi\,\mathrm{\mu s}^{-1}$ for both models, high-fidelity QGT measurements are achieved when $|v| \lesssim \min_{\boldsymbol{\lambda}} |\Delta E| / 40$.
This provides a practical guideline for selecting ramp velocities that ensure the system remains within the linear response regime, where higher-order nonadiabatic corrections are negligible.

\begin{table}
\caption{Critical ramp velocities (in $\pi\,\mathrm{\mu s}^{-1}$) for the real and imaginary parts of QGT components in models I and II, extracted using the energy fluctuation ($\Delta^2 H$) and generalized force ($f_\mu$) schemes. Each value indicates the ramp velocity at which the absolute deviation between the numerically extracted and theoretical QGT component first exceeds $0.005$ for model I or $0.01$ for model II (see circles in Fig.~\ref{fig: pH_robust}). A dash (--) denotes that the deviation remains below the threshold for all tested velocities ($v < 2\pi\,\mathrm{\mu s}^{-1}$).}
\label{table: critical velocities}
\begin{ruledtabular}
\begin{tabular}{lcccccc}
 & \multicolumn{3}{c}{Model I} & \multicolumn{3}{c}{Model II} \\
\cline{2-4} \cline{5-7}
 & $Q_{\theta\theta}$ & $Q_{\phi\theta}$ & $Q_{\phi\phi}$ & $Q_{xx}$ & $Q_{yx}$ & $Q_{yy}$ \\
\hline
Re[$Q_{\mu\nu}$] using $\Delta^2 H$ & 0.48 & 0.83 & 0.90 & 1.75 & 0.83 & 0.46 \\
Im[$Q_{\mu\nu}$] using $\Delta^2 H$ & --   & 0.59 & --   & --   & 0.90 & --    \\
Re[$Q_{\mu\nu}$] using $f_\mu$      & 0.62 & --   & 1.66 & 1.38 & 1.53 & 1.01 \\
Im[$Q_{\mu\nu}$] using $f_\mu$      & 0.44 & 0.99 & 0.83 & 0.62 & 0.43 & 0.27 \\
\end{tabular}
\end{ruledtabular}
\end{table}


\begin{thebibliography}{111}%
\makeatletter
\providecommand \@ifxundefined [1]{%
 \@ifx{#1\undefined}
}%
\providecommand \@ifnum [1]{%
 \ifnum #1\expandafter \@firstoftwo
 \else \expandafter \@secondoftwo
 \fi
}%
\providecommand \@ifx [1]{%
 \ifx #1\expandafter \@firstoftwo
 \else \expandafter \@secondoftwo
 \fi
}%
\providecommand \natexlab [1]{#1}%
\providecommand \enquote  [1]{``#1''}%
\providecommand \bibnamefont  [1]{#1}%
\providecommand \bibfnamefont [1]{#1}%
\providecommand \citenamefont [1]{#1}%
\providecommand \href@noop [0]{\@secondoftwo}%
\providecommand \href [0]{\begingroup \@sanitize@url \@href}%
\providecommand \@href[1]{\@@startlink{#1}\@@href}%
\providecommand \@@href[1]{\endgroup#1\@@endlink}%
\providecommand \@sanitize@url [0]{\catcode `\\12\catcode `\$12\catcode `\&12\catcode `\#12\catcode `\^12\catcode `\_12\catcode `\%12\relax}%
\providecommand \@@startlink[1]{}%
\providecommand \@@endlink[0]{}%
\providecommand \url  [0]{\begingroup\@sanitize@url \@url }%
\providecommand \@url [1]{\endgroup\@href {#1}{\urlprefix }}%
\providecommand \urlprefix  [0]{URL }%
\providecommand \Eprint [0]{\href }%
\providecommand \doibase [0]{https://doi.org/}%
\providecommand \selectlanguage [0]{\@gobble}%
\providecommand \bibinfo  [0]{\@secondoftwo}%
\providecommand \bibfield  [0]{\@secondoftwo}%
\providecommand \translation [1]{[#1]}%
\providecommand \BibitemOpen [0]{}%
\providecommand \bibitemStop [0]{}%
\providecommand \bibitemNoStop [0]{.\EOS\space}%
\providecommand \EOS [0]{\spacefactor3000\relax}%
\providecommand \BibitemShut  [1]{\csname bibitem#1\endcsname}%
\let\auto@bib@innerbib\@empty
\bibitem [{\citenamefont {T\"{o}rm\"{a}}\ \emph {et~al.}(2022)\citenamefont {T\"{o}rm\"{a}}, \citenamefont {Peotta},\ and\ \citenamefont {Bernevig}}]{Torma2022}%
  \BibitemOpen
  \bibfield  {author} {\bibinfo {author} {\bibfnamefont {P.}~\bibnamefont {T\"{o}rm\"{a}}}, \bibinfo {author} {\bibfnamefont {S.}~\bibnamefont {Peotta}},\ and\ \bibinfo {author} {\bibfnamefont {B.~A.}\ \bibnamefont {Bernevig}},\ }\bibfield  {title} {\bibinfo {title} {Superconductivity, superfluidity and quantum geometry in twisted multilayer systems},\ }\href {https://doi.org/10.1038/s42254-022-00466-y} {\bibfield  {journal} {\bibinfo  {journal} {Nat. Rev. Phys.}\ }\textbf {\bibinfo {volume} {4}},\ \bibinfo {pages} {528} (\bibinfo {year} {2022})}\BibitemShut {NoStop}%
\bibitem [{\citenamefont {T\"orm\"a}(2023)}]{Torma2023}%
  \BibitemOpen
  \bibfield  {author} {\bibinfo {author} {\bibfnamefont {P.}~\bibnamefont {T\"orm\"a}},\ }\bibfield  {title} {\bibinfo {title} {Essay: Where can quantum geometry lead us?},\ }\href {https://doi.org/10.1103/PhysRevLett.131.240001} {\bibfield  {journal} {\bibinfo  {journal} {Phys. Rev. Lett.}\ }\textbf {\bibinfo {volume} {131}},\ \bibinfo {pages} {240001} (\bibinfo {year} {2023})}\BibitemShut {NoStop}%
\bibitem [{\citenamefont {Zhu}(2008)}]{SLZhu2008}%
  \BibitemOpen
  \bibfield  {author} {\bibinfo {author} {\bibfnamefont {S.-L.}\ \bibnamefont {Zhu}},\ }\bibfield  {title} {\bibinfo {title} {Geometric phases and quantum phase transitions},\ }\href {https://doi.org/10.1142/S0217979208038855} {\bibfield  {journal} {\bibinfo  {journal} {Int. J. Mod. Phys. B}\ }\textbf {\bibinfo {volume} {22}},\ \bibinfo {pages} {561} (\bibinfo {year} {2008})}\BibitemShut {NoStop}%
\bibitem [{\citenamefont {Provost}\ and\ \citenamefont {Vallee}(1980)}]{Provost1980}%
  \BibitemOpen
  \bibfield  {author} {\bibinfo {author} {\bibfnamefont {J.~P.}\ \bibnamefont {Provost}}\ and\ \bibinfo {author} {\bibfnamefont {G.}~\bibnamefont {Vallee}},\ }\bibfield  {title} {\bibinfo {title} {Riemannian structure on manifolds of quantum states},\ }\href {https://doi.org/10.1007/BF02193559} {\bibfield  {journal} {\bibinfo  {journal} {Commun. Math. Phys.}\ }\textbf {\bibinfo {volume} {76}},\ \bibinfo {pages} {289} (\bibinfo {year} {1980})}\BibitemShut {NoStop}%
\bibitem [{\citenamefont {Braunstein}\ and\ \citenamefont {Caves}(1994)}]{Braunstein1994}%
  \BibitemOpen
  \bibfield  {author} {\bibinfo {author} {\bibfnamefont {S.~L.}\ \bibnamefont {Braunstein}}\ and\ \bibinfo {author} {\bibfnamefont {C.~M.}\ \bibnamefont {Caves}},\ }\bibfield  {title} {\bibinfo {title} {Statistical distance and the geometry of quantum states},\ }\href {https://doi.org/10.1103/PhysRevLett.72.3439} {\bibfield  {journal} {\bibinfo  {journal} {Phys. Rev. Lett.}\ }\textbf {\bibinfo {volume} {72}},\ \bibinfo {pages} {3439} (\bibinfo {year} {1994})}\BibitemShut {NoStop}%
\bibitem [{\citenamefont {Peotta}\ and\ \citenamefont {T{\"o}rm{\"a}}(2015)}]{Peotta2015}%
  \BibitemOpen
  \bibfield  {author} {\bibinfo {author} {\bibfnamefont {S.}~\bibnamefont {Peotta}}\ and\ \bibinfo {author} {\bibfnamefont {P.}~\bibnamefont {T{\"o}rm{\"a}}},\ }\bibfield  {title} {\bibinfo {title} {Superfluidity in topologically nontrivial flat bands},\ }\href {https://doi.org/10.1038/ncomms9944} {\bibfield  {journal} {\bibinfo  {journal} {Nat. Commun.}\ }\textbf {\bibinfo {volume} {6}},\ \bibinfo {pages} {8944} (\bibinfo {year} {2015})}\BibitemShut {NoStop}%
\bibitem [{\citenamefont {Julku}\ \emph {et~al.}(2016)\citenamefont {Julku}, \citenamefont {Peotta}, \citenamefont {Vanhala}, \citenamefont {Kim},\ and\ \citenamefont {T\"orm\"a}}]{Julku2016}%
  \BibitemOpen
  \bibfield  {author} {\bibinfo {author} {\bibfnamefont {A.}~\bibnamefont {Julku}}, \bibinfo {author} {\bibfnamefont {S.}~\bibnamefont {Peotta}}, \bibinfo {author} {\bibfnamefont {T.~I.}\ \bibnamefont {Vanhala}}, \bibinfo {author} {\bibfnamefont {D.-H.}\ \bibnamefont {Kim}},\ and\ \bibinfo {author} {\bibfnamefont {P.}~\bibnamefont {T\"orm\"a}},\ }\bibfield  {title} {\bibinfo {title} {Geometric origin of superfluidity in the {Lieb}-lattice flat band},\ }\href {https://doi.org/10.1103/PhysRevLett.117.045303} {\bibfield  {journal} {\bibinfo  {journal} {Phys. Rev. Lett.}\ }\textbf {\bibinfo {volume} {117}},\ \bibinfo {pages} {045303} (\bibinfo {year} {2016})}\BibitemShut {NoStop}%
\bibitem [{\citenamefont {Liang}\ \emph {et~al.}(2017)\citenamefont {Liang}, \citenamefont {Vanhala}, \citenamefont {Peotta}, \citenamefont {Siro}, \citenamefont {Harju},\ and\ \citenamefont {T\"orm\"a}}]{LLiang2017}%
  \BibitemOpen
  \bibfield  {author} {\bibinfo {author} {\bibfnamefont {L.}~\bibnamefont {Liang}}, \bibinfo {author} {\bibfnamefont {T.~I.}\ \bibnamefont {Vanhala}}, \bibinfo {author} {\bibfnamefont {S.}~\bibnamefont {Peotta}}, \bibinfo {author} {\bibfnamefont {T.}~\bibnamefont {Siro}}, \bibinfo {author} {\bibfnamefont {A.}~\bibnamefont {Harju}},\ and\ \bibinfo {author} {\bibfnamefont {P.}~\bibnamefont {T\"orm\"a}},\ }\bibfield  {title} {\bibinfo {title} {Band geometry, {Berry} curvature, and superfluid weight},\ }\href {https://doi.org/10.1103/PhysRevB.95.024515} {\bibfield  {journal} {\bibinfo  {journal} {Phys. Rev. B}\ }\textbf {\bibinfo {volume} {95}},\ \bibinfo {pages} {024515} (\bibinfo {year} {2017})}\BibitemShut {NoStop}%
\bibitem [{\citenamefont {Gao}\ \emph {et~al.}(2015)\citenamefont {Gao}, \citenamefont {Yang},\ and\ \citenamefont {Niu}}]{YGao2015}%
  \BibitemOpen
  \bibfield  {author} {\bibinfo {author} {\bibfnamefont {Y.}~\bibnamefont {Gao}}, \bibinfo {author} {\bibfnamefont {S.~A.}\ \bibnamefont {Yang}},\ and\ \bibinfo {author} {\bibfnamefont {Q.}~\bibnamefont {Niu}},\ }\bibfield  {title} {\bibinfo {title} {Geometrical effects in orbital magnetic susceptibility},\ }\href {https://doi.org/10.1103/PhysRevB.91.214405} {\bibfield  {journal} {\bibinfo  {journal} {Phys. Rev. B}\ }\textbf {\bibinfo {volume} {91}},\ \bibinfo {pages} {214405} (\bibinfo {year} {2015})}\BibitemShut {NoStop}%
\bibitem [{\citenamefont {Pi\'echon}\ \emph {et~al.}(2016)\citenamefont {Pi\'echon}, \citenamefont {Raoux}, \citenamefont {Fuchs},\ and\ \citenamefont {Montambaux}}]{Piechon2016}%
  \BibitemOpen
  \bibfield  {author} {\bibinfo {author} {\bibfnamefont {F.}~\bibnamefont {Pi\'echon}}, \bibinfo {author} {\bibfnamefont {A.}~\bibnamefont {Raoux}}, \bibinfo {author} {\bibfnamefont {J.-N.}\ \bibnamefont {Fuchs}},\ and\ \bibinfo {author} {\bibfnamefont {G.}~\bibnamefont {Montambaux}},\ }\bibfield  {title} {\bibinfo {title} {Geometric orbital susceptibility: Quantum metric without {Berry} curvature},\ }\href {https://doi.org/10.1103/PhysRevB.94.134423} {\bibfield  {journal} {\bibinfo  {journal} {Phys. Rev. B}\ }\textbf {\bibinfo {volume} {94}},\ \bibinfo {pages} {134423} (\bibinfo {year} {2016})}\BibitemShut {NoStop}%
\bibitem [{\citenamefont {Du}\ \emph {et~al.}(2021)\citenamefont {Du}, \citenamefont {Lu},\ and\ \citenamefont {Xie}}]{ZZDu2021}%
  \BibitemOpen
  \bibfield  {author} {\bibinfo {author} {\bibfnamefont {Z.~Z.}\ \bibnamefont {Du}}, \bibinfo {author} {\bibfnamefont {H.-Z.}\ \bibnamefont {Lu}},\ and\ \bibinfo {author} {\bibfnamefont {X.~C.}\ \bibnamefont {Xie}},\ }\bibfield  {title} {\bibinfo {title} {Nonlinear {{Hall}} effects},\ }\href {https://doi.org/10.1038/s42254-021-00359-6} {\bibfield  {journal} {\bibinfo  {journal} {Nat. Rev. Phys.}\ }\textbf {\bibinfo {volume} {3}},\ \bibinfo {pages} {744} (\bibinfo {year} {2021})}\BibitemShut {NoStop}%
\bibitem [{\citenamefont {Gao}\ \emph {et~al.}(2023)\citenamefont {Gao}, \citenamefont {Liu}, \citenamefont {Qiu}, \citenamefont {Ghosh}, \citenamefont {V.~Trevisan}, \citenamefont {Onishi}, \citenamefont {Hu}, \citenamefont {Qian}, \citenamefont {Tien}, \citenamefont {Chen}, \citenamefont {Huang}, \citenamefont {B{\'e}rub{\'e}}, \citenamefont {Li}, \citenamefont {Tzschaschel}, \citenamefont {Dinh}, \citenamefont {Sun}, \citenamefont {Ho}, \citenamefont {Lien}, \citenamefont {Singh}, \citenamefont {Watanabe}, \citenamefont {Taniguchi}, \citenamefont {Bell}, \citenamefont {Lin}, \citenamefont {Chang}, \citenamefont {Du}, \citenamefont {Bansil}, \citenamefont {Fu}, \citenamefont {Ni}, \citenamefont {Orth}, \citenamefont {Ma},\ and\ \citenamefont {Xu}}]{AYGao2023}%
  \BibitemOpen
  \bibfield  {author} {\bibinfo {author} {\bibfnamefont {A.}~\bibnamefont {Gao}}, \bibinfo {author} {\bibfnamefont {Y.-F.}\ \bibnamefont {Liu}}, \bibinfo {author} {\bibfnamefont {J.-X.}\ \bibnamefont {Qiu}}, \bibinfo {author} {\bibfnamefont {B.}~\bibnamefont {Ghosh}}, \bibinfo {author} {\bibfnamefont {T.}~\bibnamefont {V.~Trevisan}}, \bibinfo {author} {\bibfnamefont {Y.}~\bibnamefont {Onishi}}, \bibinfo {author} {\bibfnamefont {C.}~\bibnamefont {Hu}}, \bibinfo {author} {\bibfnamefont {T.}~\bibnamefont {Qian}}, \bibinfo {author} {\bibfnamefont {H.-J.}\ \bibnamefont {Tien}}, \bibinfo {author} {\bibfnamefont {S.-W.}\ \bibnamefont {Chen}}, \bibinfo {author} {\bibfnamefont {M.}~\bibnamefont {Huang}}, \bibinfo {author} {\bibfnamefont {D.}~\bibnamefont {B{\'e}rub{\'e}}}, \bibinfo {author} {\bibfnamefont {H.}~\bibnamefont {Li}}, \bibinfo {author} {\bibfnamefont {C.}~\bibnamefont {Tzschaschel}}, \bibinfo {author} {\bibfnamefont {T.}~\bibnamefont {Dinh}}, \bibinfo {author} {\bibfnamefont {Z.}~\bibnamefont {Sun}}, \bibinfo {author} {\bibfnamefont {S.-C.}\ \bibnamefont {Ho}}, \bibinfo {author} {\bibfnamefont {S.-W.}\ \bibnamefont {Lien}}, \bibinfo {author} {\bibfnamefont {B.}~\bibnamefont {Singh}}, \bibinfo {author} {\bibfnamefont {K.}~\bibnamefont {Watanabe}}, \bibinfo {author} {\bibfnamefont {T.}~\bibnamefont {Taniguchi}}, \bibinfo {author} {\bibfnamefont {D.~C.}\ \bibnamefont {Bell}}, \bibinfo {author} {\bibfnamefont {H.}~\bibnamefont {Lin}}, \bibinfo {author} {\bibfnamefont {T.-R.}\ \bibnamefont {Chang}}, \bibinfo {author} {\bibfnamefont {C.~R.}\ \bibnamefont {Du}}, \bibinfo {author} {\bibfnamefont {A.}~\bibnamefont {Bansil}}, \bibinfo {author} {\bibfnamefont {L.}~\bibnamefont {Fu}}, \bibinfo {author} {\bibfnamefont {N.}~\bibnamefont {Ni}}, \bibinfo {author} {\bibfnamefont {P.~P.}\ \bibnamefont {Orth}}, \bibinfo {author} {\bibfnamefont {Q.}~\bibnamefont {Ma}},\ and\ \bibinfo {author} {\bibfnamefont {S.-Y.}\ \bibnamefont {Xu}},\ }\bibfield  {title} {\bibinfo {title} {Quantum metric nonlinear {{Hall}} effect in a topological antiferromagnetic heterostructure},\ }\href {https://doi.org/10.1126/science.adf1506} {\bibfield  {journal} {\bibinfo  {journal} {Science}\ }\textbf {\bibinfo {volume} {381}},\ \bibinfo {pages} {181} (\bibinfo {year} {2023})}\BibitemShut {NoStop}%
\bibitem [{\citenamefont {Wang}\ \emph {et~al.}(2023)\citenamefont {Wang}, \citenamefont {Kaplan}, \citenamefont {Zhang}, \citenamefont {Holder}, \citenamefont {Cao}, \citenamefont {Wang}, \citenamefont {Zhou}, \citenamefont {Zhou}, \citenamefont {Jiang}, \citenamefont {Zhang}, \citenamefont {Ru}, \citenamefont {Cai}, \citenamefont {Watanabe}, \citenamefont {Taniguchi}, \citenamefont {Yan},\ and\ \citenamefont {Gao}}]{NZWang2023}%
  \BibitemOpen
  \bibfield  {author} {\bibinfo {author} {\bibfnamefont {N.}~\bibnamefont {Wang}}, \bibinfo {author} {\bibfnamefont {D.}~\bibnamefont {Kaplan}}, \bibinfo {author} {\bibfnamefont {Z.}~\bibnamefont {Zhang}}, \bibinfo {author} {\bibfnamefont {T.}~\bibnamefont {Holder}}, \bibinfo {author} {\bibfnamefont {N.}~\bibnamefont {Cao}}, \bibinfo {author} {\bibfnamefont {A.}~\bibnamefont {Wang}}, \bibinfo {author} {\bibfnamefont {X.}~\bibnamefont {Zhou}}, \bibinfo {author} {\bibfnamefont {F.}~\bibnamefont {Zhou}}, \bibinfo {author} {\bibfnamefont {Z.}~\bibnamefont {Jiang}}, \bibinfo {author} {\bibfnamefont {C.}~\bibnamefont {Zhang}}, \bibinfo {author} {\bibfnamefont {S.}~\bibnamefont {Ru}}, \bibinfo {author} {\bibfnamefont {H.}~\bibnamefont {Cai}}, \bibinfo {author} {\bibfnamefont {K.}~\bibnamefont {Watanabe}}, \bibinfo {author} {\bibfnamefont {T.}~\bibnamefont {Taniguchi}}, \bibinfo {author} {\bibfnamefont {B.}~\bibnamefont {Yan}},\ and\ \bibinfo {author} {\bibfnamefont {W.}~\bibnamefont {Gao}},\ }\bibfield  {title} {\bibinfo {title} {Quantum-metric-induced nonlinear transport in a topological antiferromagnet},\ }\href {https://doi.org/10.1038/s41586-023-06363-3} {\bibfield  {journal} {\bibinfo  {journal} {Nature (London)}\ }\textbf {\bibinfo {volume} {621}},\ \bibinfo {pages} {487} (\bibinfo {year} {2023})}\BibitemShut {NoStop}%
\bibitem [{\citenamefont {Xiao}\ \emph {et~al.}(2010)\citenamefont {Xiao}, \citenamefont {Chang},\ and\ \citenamefont {Niu}}]{DXiao2010}%
  \BibitemOpen
  \bibfield  {author} {\bibinfo {author} {\bibfnamefont {D.}~\bibnamefont {Xiao}}, \bibinfo {author} {\bibfnamefont {M.-C.}\ \bibnamefont {Chang}},\ and\ \bibinfo {author} {\bibfnamefont {Q.}~\bibnamefont {Niu}},\ }\bibfield  {title} {\bibinfo {title} {Berry phase effects on electronic properties},\ }\href {https://doi.org/10.1103/RevModPhys.82.1959} {\bibfield  {journal} {\bibinfo  {journal} {Rev. Mod. Phys.}\ }\textbf {\bibinfo {volume} {82}},\ \bibinfo {pages} {1959} (\bibinfo {year} {2010})}\BibitemShut {NoStop}%
\bibitem [{\citenamefont {Simon}(1983)}]{Simon1983}%
  \BibitemOpen
  \bibfield  {author} {\bibinfo {author} {\bibfnamefont {B.}~\bibnamefont {Simon}},\ }\bibfield  {title} {\bibinfo {title} {Holonomy, the quantum adiabatic theorem, and {Berry}'s phase},\ }\href {https://doi.org/10.1103/PhysRevLett.51.2167} {\bibfield  {journal} {\bibinfo  {journal} {Phys. Rev. Lett.}\ }\textbf {\bibinfo {volume} {51}},\ \bibinfo {pages} {2167} (\bibinfo {year} {1983})}\BibitemShut {NoStop}%
\bibitem [{\citenamefont {Berry}(1984)}]{Berry1984}%
  \BibitemOpen
  \bibfield  {author} {\bibinfo {author} {\bibfnamefont {M.~V.}\ \bibnamefont {Berry}},\ }\bibfield  {title} {\bibinfo {title} {Quantal phase factors accompanying adiabatic changes},\ }\href {https://doi.org/10.1098/rspa.1984.0023} {\bibfield  {journal} {\bibinfo  {journal} {Proc. R. Soc. London, Ser. A}\ }\textbf {\bibinfo {volume} {392}},\ \bibinfo {pages} {45} (\bibinfo {year} {1984})}\BibitemShut {NoStop}%
\bibitem [{\citenamefont {Thouless}\ \emph {et~al.}(1982)\citenamefont {Thouless}, \citenamefont {Kohmoto}, \citenamefont {Nightingale},\ and\ \citenamefont {den Nijs}}]{Thouless1982}%
  \BibitemOpen
  \bibfield  {author} {\bibinfo {author} {\bibfnamefont {D.~J.}\ \bibnamefont {Thouless}}, \bibinfo {author} {\bibfnamefont {M.}~\bibnamefont {Kohmoto}}, \bibinfo {author} {\bibfnamefont {M.~P.}\ \bibnamefont {Nightingale}},\ and\ \bibinfo {author} {\bibfnamefont {M.}~\bibnamefont {den Nijs}},\ }\bibfield  {title} {\bibinfo {title} {Quantized {Hall} conductance in a two-dimensional periodic potential},\ }\href {https://doi.org/10.1103/PhysRevLett.49.405} {\bibfield  {journal} {\bibinfo  {journal} {Phys. Rev. Lett.}\ }\textbf {\bibinfo {volume} {49}},\ \bibinfo {pages} {405} (\bibinfo {year} {1982})}\BibitemShut {NoStop}%
\bibitem [{\citenamefont {Hasan}\ and\ \citenamefont {Kane}(2010)}]{Hasan2010}%
  \BibitemOpen
  \bibfield  {author} {\bibinfo {author} {\bibfnamefont {M.~Z.}\ \bibnamefont {Hasan}}\ and\ \bibinfo {author} {\bibfnamefont {C.~L.}\ \bibnamefont {Kane}},\ }\bibfield  {title} {\bibinfo {title} {Colloquium: Topological insulators},\ }\href {https://doi.org/10.1103/RevModPhys.82.3045} {\bibfield  {journal} {\bibinfo  {journal} {Rev. Mod. Phys.}\ }\textbf {\bibinfo {volume} {82}},\ \bibinfo {pages} {3045} (\bibinfo {year} {2010})}\BibitemShut {NoStop}%
\bibitem [{\citenamefont {Qi}\ and\ \citenamefont {Zhang}(2011)}]{XLQi2011}%
  \BibitemOpen
  \bibfield  {author} {\bibinfo {author} {\bibfnamefont {X.-L.}\ \bibnamefont {Qi}}\ and\ \bibinfo {author} {\bibfnamefont {S.-C.}\ \bibnamefont {Zhang}},\ }\bibfield  {title} {\bibinfo {title} {Topological insulators and superconductors},\ }\href {https://doi.org/10.1103/RevModPhys.83.1057} {\bibfield  {journal} {\bibinfo  {journal} {Rev. Mod. Phys.}\ }\textbf {\bibinfo {volume} {83}},\ \bibinfo {pages} {1057} (\bibinfo {year} {2011})}\BibitemShut {NoStop}%
\bibitem [{\citenamefont {Armitage}\ \emph {et~al.}(2018)\citenamefont {Armitage}, \citenamefont {Mele},\ and\ \citenamefont {Vishwanath}}]{Armitage2018}%
  \BibitemOpen
  \bibfield  {author} {\bibinfo {author} {\bibfnamefont {N.~P.}\ \bibnamefont {Armitage}}, \bibinfo {author} {\bibfnamefont {E.~J.}\ \bibnamefont {Mele}},\ and\ \bibinfo {author} {\bibfnamefont {A.}~\bibnamefont {Vishwanath}},\ }\bibfield  {title} {\bibinfo {title} {{Weyl} and {Dirac} semimetals in three-dimensional solids},\ }\href {https://doi.org/10.1103/RevModPhys.90.015001} {\bibfield  {journal} {\bibinfo  {journal} {Rev. Mod. Phys.}\ }\textbf {\bibinfo {volume} {90}},\ \bibinfo {pages} {015001} (\bibinfo {year} {2018})}\BibitemShut {NoStop}%
\bibitem [{\citenamefont {Zhu}\ \emph {et~al.}(2006)\citenamefont {Zhu}, \citenamefont {Fu}, \citenamefont {Wu}, \citenamefont {Zhang},\ and\ \citenamefont {Duan}}]{SLZhu2006SHE}%
  \BibitemOpen
  \bibfield  {author} {\bibinfo {author} {\bibfnamefont {S.-L.}\ \bibnamefont {Zhu}}, \bibinfo {author} {\bibfnamefont {H.}~\bibnamefont {Fu}}, \bibinfo {author} {\bibfnamefont {C.-J.}\ \bibnamefont {Wu}}, \bibinfo {author} {\bibfnamefont {S.-C.}\ \bibnamefont {Zhang}},\ and\ \bibinfo {author} {\bibfnamefont {L.-M.}\ \bibnamefont {Duan}},\ }\bibfield  {title} {\bibinfo {title} {Spin {Hall} effects for cold atoms in a light-induced gauge potential},\ }\href {https://doi.org/10.1103/PhysRevLett.97.240401} {\bibfield  {journal} {\bibinfo  {journal} {Phys. Rev. Lett.}\ }\textbf {\bibinfo {volume} {97}},\ \bibinfo {pages} {240401} (\bibinfo {year} {2006})}\BibitemShut {NoStop}%
\bibitem [{\citenamefont {Zhang}\ \emph {et~al.}(2018)\citenamefont {Zhang}, \citenamefont {Zhu}, \citenamefont {Zhao}, \citenamefont {Yan},\ and\ \citenamefont {Zhu}}]{DWZhang2018}%
  \BibitemOpen
  \bibfield  {author} {\bibinfo {author} {\bibfnamefont {D.-W.}\ \bibnamefont {Zhang}}, \bibinfo {author} {\bibfnamefont {Y.-Q.}\ \bibnamefont {Zhu}}, \bibinfo {author} {\bibfnamefont {Y.~X.}\ \bibnamefont {Zhao}}, \bibinfo {author} {\bibfnamefont {H.}~\bibnamefont {Yan}},\ and\ \bibinfo {author} {\bibfnamefont {S.-L.}\ \bibnamefont {Zhu}},\ }\bibfield  {title} {\bibinfo {title} {Topological quantum matter with cold atoms},\ }\href {https://doi.org/10.1080/00018732.2019.1594094} {\bibfield  {journal} {\bibinfo  {journal} {Adv. Phys.}\ }\textbf {\bibinfo {volume} {67}},\ \bibinfo {pages} {253} (\bibinfo {year} {2018})}\BibitemShut {NoStop}%
\bibitem [{\citenamefont {Carollo}\ and\ \citenamefont {Pachos}(2005)}]{Carollo2005}%
  \BibitemOpen
  \bibfield  {author} {\bibinfo {author} {\bibfnamefont {A.~C.~M.}\ \bibnamefont {Carollo}}\ and\ \bibinfo {author} {\bibfnamefont {J.~K.}\ \bibnamefont {Pachos}},\ }\bibfield  {title} {\bibinfo {title} {Geometric phases and criticality in spin-chain systems},\ }\href {https://doi.org/10.1103/PhysRevLett.95.157203} {\bibfield  {journal} {\bibinfo  {journal} {Phys. Rev. Lett.}\ }\textbf {\bibinfo {volume} {95}},\ \bibinfo {pages} {157203} (\bibinfo {year} {2005})}\BibitemShut {NoStop}%
\bibitem [{\citenamefont {Zhu}(2006)}]{SLZhu2006}%
  \BibitemOpen
  \bibfield  {author} {\bibinfo {author} {\bibfnamefont {S.-L.}\ \bibnamefont {Zhu}},\ }\bibfield  {title} {\bibinfo {title} {Scaling of geometric phases close to the quantum phase transition in the {$XY$} spin chain},\ }\href {https://doi.org/10.1103/PhysRevLett.96.077206} {\bibfield  {journal} {\bibinfo  {journal} {Phys. Rev. Lett.}\ }\textbf {\bibinfo {volume} {96}},\ \bibinfo {pages} {077206} (\bibinfo {year} {2006})}\BibitemShut {NoStop}%
\bibitem [{\citenamefont {Campos~Venuti}\ and\ \citenamefont {Zanardi}(2007)}]{Venuti2007}%
  \BibitemOpen
  \bibfield  {author} {\bibinfo {author} {\bibfnamefont {L.}~\bibnamefont {Campos~Venuti}}\ and\ \bibinfo {author} {\bibfnamefont {P.}~\bibnamefont {Zanardi}},\ }\bibfield  {title} {\bibinfo {title} {Quantum critical scaling of the geometric tensors},\ }\href {https://doi.org/10.1103/PhysRevLett.99.095701} {\bibfield  {journal} {\bibinfo  {journal} {Phys. Rev. Lett.}\ }\textbf {\bibinfo {volume} {99}},\ \bibinfo {pages} {095701} (\bibinfo {year} {2007})}\BibitemShut {NoStop}%
\bibitem [{\citenamefont {Gu}\ \emph {et~al.}(2008)\citenamefont {Gu}, \citenamefont {Kwok}, \citenamefont {Ning},\ and\ \citenamefont {Lin}}]{SJGu2008}%
  \BibitemOpen
  \bibfield  {author} {\bibinfo {author} {\bibfnamefont {S.-J.}\ \bibnamefont {Gu}}, \bibinfo {author} {\bibfnamefont {H.-M.}\ \bibnamefont {Kwok}}, \bibinfo {author} {\bibfnamefont {W.-Q.}\ \bibnamefont {Ning}},\ and\ \bibinfo {author} {\bibfnamefont {H.-Q.}\ \bibnamefont {Lin}},\ }\bibfield  {title} {\bibinfo {title} {Fidelity susceptibility, scaling, and universality in quantum critical phenomena},\ }\href {https://doi.org/10.1103/PhysRevB.77.245109} {\bibfield  {journal} {\bibinfo  {journal} {Phys. Rev. B}\ }\textbf {\bibinfo {volume} {77}},\ \bibinfo {pages} {245109} (\bibinfo {year} {2008})}\BibitemShut {NoStop}%
\bibitem [{\citenamefont {Ma}\ \emph {et~al.}(2010)\citenamefont {Ma}, \citenamefont {Chen}, \citenamefont {Fan},\ and\ \citenamefont {Liu}}]{YQMa2010}%
  \BibitemOpen
  \bibfield  {author} {\bibinfo {author} {\bibfnamefont {Y.-Q.}\ \bibnamefont {Ma}}, \bibinfo {author} {\bibfnamefont {S.}~\bibnamefont {Chen}}, \bibinfo {author} {\bibfnamefont {H.}~\bibnamefont {Fan}},\ and\ \bibinfo {author} {\bibfnamefont {W.-M.}\ \bibnamefont {Liu}},\ }\bibfield  {title} {\bibinfo {title} {{Abelian} and non-{Abelian} quantum geometric tensor},\ }\href {https://doi.org/10.1103/PhysRevB.81.245129} {\bibfield  {journal} {\bibinfo  {journal} {Phys. Rev. B}\ }\textbf {\bibinfo {volume} {81}},\ \bibinfo {pages} {245129} (\bibinfo {year} {2010})}\BibitemShut {NoStop}%
\bibitem [{\citenamefont {Ding}\ \emph {et~al.}(2024)\citenamefont {Ding}, \citenamefont {Zhang}, \citenamefont {Liu}, \citenamefont {Wang}, \citenamefont {Zhang},\ and\ \citenamefont {Zhu}}]{HTDing2024}%
  \BibitemOpen
  \bibfield  {author} {\bibinfo {author} {\bibfnamefont {H.-T.}\ \bibnamefont {Ding}}, \bibinfo {author} {\bibfnamefont {C.-X.}\ \bibnamefont {Zhang}}, \bibinfo {author} {\bibfnamefont {J.-X.}\ \bibnamefont {Liu}}, \bibinfo {author} {\bibfnamefont {J.-T.}\ \bibnamefont {Wang}}, \bibinfo {author} {\bibfnamefont {D.-W.}\ \bibnamefont {Zhang}},\ and\ \bibinfo {author} {\bibfnamefont {S.-L.}\ \bibnamefont {Zhu}},\ }\bibfield  {title} {\bibinfo {title} {Non-{Abelian} quantum geometric tensor in degenerate topological semimetals},\ }\href {https://doi.org/10.1103/PhysRevA.109.043305} {\bibfield  {journal} {\bibinfo  {journal} {Phys. Rev. A}\ }\textbf {\bibinfo {volume} {109}},\ \bibinfo {pages} {043305} (\bibinfo {year} {2024})}\BibitemShut {NoStop}%
\bibitem [{\citenamefont {Hou}\ \emph {et~al.}(2024)\citenamefont {Hou}, \citenamefont {Zhou}, \citenamefont {Wang}, \citenamefont {Guo},\ and\ \citenamefont {Chien}}]{XYHou2024}%
  \BibitemOpen
  \bibfield  {author} {\bibinfo {author} {\bibfnamefont {X.-Y.}\ \bibnamefont {Hou}}, \bibinfo {author} {\bibfnamefont {Z.}~\bibnamefont {Zhou}}, \bibinfo {author} {\bibfnamefont {X.}~\bibnamefont {Wang}}, \bibinfo {author} {\bibfnamefont {H.}~\bibnamefont {Guo}},\ and\ \bibinfo {author} {\bibfnamefont {C.-C.}\ \bibnamefont {Chien}},\ }\bibfield  {title} {\bibinfo {title} {Local geometry and quantum geometric tensor of mixed states},\ }\href {https://doi.org/10.1103/PhysRevB.110.035144} {\bibfield  {journal} {\bibinfo  {journal} {Phys. Rev. B}\ }\textbf {\bibinfo {volume} {110}},\ \bibinfo {pages} {035144} (\bibinfo {year} {2024})}\BibitemShut {NoStop}%
\bibitem [{\citenamefont {Zhou}\ \emph {et~al.}(2024)\citenamefont {Zhou}, \citenamefont {Hou}, \citenamefont {Wang}, \citenamefont {Tang}, \citenamefont {Guo},\ and\ \citenamefont {Chien}}]{ZZhou2024}%
  \BibitemOpen
  \bibfield  {author} {\bibinfo {author} {\bibfnamefont {Z.}~\bibnamefont {Zhou}}, \bibinfo {author} {\bibfnamefont {X.-Y.}\ \bibnamefont {Hou}}, \bibinfo {author} {\bibfnamefont {X.}~\bibnamefont {Wang}}, \bibinfo {author} {\bibfnamefont {J.-C.}\ \bibnamefont {Tang}}, \bibinfo {author} {\bibfnamefont {H.}~\bibnamefont {Guo}},\ and\ \bibinfo {author} {\bibfnamefont {C.-C.}\ \bibnamefont {Chien}},\ }\bibfield  {title} {\bibinfo {title} {Sj\"oqvist quantum geometric tensor of finite-temperature mixed states},\ }\href {https://doi.org/10.1103/PhysRevB.110.035404} {\bibfield  {journal} {\bibinfo  {journal} {Phys. Rev. B}\ }\textbf {\bibinfo {volume} {110}},\ \bibinfo {pages} {035404} (\bibinfo {year} {2024})}\BibitemShut {NoStop}%
\bibitem [{\citenamefont {Wang}\ \emph {et~al.}(2025)\citenamefont {Wang}, \citenamefont {Wang}, \citenamefont {Wang},\ and\ \citenamefont {Zhang}}]{QYWang2025}%
  \BibitemOpen
  \bibfield  {author} {\bibinfo {author} {\bibfnamefont {Q.}~\bibnamefont {Wang}}, \bibinfo {author} {\bibfnamefont {B.}~\bibnamefont {Wang}}, \bibinfo {author} {\bibfnamefont {J.}~\bibnamefont {Wang}},\ and\ \bibinfo {author} {\bibfnamefont {L.}~\bibnamefont {Zhang}},\ }\bibfield  {title} {\bibinfo {title} {Quantum geometric tensor for mixed states based on the covariant derivative},\ }\href {https://doi.org/10.1088/0256-307X/42/7/070603} {\bibfield  {journal} {\bibinfo  {journal} {Chin. Phys. Lett.}\ }\textbf {\bibinfo {volume} {42}},\ \bibinfo {pages} {070603} (\bibinfo {year} {2025})}\BibitemShut {NoStop}%
\bibitem [{\citenamefont {Schroer}\ \emph {et~al.}(2014)\citenamefont {Schroer}, \citenamefont {Kolodrubetz}, \citenamefont {Kindel}, \citenamefont {Sandberg}, \citenamefont {Gao}, \citenamefont {Vissers}, \citenamefont {Pappas}, \citenamefont {Polkovnikov},\ and\ \citenamefont {Lehnert}}]{Schroer2014}%
  \BibitemOpen
  \bibfield  {author} {\bibinfo {author} {\bibfnamefont {M.~D.}\ \bibnamefont {Schroer}}, \bibinfo {author} {\bibfnamefont {M.~H.}\ \bibnamefont {Kolodrubetz}}, \bibinfo {author} {\bibfnamefont {W.~F.}\ \bibnamefont {Kindel}}, \bibinfo {author} {\bibfnamefont {M.}~\bibnamefont {Sandberg}}, \bibinfo {author} {\bibfnamefont {J.}~\bibnamefont {Gao}}, \bibinfo {author} {\bibfnamefont {M.~R.}\ \bibnamefont {Vissers}}, \bibinfo {author} {\bibfnamefont {D.~P.}\ \bibnamefont {Pappas}}, \bibinfo {author} {\bibfnamefont {A.}~\bibnamefont {Polkovnikov}},\ and\ \bibinfo {author} {\bibfnamefont {K.~W.}\ \bibnamefont {Lehnert}},\ }\bibfield  {title} {\bibinfo {title} {Measuring a topological transition in an artificial spin-$1/2$ system},\ }\href {https://doi.org/10.1103/PhysRevLett.113.050402} {\bibfield  {journal} {\bibinfo  {journal} {Phys. Rev. Lett.}\ }\textbf {\bibinfo {volume} {113}},\ \bibinfo {pages} {050402} (\bibinfo {year} {2014})}\BibitemShut {NoStop}%
\bibitem [{\citenamefont {Roushan}\ \emph {et~al.}(2014)\citenamefont {Roushan}, \citenamefont {Neill}, \citenamefont {Chen}, \citenamefont {Kolodrubetz}, \citenamefont {Quintana}, \citenamefont {Leung}, \citenamefont {Fang}, \citenamefont {Barends}, \citenamefont {Campbell}, \citenamefont {Chen}, \citenamefont {Chiaro}, \citenamefont {Dunsworth}, \citenamefont {Jeffrey}, \citenamefont {Kelly}, \citenamefont {Megrant}, \citenamefont {Mutus}, \citenamefont {O'Malley}, \citenamefont {Sank}, \citenamefont {Vainsencher}, \citenamefont {Wenner}, \citenamefont {White}, \citenamefont {Polkovnikov}, \citenamefont {Cleland},\ and\ \citenamefont {Martinis}}]{Roushan2014}%
  \BibitemOpen
  \bibfield  {author} {\bibinfo {author} {\bibfnamefont {P.}~\bibnamefont {Roushan}}, \bibinfo {author} {\bibfnamefont {C.}~\bibnamefont {Neill}}, \bibinfo {author} {\bibfnamefont {Y.}~\bibnamefont {Chen}}, \bibinfo {author} {\bibfnamefont {M.}~\bibnamefont {Kolodrubetz}}, \bibinfo {author} {\bibfnamefont {C.}~\bibnamefont {Quintana}}, \bibinfo {author} {\bibfnamefont {N.}~\bibnamefont {Leung}}, \bibinfo {author} {\bibfnamefont {M.}~\bibnamefont {Fang}}, \bibinfo {author} {\bibfnamefont {R.}~\bibnamefont {Barends}}, \bibinfo {author} {\bibfnamefont {B.}~\bibnamefont {Campbell}}, \bibinfo {author} {\bibfnamefont {Z.}~\bibnamefont {Chen}}, \bibinfo {author} {\bibfnamefont {B.}~\bibnamefont {Chiaro}}, \bibinfo {author} {\bibfnamefont {A.}~\bibnamefont {Dunsworth}}, \bibinfo {author} {\bibfnamefont {E.}~\bibnamefont {Jeffrey}}, \bibinfo {author} {\bibfnamefont {J.}~\bibnamefont {Kelly}}, \bibinfo {author} {\bibfnamefont {A.}~\bibnamefont {Megrant}}, \bibinfo {author} {\bibfnamefont {J.}~\bibnamefont {Mutus}}, \bibinfo {author} {\bibfnamefont {P.~J.~J.}\ \bibnamefont {O'Malley}}, \bibinfo {author} {\bibfnamefont {D.}~\bibnamefont {Sank}}, \bibinfo {author} {\bibfnamefont {A.}~\bibnamefont {Vainsencher}}, \bibinfo {author} {\bibfnamefont {J.}~\bibnamefont {Wenner}}, \bibinfo {author} {\bibfnamefont {T.}~\bibnamefont {White}}, \bibinfo {author} {\bibfnamefont {A.}~\bibnamefont {Polkovnikov}}, \bibinfo {author} {\bibfnamefont {A.~N.}\ \bibnamefont {Cleland}},\ and\ \bibinfo {author} {\bibfnamefont {J.~M.}\ \bibnamefont {Martinis}},\ }\bibfield  {title} {\bibinfo {title} {Observation of topological transitions in interacting quantum circuits},\ }\href {https://doi.org/10.1038/nature13891} {\bibfield  {journal} {\bibinfo  {journal} {Nature (London)}\ }\textbf {\bibinfo {volume} {515}},\ \bibinfo {pages} {241} (\bibinfo {year} {2014})}\BibitemShut {NoStop}%
\bibitem [{\citenamefont {Tan}\ \emph {et~al.}(2018)\citenamefont {Tan}, \citenamefont {Zhang}, \citenamefont {Liu}, \citenamefont {Xue}, \citenamefont {Yu}, \citenamefont {Zhu}, \citenamefont {Yan}, \citenamefont {Zhu},\ and\ \citenamefont {Yu}}]{XSTan2018}%
  \BibitemOpen
  \bibfield  {author} {\bibinfo {author} {\bibfnamefont {X.}~\bibnamefont {Tan}}, \bibinfo {author} {\bibfnamefont {D.-W.}\ \bibnamefont {Zhang}}, \bibinfo {author} {\bibfnamefont {Q.}~\bibnamefont {Liu}}, \bibinfo {author} {\bibfnamefont {G.}~\bibnamefont {Xue}}, \bibinfo {author} {\bibfnamefont {H.-F.}\ \bibnamefont {Yu}}, \bibinfo {author} {\bibfnamefont {Y.-Q.}\ \bibnamefont {Zhu}}, \bibinfo {author} {\bibfnamefont {H.}~\bibnamefont {Yan}}, \bibinfo {author} {\bibfnamefont {S.-L.}\ \bibnamefont {Zhu}},\ and\ \bibinfo {author} {\bibfnamefont {Y.}~\bibnamefont {Yu}},\ }\bibfield  {title} {\bibinfo {title} {Topological {Maxwell} metal bands in a superconducting qutrit},\ }\href {https://doi.org/10.1103/PhysRevLett.120.130503} {\bibfield  {journal} {\bibinfo  {journal} {Phys. Rev. Lett.}\ }\textbf {\bibinfo {volume} {120}},\ \bibinfo {pages} {130503} (\bibinfo {year} {2018})}\BibitemShut {NoStop}%
\bibitem [{\citenamefont {Tan}\ \emph {et~al.}(2019{\natexlab{a}})\citenamefont {Tan}, \citenamefont {Zhao}, \citenamefont {Liu}, \citenamefont {Xue}, \citenamefont {Yu}, \citenamefont {Wang},\ and\ \citenamefont {Yu}}]{XSTan2019}%
  \BibitemOpen
  \bibfield  {author} {\bibinfo {author} {\bibfnamefont {X.}~\bibnamefont {Tan}}, \bibinfo {author} {\bibfnamefont {Y.~X.}\ \bibnamefont {Zhao}}, \bibinfo {author} {\bibfnamefont {Q.}~\bibnamefont {Liu}}, \bibinfo {author} {\bibfnamefont {G.}~\bibnamefont {Xue}}, \bibinfo {author} {\bibfnamefont {H.-F.}\ \bibnamefont {Yu}}, \bibinfo {author} {\bibfnamefont {Z.~D.}\ \bibnamefont {Wang}},\ and\ \bibinfo {author} {\bibfnamefont {Y.}~\bibnamefont {Yu}},\ }\bibfield  {title} {\bibinfo {title} {Simulation and manipulation of tunable {Weyl}-semimetal bands using superconducting quantum circuits},\ }\href {https://doi.org/10.1103/PhysRevLett.122.010501} {\bibfield  {journal} {\bibinfo  {journal} {Phys. Rev. Lett.}\ }\textbf {\bibinfo {volume} {122}},\ \bibinfo {pages} {010501} (\bibinfo {year} {2019}{\natexlab{a}})}\BibitemShut {NoStop}%
\bibitem [{\citenamefont {Tan}\ \emph {et~al.}(2019{\natexlab{b}})\citenamefont {Tan}, \citenamefont {Zhang}, \citenamefont {Yang}, \citenamefont {Chu}, \citenamefont {Zhu}, \citenamefont {Li}, \citenamefont {Yang}, \citenamefont {Song}, \citenamefont {Han}, \citenamefont {Li}, \citenamefont {Dong}, \citenamefont {Yu}, \citenamefont {Yan}, \citenamefont {Zhu},\ and\ \citenamefont {Yu}}]{XSTan2019b}%
  \BibitemOpen
  \bibfield  {author} {\bibinfo {author} {\bibfnamefont {X.}~\bibnamefont {Tan}}, \bibinfo {author} {\bibfnamefont {D.-W.}\ \bibnamefont {Zhang}}, \bibinfo {author} {\bibfnamefont {Z.}~\bibnamefont {Yang}}, \bibinfo {author} {\bibfnamefont {J.}~\bibnamefont {Chu}}, \bibinfo {author} {\bibfnamefont {Y.-Q.}\ \bibnamefont {Zhu}}, \bibinfo {author} {\bibfnamefont {D.}~\bibnamefont {Li}}, \bibinfo {author} {\bibfnamefont {X.}~\bibnamefont {Yang}}, \bibinfo {author} {\bibfnamefont {S.}~\bibnamefont {Song}}, \bibinfo {author} {\bibfnamefont {Z.}~\bibnamefont {Han}}, \bibinfo {author} {\bibfnamefont {Z.}~\bibnamefont {Li}}, \bibinfo {author} {\bibfnamefont {Y.}~\bibnamefont {Dong}}, \bibinfo {author} {\bibfnamefont {H.-F.}\ \bibnamefont {Yu}}, \bibinfo {author} {\bibfnamefont {H.}~\bibnamefont {Yan}}, \bibinfo {author} {\bibfnamefont {S.-L.}\ \bibnamefont {Zhu}},\ and\ \bibinfo {author} {\bibfnamefont {Y.}~\bibnamefont {Yu}},\ }\bibfield  {title} {\bibinfo {title} {Experimental measurement of the quantum metric tensor and related topological phase transition with a superconducting qubit},\ }\href {https://doi.org/10.1103/PhysRevLett.122.210401} {\bibfield  {journal} {\bibinfo  {journal} {Phys. Rev. Lett.}\ }\textbf {\bibinfo {volume} {122}},\ \bibinfo {pages} {210401} (\bibinfo {year} {2019}{\natexlab{b}})}\BibitemShut {NoStop}%
\bibitem [{\citenamefont {Tan}\ \emph {et~al.}(2021)\citenamefont {Tan}, \citenamefont {Zhang}, \citenamefont {Zheng}, \citenamefont {Yang}, \citenamefont {Song}, \citenamefont {Han}, \citenamefont {Dong}, \citenamefont {Wang}, \citenamefont {Lan}, \citenamefont {Yan}, \citenamefont {Zhu},\ and\ \citenamefont {Yu}}]{XSTan2021}%
  \BibitemOpen
  \bibfield  {author} {\bibinfo {author} {\bibfnamefont {X.}~\bibnamefont {Tan}}, \bibinfo {author} {\bibfnamefont {D.-W.}\ \bibnamefont {Zhang}}, \bibinfo {author} {\bibfnamefont {W.}~\bibnamefont {Zheng}}, \bibinfo {author} {\bibfnamefont {X.}~\bibnamefont {Yang}}, \bibinfo {author} {\bibfnamefont {S.}~\bibnamefont {Song}}, \bibinfo {author} {\bibfnamefont {Z.}~\bibnamefont {Han}}, \bibinfo {author} {\bibfnamefont {Y.}~\bibnamefont {Dong}}, \bibinfo {author} {\bibfnamefont {Z.}~\bibnamefont {Wang}}, \bibinfo {author} {\bibfnamefont {D.}~\bibnamefont {Lan}}, \bibinfo {author} {\bibfnamefont {H.}~\bibnamefont {Yan}}, \bibinfo {author} {\bibfnamefont {S.-L.}\ \bibnamefont {Zhu}},\ and\ \bibinfo {author} {\bibfnamefont {Y.}~\bibnamefont {Yu}},\ }\bibfield  {title} {\bibinfo {title} {Experimental observation of tensor monopoles with a superconducting qudit},\ }\href {https://doi.org/10.1103/PhysRevLett.126.017702} {\bibfield  {journal} {\bibinfo  {journal} {Phys. Rev. Lett.}\ }\textbf {\bibinfo {volume} {126}},\ \bibinfo {pages} {017702} (\bibinfo {year} {2021})}\BibitemShut {NoStop}%
\bibitem [{\citenamefont {Chen}\ \emph {et~al.}(2024)\citenamefont {Chen}, \citenamefont {Ding}, \citenamefont {Shen}, \citenamefont {Zhu},\ and\ \citenamefont {Gong}}]{TQChen2024}%
  \BibitemOpen
  \bibfield  {author} {\bibinfo {author} {\bibfnamefont {T.}~\bibnamefont {Chen}}, \bibinfo {author} {\bibfnamefont {H.-T.}\ \bibnamefont {Ding}}, \bibinfo {author} {\bibfnamefont {R.}~\bibnamefont {Shen}}, \bibinfo {author} {\bibfnamefont {S.-L.}\ \bibnamefont {Zhu}},\ and\ \bibinfo {author} {\bibfnamefont {J.}~\bibnamefont {Gong}},\ }\bibfield  {title} {\bibinfo {title} {Direct probe of topology and geometry of quantum states on the {IBM Q} quantum processor},\ }\href {https://doi.org/10.1103/PhysRevB.110.205402} {\bibfield  {journal} {\bibinfo  {journal} {Phys. Rev. B}\ }\textbf {\bibinfo {volume} {110}},\ \bibinfo {pages} {205402} (\bibinfo {year} {2024})}\BibitemShut {NoStop}%
\bibitem [{\citenamefont {Tran}\ \emph {et~al.}(2017)\citenamefont {Tran}, \citenamefont {Dauphin}, \citenamefont {Grushin}, \citenamefont {Zoller},\ and\ \citenamefont {Goldman}}]{Tran2017}%
  \BibitemOpen
  \bibfield  {author} {\bibinfo {author} {\bibfnamefont {D.~T.}\ \bibnamefont {Tran}}, \bibinfo {author} {\bibfnamefont {A.}~\bibnamefont {Dauphin}}, \bibinfo {author} {\bibfnamefont {A.~G.}\ \bibnamefont {Grushin}}, \bibinfo {author} {\bibfnamefont {P.}~\bibnamefont {Zoller}},\ and\ \bibinfo {author} {\bibfnamefont {N.}~\bibnamefont {Goldman}},\ }\bibfield  {title} {\bibinfo {title} {Probing topology by "heating": Quantized circular dichroism in ultracold atoms},\ }\href {https://doi.org/10.1126/sciadv.1701207} {\bibfield  {journal} {\bibinfo  {journal} {Sci. Adv.}\ }\textbf {\bibinfo {volume} {3}},\ \bibinfo {pages} {e1701207} (\bibinfo {year} {2017})}\BibitemShut {NoStop}%
\bibitem [{\citenamefont {Ozawa}\ and\ \citenamefont {Goldman}(2018)}]{Ozawa2018}%
  \BibitemOpen
  \bibfield  {author} {\bibinfo {author} {\bibfnamefont {T.}~\bibnamefont {Ozawa}}\ and\ \bibinfo {author} {\bibfnamefont {N.}~\bibnamefont {Goldman}},\ }\bibfield  {title} {\bibinfo {title} {Extracting the quantum metric tensor through periodic driving},\ }\href {https://doi.org/10.1103/PhysRevB.97.201117} {\bibfield  {journal} {\bibinfo  {journal} {Phys. Rev. B}\ }\textbf {\bibinfo {volume} {97}},\ \bibinfo {pages} {201117} (\bibinfo {year} {2018})}\BibitemShut {NoStop}%
\bibitem [{\citenamefont {Asteria}\ \emph {et~al.}(2019)\citenamefont {Asteria}, \citenamefont {Tran}, \citenamefont {Ozawa}, \citenamefont {Tarnowski}, \citenamefont {Rem}, \citenamefont {Fl{\"a}schner}, \citenamefont {Sengstock}, \citenamefont {Goldman},\ and\ \citenamefont {Weitenberg}}]{Asteria2019}%
  \BibitemOpen
  \bibfield  {author} {\bibinfo {author} {\bibfnamefont {L.}~\bibnamefont {Asteria}}, \bibinfo {author} {\bibfnamefont {D.~T.}\ \bibnamefont {Tran}}, \bibinfo {author} {\bibfnamefont {T.}~\bibnamefont {Ozawa}}, \bibinfo {author} {\bibfnamefont {M.}~\bibnamefont {Tarnowski}}, \bibinfo {author} {\bibfnamefont {B.~S.}\ \bibnamefont {Rem}}, \bibinfo {author} {\bibfnamefont {N.}~\bibnamefont {Fl{\"a}schner}}, \bibinfo {author} {\bibfnamefont {K.}~\bibnamefont {Sengstock}}, \bibinfo {author} {\bibfnamefont {N.}~\bibnamefont {Goldman}},\ and\ \bibinfo {author} {\bibfnamefont {C.}~\bibnamefont {Weitenberg}},\ }\bibfield  {title} {\bibinfo {title} {Measuring quantized circular dichroism in ultracold topological matter},\ }\href {https://doi.org/10.1038/s41567-019-0417-8} {\bibfield  {journal} {\bibinfo  {journal} {Nat. Phys.}\ }\textbf {\bibinfo {volume} {15}},\ \bibinfo {pages} {449} (\bibinfo {year} {2019})}\BibitemShut {NoStop}%
\bibitem [{\citenamefont {Lv}\ \emph {et~al.}(2021)\citenamefont {Lv}, \citenamefont {Du}, \citenamefont {Liang}, \citenamefont {Liu}, \citenamefont {Liang}, \citenamefont {Chen}, \citenamefont {Zhou}, \citenamefont {Zhang}, \citenamefont {Zhang}, \citenamefont {Ai}, \citenamefont {Yan},\ and\ \citenamefont {Zhu}}]{QXLv2021}%
  \BibitemOpen
  \bibfield  {author} {\bibinfo {author} {\bibfnamefont {Q.-X.}\ \bibnamefont {Lv}}, \bibinfo {author} {\bibfnamefont {Y.-X.}\ \bibnamefont {Du}}, \bibinfo {author} {\bibfnamefont {Z.-T.}\ \bibnamefont {Liang}}, \bibinfo {author} {\bibfnamefont {H.-Z.}\ \bibnamefont {Liu}}, \bibinfo {author} {\bibfnamefont {J.-H.}\ \bibnamefont {Liang}}, \bibinfo {author} {\bibfnamefont {L.-Q.}\ \bibnamefont {Chen}}, \bibinfo {author} {\bibfnamefont {L.-M.}\ \bibnamefont {Zhou}}, \bibinfo {author} {\bibfnamefont {S.-C.}\ \bibnamefont {Zhang}}, \bibinfo {author} {\bibfnamefont {D.-W.}\ \bibnamefont {Zhang}}, \bibinfo {author} {\bibfnamefont {B.-Q.}\ \bibnamefont {Ai}}, \bibinfo {author} {\bibfnamefont {H.}~\bibnamefont {Yan}},\ and\ \bibinfo {author} {\bibfnamefont {S.-L.}\ \bibnamefont {Zhu}},\ }\bibfield  {title} {\bibinfo {title} {Measurement of spin {Chern} numbers in quantum simulated topological insulators},\ }\href {https://doi.org/10.1103/PhysRevLett.127.136802} {\bibfield  {journal} {\bibinfo  {journal} {Phys. Rev. Lett.}\ }\textbf {\bibinfo {volume} {127}},\ \bibinfo {pages} {136802} (\bibinfo {year} {2021})}\BibitemShut {NoStop}%
\bibitem [{\citenamefont {Ding}\ \emph {et~al.}(2022)\citenamefont {Ding}, \citenamefont {Zhu}, \citenamefont {He}, \citenamefont {Liu}, \citenamefont {Wang}, \citenamefont {Zhang},\ and\ \citenamefont {Zhu}}]{HTDing2022}%
  \BibitemOpen
  \bibfield  {author} {\bibinfo {author} {\bibfnamefont {H.-T.}\ \bibnamefont {Ding}}, \bibinfo {author} {\bibfnamefont {Y.-Q.}\ \bibnamefont {Zhu}}, \bibinfo {author} {\bibfnamefont {P.}~\bibnamefont {He}}, \bibinfo {author} {\bibfnamefont {Y.-G.}\ \bibnamefont {Liu}}, \bibinfo {author} {\bibfnamefont {J.-T.}\ \bibnamefont {Wang}}, \bibinfo {author} {\bibfnamefont {D.-W.}\ \bibnamefont {Zhang}},\ and\ \bibinfo {author} {\bibfnamefont {S.-L.}\ \bibnamefont {Zhu}},\ }\bibfield  {title} {\bibinfo {title} {Extracting non-{Abelian} quantum metric tensor and its related {Chern} numbers},\ }\href {https://doi.org/10.1103/PhysRevA.105.012210} {\bibfield  {journal} {\bibinfo  {journal} {Phys. Rev. A}\ }\textbf {\bibinfo {volume} {105}},\ \bibinfo {pages} {012210} (\bibinfo {year} {2022})}\BibitemShut {NoStop}%
\bibitem [{\citenamefont {Yi}\ \emph {et~al.}(2023)\citenamefont {Yi}, \citenamefont {Yu}, \citenamefont {Yuan}, \citenamefont {Jiao}, \citenamefont {Yang}, \citenamefont {Jiang}, \citenamefont {Zhang}, \citenamefont {Chen},\ and\ \citenamefont {Pan}}]{CRYi2023}%
  \BibitemOpen
  \bibfield  {author} {\bibinfo {author} {\bibfnamefont {C.-R.}\ \bibnamefont {Yi}}, \bibinfo {author} {\bibfnamefont {J.}~\bibnamefont {Yu}}, \bibinfo {author} {\bibfnamefont {H.}~\bibnamefont {Yuan}}, \bibinfo {author} {\bibfnamefont {R.-H.}\ \bibnamefont {Jiao}}, \bibinfo {author} {\bibfnamefont {Y.-M.}\ \bibnamefont {Yang}}, \bibinfo {author} {\bibfnamefont {X.}~\bibnamefont {Jiang}}, \bibinfo {author} {\bibfnamefont {J.-Y.}\ \bibnamefont {Zhang}}, \bibinfo {author} {\bibfnamefont {S.}~\bibnamefont {Chen}},\ and\ \bibinfo {author} {\bibfnamefont {J.-W.}\ \bibnamefont {Pan}},\ }\bibfield  {title} {\bibinfo {title} {Extracting the quantum geometric tensor of an optical raman lattice by {Bloch}-state tomography},\ }\href {https://doi.org/10.1103/PhysRevResearch.5.L032016} {\bibfield  {journal} {\bibinfo  {journal} {Phys. Rev. Res.}\ }\textbf {\bibinfo {volume} {5}},\ \bibinfo {pages} {L032016} (\bibinfo {year} {2023})}\BibitemShut {NoStop}%
\bibitem [{\citenamefont {Zhu}\ \emph {et~al.}(2007)\citenamefont {Zhu}, \citenamefont {Wang},\ and\ \citenamefont {Duan}}]{SLZhu2007}%
  \BibitemOpen
  \bibfield  {author} {\bibinfo {author} {\bibfnamefont {S.-L.}\ \bibnamefont {Zhu}}, \bibinfo {author} {\bibfnamefont {B.}~\bibnamefont {Wang}},\ and\ \bibinfo {author} {\bibfnamefont {L.-M.}\ \bibnamefont {Duan}},\ }\bibfield  {title} {\bibinfo {title} {Simulation and detection of {Dirac} fermions with cold atoms in an optical lattice},\ }\href {https://doi.org/10.1103/PhysRevLett.98.260402} {\bibfield  {journal} {\bibinfo  {journal} {Phys. Rev. Lett.}\ }\textbf {\bibinfo {volume} {98}},\ \bibinfo {pages} {260402} (\bibinfo {year} {2007})}\BibitemShut {NoStop}%
\bibitem [{\citenamefont {Yu}\ \emph {et~al.}(2019)\citenamefont {Yu}, \citenamefont {Yang}, \citenamefont {Gong}, \citenamefont {Cao}, \citenamefont {Lu}, \citenamefont {Liu}, \citenamefont {Zhang}, \citenamefont {Plenio}, \citenamefont {Jelezko}, \citenamefont {Ozawa}, \citenamefont {Goldman},\ and\ \citenamefont {Cai}}]{MYu2020}%
  \BibitemOpen
  \bibfield  {author} {\bibinfo {author} {\bibfnamefont {M.}~\bibnamefont {Yu}}, \bibinfo {author} {\bibfnamefont {P.}~\bibnamefont {Yang}}, \bibinfo {author} {\bibfnamefont {M.}~\bibnamefont {Gong}}, \bibinfo {author} {\bibfnamefont {Q.}~\bibnamefont {Cao}}, \bibinfo {author} {\bibfnamefont {Q.}~\bibnamefont {Lu}}, \bibinfo {author} {\bibfnamefont {H.}~\bibnamefont {Liu}}, \bibinfo {author} {\bibfnamefont {S.}~\bibnamefont {Zhang}}, \bibinfo {author} {\bibfnamefont {M.~B.}\ \bibnamefont {Plenio}}, \bibinfo {author} {\bibfnamefont {F.}~\bibnamefont {Jelezko}}, \bibinfo {author} {\bibfnamefont {T.}~\bibnamefont {Ozawa}}, \bibinfo {author} {\bibfnamefont {N.}~\bibnamefont {Goldman}},\ and\ \bibinfo {author} {\bibfnamefont {J.}~\bibnamefont {Cai}},\ }\bibfield  {title} {\bibinfo {title} {Experimental measurement of the quantum geometric tensor using coupled qubits in diamond},\ }\href {https://doi.org/10.1093/nsr/nwz193} {\bibfield  {journal} {\bibinfo  {journal} {Natl. Sci. Rev.}\ }\textbf {\bibinfo {volume} {7}},\ \bibinfo {pages} {254} (\bibinfo {year} {2019})}\BibitemShut {NoStop}%
\bibitem [{\citenamefont {Chen}\ \emph {et~al.}(2022)\citenamefont {Chen}, \citenamefont {Li}, \citenamefont {Palumbo}, \citenamefont {Zhu}, \citenamefont {Goldman},\ and\ \citenamefont {Cappellaro}}]{MChen2022}%
  \BibitemOpen
  \bibfield  {author} {\bibinfo {author} {\bibfnamefont {M.}~\bibnamefont {Chen}}, \bibinfo {author} {\bibfnamefont {C.}~\bibnamefont {Li}}, \bibinfo {author} {\bibfnamefont {G.}~\bibnamefont {Palumbo}}, \bibinfo {author} {\bibfnamefont {Y.-Q.}\ \bibnamefont {Zhu}}, \bibinfo {author} {\bibfnamefont {N.}~\bibnamefont {Goldman}},\ and\ \bibinfo {author} {\bibfnamefont {P.}~\bibnamefont {Cappellaro}},\ }\bibfield  {title} {\bibinfo {title} {A synthetic monopole source of {Kalb-Ramond} field in diamond},\ }\href {https://doi.org/10.1126/science.abe6437} {\bibfield  {journal} {\bibinfo  {journal} {Science}\ }\textbf {\bibinfo {volume} {375}},\ \bibinfo {pages} {1017} (\bibinfo {year} {2022})}\BibitemShut {NoStop}%
\bibitem [{\citenamefont {Zhang}\ \emph {et~al.}(2023)\citenamefont {Zhang}, \citenamefont {Lu}, \citenamefont {Liu}, \citenamefont {Ding},\ and\ \citenamefont {Wang}}]{XYZhang2023}%
  \BibitemOpen
  \bibfield  {author} {\bibinfo {author} {\bibfnamefont {X.}~\bibnamefont {Zhang}}, \bibinfo {author} {\bibfnamefont {X.-M.}\ \bibnamefont {Lu}}, \bibinfo {author} {\bibfnamefont {J.}~\bibnamefont {Liu}}, \bibinfo {author} {\bibfnamefont {W.}~\bibnamefont {Ding}},\ and\ \bibinfo {author} {\bibfnamefont {X.}~\bibnamefont {Wang}},\ }\bibfield  {title} {\bibinfo {title} {Direct measurement of quantum {Fisher} information},\ }\href {https://doi.org/10.1103/PhysRevA.107.012414} {\bibfield  {journal} {\bibinfo  {journal} {Phys. Rev. A}\ }\textbf {\bibinfo {volume} {107}},\ \bibinfo {pages} {012414} (\bibinfo {year} {2023})}\BibitemShut {NoStop}%
\bibitem [{\citenamefont {Gianfrate}\ \emph {et~al.}(2020)\citenamefont {Gianfrate}, \citenamefont {Bleu}, \citenamefont {Dominici}, \citenamefont {Ardizzone}, \citenamefont {De~Giorgi}, \citenamefont {Ballarini}, \citenamefont {Lerario}, \citenamefont {West}, \citenamefont {Pfeiffer}, \citenamefont {Solnyshkov}, \citenamefont {Sanvitto},\ and\ \citenamefont {Malpuech}}]{Gianfrate2020}%
  \BibitemOpen
  \bibfield  {author} {\bibinfo {author} {\bibfnamefont {A.}~\bibnamefont {Gianfrate}}, \bibinfo {author} {\bibfnamefont {O.}~\bibnamefont {Bleu}}, \bibinfo {author} {\bibfnamefont {L.}~\bibnamefont {Dominici}}, \bibinfo {author} {\bibfnamefont {V.}~\bibnamefont {Ardizzone}}, \bibinfo {author} {\bibfnamefont {M.}~\bibnamefont {De~Giorgi}}, \bibinfo {author} {\bibfnamefont {D.}~\bibnamefont {Ballarini}}, \bibinfo {author} {\bibfnamefont {G.}~\bibnamefont {Lerario}}, \bibinfo {author} {\bibfnamefont {K.~W.}\ \bibnamefont {West}}, \bibinfo {author} {\bibfnamefont {L.~N.}\ \bibnamefont {Pfeiffer}}, \bibinfo {author} {\bibfnamefont {D.~D.}\ \bibnamefont {Solnyshkov}}, \bibinfo {author} {\bibfnamefont {D.}~\bibnamefont {Sanvitto}},\ and\ \bibinfo {author} {\bibfnamefont {G.}~\bibnamefont {Malpuech}},\ }\bibfield  {title} {\bibinfo {title} {Measurement of the quantum geometric tensor and of the anomalous {{Hall}} drift},\ }\href {https://doi.org/10.1038/s41586-020-1989-2} {\bibfield  {journal} {\bibinfo  {journal} {Nature (London)}\ }\textbf {\bibinfo {volume} {578}},\ \bibinfo {pages} {381} (\bibinfo {year} {2020})}\BibitemShut {NoStop}%
\bibitem [{\citenamefont {Kim}\ \emph {et~al.}(2025)\citenamefont {Kim}, \citenamefont {Chung}, \citenamefont {Qian}, \citenamefont {Park}, \citenamefont {Jozwiak}, \citenamefont {Rotenberg}, \citenamefont {Bostwick}, \citenamefont {Kim},\ and\ \citenamefont {Yang}}]{Kim2025}%
  \BibitemOpen
  \bibfield  {author} {\bibinfo {author} {\bibfnamefont {S.}~\bibnamefont {Kim}}, \bibinfo {author} {\bibfnamefont {Y.}~\bibnamefont {Chung}}, \bibinfo {author} {\bibfnamefont {Y.}~\bibnamefont {Qian}}, \bibinfo {author} {\bibfnamefont {S.}~\bibnamefont {Park}}, \bibinfo {author} {\bibfnamefont {C.}~\bibnamefont {Jozwiak}}, \bibinfo {author} {\bibfnamefont {E.}~\bibnamefont {Rotenberg}}, \bibinfo {author} {\bibfnamefont {A.}~\bibnamefont {Bostwick}}, \bibinfo {author} {\bibfnamefont {K.~S.}\ \bibnamefont {Kim}},\ and\ \bibinfo {author} {\bibfnamefont {B.-J.}\ \bibnamefont {Yang}},\ }\bibfield  {title} {\bibinfo {title} {Direct measurement of the quantum metric tensor in solids},\ }\href {https://doi.org/10.1126/science.ado6049} {\bibfield  {journal} {\bibinfo  {journal} {Science}\ }\textbf {\bibinfo {volume} {388}},\ \bibinfo {pages} {1050} (\bibinfo {year} {2025})}\BibitemShut {NoStop}%
\bibitem [{\citenamefont {Ghatak}\ and\ \citenamefont {Das}(2019)}]{Ghatak2019}%
  \BibitemOpen
  \bibfield  {author} {\bibinfo {author} {\bibfnamefont {A.}~\bibnamefont {Ghatak}}\ and\ \bibinfo {author} {\bibfnamefont {T.}~\bibnamefont {Das}},\ }\bibfield  {title} {\bibinfo {title} {New topological invariants in non-{Hermitian} systems},\ }\href {https://doi.org/10.1088/1361-648x/ab11b3} {\bibfield  {journal} {\bibinfo  {journal} {J. Phys.: Condens. Matter}\ }\textbf {\bibinfo {volume} {31}},\ \bibinfo {pages} {263001} (\bibinfo {year} {2019})}\BibitemShut {NoStop}%
\bibitem [{\citenamefont {Ashida}\ \emph {et~al.}(2020)\citenamefont {Ashida}, \citenamefont {Gong},\ and\ \citenamefont {Ueda}}]{Ashida2020}%
  \BibitemOpen
  \bibfield  {author} {\bibinfo {author} {\bibfnamefont {Y.}~\bibnamefont {Ashida}}, \bibinfo {author} {\bibfnamefont {Z.}~\bibnamefont {Gong}},\ and\ \bibinfo {author} {\bibfnamefont {M.}~\bibnamefont {Ueda}},\ }\bibfield  {title} {\bibinfo {title} {Non-{Hermitian} physics},\ }\href {https://doi.org/10.1080/00018732.2021.1876991} {\bibfield  {journal} {\bibinfo  {journal} {Adv. Phys.}\ }\textbf {\bibinfo {volume} {69}},\ \bibinfo {pages} {249} (\bibinfo {year} {2020})}\BibitemShut {NoStop}%
\bibitem [{\citenamefont {Bergholtz}\ \emph {et~al.}(2021)\citenamefont {Bergholtz}, \citenamefont {Budich},\ and\ \citenamefont {Kunst}}]{Bergholtz2021}%
  \BibitemOpen
  \bibfield  {author} {\bibinfo {author} {\bibfnamefont {E.~J.}\ \bibnamefont {Bergholtz}}, \bibinfo {author} {\bibfnamefont {J.~C.}\ \bibnamefont {Budich}},\ and\ \bibinfo {author} {\bibfnamefont {F.~K.}\ \bibnamefont {Kunst}},\ }\bibfield  {title} {\bibinfo {title} {Exceptional topology of non-{Hermitian} systems},\ }\href {https://doi.org/10.1103/RevModPhys.93.015005} {\bibfield  {journal} {\bibinfo  {journal} {Rev. Mod. Phys.}\ }\textbf {\bibinfo {volume} {93}},\ \bibinfo {pages} {015005} (\bibinfo {year} {2021})}\BibitemShut {NoStop}%
\bibitem [{\citenamefont {Solnyshkov}\ \emph {et~al.}(2021)\citenamefont {Solnyshkov}, \citenamefont {Leblanc}, \citenamefont {Bessonart}, \citenamefont {Nalitov}, \citenamefont {Ren}, \citenamefont {Liao}, \citenamefont {Li},\ and\ \citenamefont {Malpuech}}]{Solnyshkov2021}%
  \BibitemOpen
  \bibfield  {author} {\bibinfo {author} {\bibfnamefont {D.~D.}\ \bibnamefont {Solnyshkov}}, \bibinfo {author} {\bibfnamefont {C.}~\bibnamefont {Leblanc}}, \bibinfo {author} {\bibfnamefont {L.}~\bibnamefont {Bessonart}}, \bibinfo {author} {\bibfnamefont {A.}~\bibnamefont {Nalitov}}, \bibinfo {author} {\bibfnamefont {J.}~\bibnamefont {Ren}}, \bibinfo {author} {\bibfnamefont {Q.}~\bibnamefont {Liao}}, \bibinfo {author} {\bibfnamefont {F.}~\bibnamefont {Li}},\ and\ \bibinfo {author} {\bibfnamefont {G.}~\bibnamefont {Malpuech}},\ }\bibfield  {title} {\bibinfo {title} {Quantum metric and wave packets at exceptional points in non-{Hermitian} systems},\ }\href {https://doi.org/10.1103/PhysRevB.103.125302} {\bibfield  {journal} {\bibinfo  {journal} {Phys. Rev. B}\ }\textbf {\bibinfo {volume} {103}},\ \bibinfo {pages} {125302} (\bibinfo {year} {2021})}\BibitemShut {NoStop}%
\bibitem [{\citenamefont {Alon}\ \emph {et~al.}(2024)\citenamefont {Alon}, \citenamefont {Ilan},\ and\ \citenamefont {Goldstein}}]{Alon2024}%
  \BibitemOpen
  \bibfield  {author} {\bibinfo {author} {\bibfnamefont {B.}~\bibnamefont {Alon}}, \bibinfo {author} {\bibfnamefont {R.}~\bibnamefont {Ilan}},\ and\ \bibinfo {author} {\bibfnamefont {M.}~\bibnamefont {Goldstein}},\ }\bibfield  {title} {\bibinfo {title} {Quantum metric dependent anomalous velocity in systems subject to complex electric fields},\ }\href {https://doi.org/10.1103/PhysRevB.110.245103} {\bibfield  {journal} {\bibinfo  {journal} {Phys. Rev. B}\ }\textbf {\bibinfo {volume} {110}},\ \bibinfo {pages} {245103} (\bibinfo {year} {2024})}\BibitemShut {NoStop}%
\bibitem [{\citenamefont {Cuerda}\ \emph {et~al.}(2024{\natexlab{a}})\citenamefont {Cuerda}, \citenamefont {Taskinen}, \citenamefont {K\"allman}, \citenamefont {Grabitz},\ and\ \citenamefont {T\"orm\"a}}]{Cuerda2024b}%
  \BibitemOpen
  \bibfield  {author} {\bibinfo {author} {\bibfnamefont {J.}~\bibnamefont {Cuerda}}, \bibinfo {author} {\bibfnamefont {J.~M.}\ \bibnamefont {Taskinen}}, \bibinfo {author} {\bibfnamefont {N.}~\bibnamefont {K\"allman}}, \bibinfo {author} {\bibfnamefont {L.}~\bibnamefont {Grabitz}},\ and\ \bibinfo {author} {\bibfnamefont {P.}~\bibnamefont {T\"orm\"a}},\ }\bibfield  {title} {\bibinfo {title} {Pseudospin-orbit coupling and non-{Hermitian} effects in the quantum geometric tensor of a plasmonic lattice},\ }\href {https://doi.org/10.1103/PhysRevB.109.165439} {\bibfield  {journal} {\bibinfo  {journal} {Phys. Rev. B}\ }\textbf {\bibinfo {volume} {109}},\ \bibinfo {pages} {165439} (\bibinfo {year} {2024}{\natexlab{a}})}\BibitemShut {NoStop}%
\bibitem [{\citenamefont {Ren}\ \emph {et~al.}(2024)\citenamefont {Ren}, \citenamefont {Li}, \citenamefont {Ding},\ and\ \citenamefont {Zhang}}]{JFRen2024}%
  \BibitemOpen
  \bibfield  {author} {\bibinfo {author} {\bibfnamefont {J.-F.}\ \bibnamefont {Ren}}, \bibinfo {author} {\bibfnamefont {J.}~\bibnamefont {Li}}, \bibinfo {author} {\bibfnamefont {H.-T.}\ \bibnamefont {Ding}},\ and\ \bibinfo {author} {\bibfnamefont {D.-W.}\ \bibnamefont {Zhang}},\ }\bibfield  {title} {\bibinfo {title} {Identifying non-{Hermitian} critical points with the quantum metric},\ }\href {https://doi.org/10.1103/PhysRevA.110.052203} {\bibfield  {journal} {\bibinfo  {journal} {Phys. Rev. A}\ }\textbf {\bibinfo {volume} {110}},\ \bibinfo {pages} {052203} (\bibinfo {year} {2024})}\BibitemShut {NoStop}%
\bibitem [{\citenamefont {Brody}\ and\ \citenamefont {Graefe}(2013)}]{Brody2013}%
  \BibitemOpen
  \bibfield  {author} {\bibinfo {author} {\bibfnamefont {D.~C.}\ \bibnamefont {Brody}}\ and\ \bibinfo {author} {\bibfnamefont {E.-M.}\ \bibnamefont {Graefe}},\ }\bibfield  {title} {\bibinfo {title} {Information geometry of complex {Hamiltonians} and exceptional points},\ }\href {https://doi.org/10.3390/e15093361} {\bibfield  {journal} {\bibinfo  {journal} {Entropy}\ }\textbf {\bibinfo {volume} {15}},\ \bibinfo {pages} {3361} (\bibinfo {year} {2013})}\BibitemShut {NoStop}%
\bibitem [{\citenamefont {Zhang}\ \emph {et~al.}(2019)\citenamefont {Zhang}, \citenamefont {Wang},\ and\ \citenamefont {Gong}}]{DJZhang2019}%
  \BibitemOpen
  \bibfield  {author} {\bibinfo {author} {\bibfnamefont {D.-J.}\ \bibnamefont {Zhang}}, \bibinfo {author} {\bibfnamefont {Q.-h.}\ \bibnamefont {Wang}},\ and\ \bibinfo {author} {\bibfnamefont {J.}~\bibnamefont {Gong}},\ }\bibfield  {title} {\bibinfo {title} {Quantum geometric tensor in $\mathcal{PT}$-symmetric quantum mechanics},\ }\href {https://doi.org/10.1103/PhysRevA.99.042104} {\bibfield  {journal} {\bibinfo  {journal} {Phys. Rev. A}\ }\textbf {\bibinfo {volume} {99}},\ \bibinfo {pages} {042104} (\bibinfo {year} {2019})}\BibitemShut {NoStop}%
\bibitem [{\citenamefont {Sun}\ \emph {et~al.}(2022)\citenamefont {Sun}, \citenamefont {Tang},\ and\ \citenamefont {Kou}}]{GYSun2022}%
  \BibitemOpen
  \bibfield  {author} {\bibinfo {author} {\bibfnamefont {G.}~\bibnamefont {Sun}}, \bibinfo {author} {\bibfnamefont {J.-C.}\ \bibnamefont {Tang}},\ and\ \bibinfo {author} {\bibfnamefont {S.-P.}\ \bibnamefont {Kou}},\ }\bibfield  {title} {\bibinfo {title} {Biorthogonal quantum criticality in non-{Hermitian} many-body systems},\ }\href {https://doi.org/10.1007/s11467-021-1126-1} {\bibfield  {journal} {\bibinfo  {journal} {Front. Phys.}\ }\textbf {\bibinfo {volume} {17}},\ \bibinfo {pages} {33502} (\bibinfo {year} {2022})}\BibitemShut {NoStop}%
\bibitem [{\citenamefont {Chen~Ye}\ \emph {et~al.}(2024)\citenamefont {Chen~Ye}, \citenamefont {Vleeshouwers}, \citenamefont {Heatley}, \citenamefont {Gritsev},\ and\ \citenamefont {Morais~Smith}}]{CYC2024}%
  \BibitemOpen
  \bibfield  {author} {\bibinfo {author} {\bibfnamefont {C.}~\bibnamefont {Chen~Ye}}, \bibinfo {author} {\bibfnamefont {W.~L.}\ \bibnamefont {Vleeshouwers}}, \bibinfo {author} {\bibfnamefont {S.}~\bibnamefont {Heatley}}, \bibinfo {author} {\bibfnamefont {V.}~\bibnamefont {Gritsev}},\ and\ \bibinfo {author} {\bibfnamefont {C.}~\bibnamefont {Morais~Smith}},\ }\bibfield  {title} {\bibinfo {title} {Quantum metric of non-{Hermitian} {Su-Schrieffer-Heeger} systems},\ }\href {https://doi.org/10.1103/PhysRevResearch.6.023202} {\bibfield  {journal} {\bibinfo  {journal} {Phys. Rev. Res.}\ }\textbf {\bibinfo {volume} {6}},\ \bibinfo {pages} {023202} (\bibinfo {year} {2024})}\BibitemShut {NoStop}%
\bibitem [{\citenamefont {Tzeng}\ \emph {et~al.}(2021)\citenamefont {Tzeng}, \citenamefont {Ju}, \citenamefont {Chen},\ and\ \citenamefont {Huang}}]{Tzeng2021}%
  \BibitemOpen
  \bibfield  {author} {\bibinfo {author} {\bibfnamefont {Y.-C.}\ \bibnamefont {Tzeng}}, \bibinfo {author} {\bibfnamefont {C.-Y.}\ \bibnamefont {Ju}}, \bibinfo {author} {\bibfnamefont {G.-Y.}\ \bibnamefont {Chen}},\ and\ \bibinfo {author} {\bibfnamefont {W.-M.}\ \bibnamefont {Huang}},\ }\bibfield  {title} {\bibinfo {title} {Hunting for the non-{Hermitian} exceptional points with fidelity susceptibility},\ }\href {https://doi.org/10.1103/PhysRevResearch.3.013015} {\bibfield  {journal} {\bibinfo  {journal} {Phys. Rev. Res.}\ }\textbf {\bibinfo {volume} {3}},\ \bibinfo {pages} {013015} (\bibinfo {year} {2021})}\BibitemShut {NoStop}%
\bibitem [{\citenamefont {Tu}\ \emph {et~al.}(2023)\citenamefont {Tu}, \citenamefont {Jang}, \citenamefont {Chang},\ and\ \citenamefont {Tzeng}}]{YTTu2023}%
  \BibitemOpen
  \bibfield  {author} {\bibinfo {author} {\bibfnamefont {Y.-T.}\ \bibnamefont {Tu}}, \bibinfo {author} {\bibfnamefont {I.}~\bibnamefont {Jang}}, \bibinfo {author} {\bibfnamefont {P.-Y.}\ \bibnamefont {Chang}},\ and\ \bibinfo {author} {\bibfnamefont {Y.-C.}\ \bibnamefont {Tzeng}},\ }\bibfield  {title} {\bibinfo {title} {General properties of fidelity in non-{H}ermitian quantum systems with {PT} symmetry},\ }\href {https://doi.org/10.22331/q-2023-03-23-960} {\bibfield  {journal} {\bibinfo  {journal} {{Quantum}}\ }\textbf {\bibinfo {volume} {7}},\ \bibinfo {pages} {960} (\bibinfo {year} {2023})}\BibitemShut {NoStop}%
\bibitem [{\citenamefont {Fan}\ \emph {et~al.}(2020)\citenamefont {Fan}, \citenamefont {Huang},\ and\ \citenamefont {Liang}}]{ANFan2020}%
  \BibitemOpen
  \bibfield  {author} {\bibinfo {author} {\bibfnamefont {A.}~\bibnamefont {Fan}}, \bibinfo {author} {\bibfnamefont {G.-Y.}\ \bibnamefont {Huang}},\ and\ \bibinfo {author} {\bibfnamefont {S.-D.}\ \bibnamefont {Liang}},\ }\bibfield  {title} {\bibinfo {title} {Complex {Berry} curvature pair and quantum {Hall} admittance in non-{Hermitian} systems},\ }\href {https://doi.org/10.1088/2399-6528/abcab6} {\bibfield  {journal} {\bibinfo  {journal} {J. Phys. Commun.}\ }\textbf {\bibinfo {volume} {4}},\ \bibinfo {pages} {115006} (\bibinfo {year} {2020})}\BibitemShut {NoStop}%
\bibitem [{\citenamefont {He}\ \emph {et~al.}(2021)\citenamefont {He}, \citenamefont {Ding},\ and\ \citenamefont {Zhu}}]{PHe2021}%
  \BibitemOpen
  \bibfield  {author} {\bibinfo {author} {\bibfnamefont {P.}~\bibnamefont {He}}, \bibinfo {author} {\bibfnamefont {H.-T.}\ \bibnamefont {Ding}},\ and\ \bibinfo {author} {\bibfnamefont {S.-L.}\ \bibnamefont {Zhu}},\ }\bibfield  {title} {\bibinfo {title} {Geometry and superfluidity of the flat band in a non-{Hermitian} optical lattice},\ }\href {https://doi.org/10.1103/PhysRevA.103.043329} {\bibfield  {journal} {\bibinfo  {journal} {Phys. Rev. A}\ }\textbf {\bibinfo {volume} {103}},\ \bibinfo {pages} {043329} (\bibinfo {year} {2021})}\BibitemShut {NoStop}%
\bibitem [{\citenamefont {Zhu}\ \emph {et~al.}(2021)\citenamefont {Zhu}, \citenamefont {Zheng}, \citenamefont {Zhu},\ and\ \citenamefont {Palumbo}}]{YQZhu2021}%
  \BibitemOpen
  \bibfield  {author} {\bibinfo {author} {\bibfnamefont {Y.-Q.}\ \bibnamefont {Zhu}}, \bibinfo {author} {\bibfnamefont {W.}~\bibnamefont {Zheng}}, \bibinfo {author} {\bibfnamefont {S.-L.}\ \bibnamefont {Zhu}},\ and\ \bibinfo {author} {\bibfnamefont {G.}~\bibnamefont {Palumbo}},\ }\bibfield  {title} {\bibinfo {title} {Band topology of pseudo-{Hermitian} phases through tensor {Berry} connections and quantum metric},\ }\href {https://doi.org/10.1103/PhysRevB.104.205103} {\bibfield  {journal} {\bibinfo  {journal} {Phys. Rev. B}\ }\textbf {\bibinfo {volume} {104}},\ \bibinfo {pages} {205103} (\bibinfo {year} {2021})}\BibitemShut {NoStop}%
\bibitem [{\citenamefont {He}\ \emph {et~al.}(2023)\citenamefont {He}, \citenamefont {Zhu}, \citenamefont {Wang},\ and\ \citenamefont {Zhu}}]{PHe2023}%
  \BibitemOpen
  \bibfield  {author} {\bibinfo {author} {\bibfnamefont {P.}~\bibnamefont {He}}, \bibinfo {author} {\bibfnamefont {Y.-Q.}\ \bibnamefont {Zhu}}, \bibinfo {author} {\bibfnamefont {J.-T.}\ \bibnamefont {Wang}},\ and\ \bibinfo {author} {\bibfnamefont {S.-L.}\ \bibnamefont {Zhu}},\ }\bibfield  {title} {\bibinfo {title} {Quantum quenches in a pseudo-{Hermitian} {Chern} insulator},\ }\href {https://doi.org/10.1103/PhysRevA.107.012219} {\bibfield  {journal} {\bibinfo  {journal} {Phys. Rev. A}\ }\textbf {\bibinfo {volume} {107}},\ \bibinfo {pages} {012219} (\bibinfo {year} {2023})}\BibitemShut {NoStop}%
\bibitem [{\citenamefont {Hu}\ \emph {et~al.}(2025)\citenamefont {Hu}, \citenamefont {Ostrovskaya},\ and\ \citenamefont {Estrecho}}]{YMRobinHu2025}%
  \BibitemOpen
  \bibfield  {author} {\bibinfo {author} {\bibfnamefont {Y.-M.~R.}\ \bibnamefont {Hu}}, \bibinfo {author} {\bibfnamefont {E.~A.}\ \bibnamefont {Ostrovskaya}},\ and\ \bibinfo {author} {\bibfnamefont {E.}~\bibnamefont {Estrecho}},\ }\bibfield  {title} {\bibinfo {title} {Quantum geometric tensor and wavepacket dynamics in two-dimensional non-{Hermitian} systems},\ }\href {https://doi.org/10.1103/PhysRevResearch.7.L012067} {\bibfield  {journal} {\bibinfo  {journal} {Phys. Rev. Res.}\ }\textbf {\bibinfo {volume} {7}},\ \bibinfo {pages} {L012067} (\bibinfo {year} {2025})}\BibitemShut {NoStop}%
\bibitem [{\citenamefont {Silberstein}\ \emph {et~al.}(2020)\citenamefont {Silberstein}, \citenamefont {Behrends}, \citenamefont {Goldstein},\ and\ \citenamefont {Ilan}}]{Silberstein2020}%
  \BibitemOpen
  \bibfield  {author} {\bibinfo {author} {\bibfnamefont {N.}~\bibnamefont {Silberstein}}, \bibinfo {author} {\bibfnamefont {J.}~\bibnamefont {Behrends}}, \bibinfo {author} {\bibfnamefont {M.}~\bibnamefont {Goldstein}},\ and\ \bibinfo {author} {\bibfnamefont {R.}~\bibnamefont {Ilan}},\ }\bibfield  {title} {\bibinfo {title} {Berry connection induced anomalous wave-packet dynamics in non-{Hermitian} systems},\ }\href {https://doi.org/10.1103/PhysRevB.102.245147} {\bibfield  {journal} {\bibinfo  {journal} {Phys. Rev. B}\ }\textbf {\bibinfo {volume} {102}},\ \bibinfo {pages} {245147} (\bibinfo {year} {2020})}\BibitemShut {NoStop}%
\bibitem [{\citenamefont {Wang}\ \emph {et~al.}(2022)\citenamefont {Wang}, \citenamefont {Tao},\ and\ \citenamefont {Xu}}]{JHWang2022}%
  \BibitemOpen
  \bibfield  {author} {\bibinfo {author} {\bibfnamefont {J.-H.}\ \bibnamefont {Wang}}, \bibinfo {author} {\bibfnamefont {Y.-L.}\ \bibnamefont {Tao}},\ and\ \bibinfo {author} {\bibfnamefont {Y.}~\bibnamefont {Xu}},\ }\bibfield  {title} {\bibinfo {title} {Anomalous transport induced by non-{Hermitian} anomalous {Berry} connection in non-{Hermitian} systems},\ }\href {https://doi.org/10.1088/0256-307X/39/1/010301} {\bibfield  {journal} {\bibinfo  {journal} {Chin. Phys. Lett.}\ }\textbf {\bibinfo {volume} {39}},\ \bibinfo {pages} {010301} (\bibinfo {year} {2022})}\BibitemShut {NoStop}%
\bibitem [{\citenamefont {Shen}\ \emph {et~al.}(2018)\citenamefont {Shen}, \citenamefont {Zhen},\ and\ \citenamefont {Fu}}]{HTShen2018}%
  \BibitemOpen
  \bibfield  {author} {\bibinfo {author} {\bibfnamefont {H.}~\bibnamefont {Shen}}, \bibinfo {author} {\bibfnamefont {B.}~\bibnamefont {Zhen}},\ and\ \bibinfo {author} {\bibfnamefont {L.}~\bibnamefont {Fu}},\ }\bibfield  {title} {\bibinfo {title} {Topological band theory for non-{Hermitian} {Hamiltonians}},\ }\href {https://doi.org/10.1103/PhysRevLett.120.146402} {\bibfield  {journal} {\bibinfo  {journal} {Phys. Rev. Lett.}\ }\textbf {\bibinfo {volume} {120}},\ \bibinfo {pages} {146402} (\bibinfo {year} {2018})}\BibitemShut {NoStop}%
\bibitem [{\citenamefont {Qin}\ \emph {et~al.}(2025)\citenamefont {Qin}, \citenamefont {Shen},\ and\ \citenamefont {Lee}}]{FQin2025}%
  \BibitemOpen
  \bibfield  {author} {\bibinfo {author} {\bibfnamefont {F.}~\bibnamefont {Qin}}, \bibinfo {author} {\bibfnamefont {R.}~\bibnamefont {Shen}},\ and\ \bibinfo {author} {\bibfnamefont {C.~H.}\ \bibnamefont {Lee}},\ }\bibfield  {title} {\bibinfo {title} {Nonlinear {Hall} effects with an exceptional ring},\ }\href {https://doi.org/10.1103/8g3q-qrpg} {\bibfield  {journal} {\bibinfo  {journal} {Phys. Rev. B}\ }\textbf {\bibinfo {volume} {111}},\ \bibinfo {pages} {245413} (\bibinfo {year} {2025})}\BibitemShut {NoStop}%
\bibitem [{\citenamefont {Chen}\ and\ \citenamefont {Zhu}()}]{KChen2025}%
  \BibitemOpen
  \bibfield  {author} {\bibinfo {author} {\bibfnamefont {K.}~\bibnamefont {Chen}}\ and\ \bibinfo {author} {\bibfnamefont {J.}~\bibnamefont {Zhu}},\ }\href {https://arxiv.org/abs/2509.11765} {\bibinfo {title} {Non-{Hermitian} quantum geometric tensor and nonlinear electrical response}},\ \Eprint {https://arxiv.org/abs/2509.11765} {arXiv:2509.11765} \BibitemShut {NoStop}%
\bibitem [{\citenamefont {Liao}\ \emph {et~al.}(2021)\citenamefont {Liao}, \citenamefont {Leblanc}, \citenamefont {Ren}, \citenamefont {Li}, \citenamefont {Li}, \citenamefont {Solnyshkov}, \citenamefont {Malpuech}, \citenamefont {Yao},\ and\ \citenamefont {Fu}}]{QLiao2021}%
  \BibitemOpen
  \bibfield  {author} {\bibinfo {author} {\bibfnamefont {Q.}~\bibnamefont {Liao}}, \bibinfo {author} {\bibfnamefont {C.}~\bibnamefont {Leblanc}}, \bibinfo {author} {\bibfnamefont {J.}~\bibnamefont {Ren}}, \bibinfo {author} {\bibfnamefont {F.}~\bibnamefont {Li}}, \bibinfo {author} {\bibfnamefont {Y.}~\bibnamefont {Li}}, \bibinfo {author} {\bibfnamefont {D.}~\bibnamefont {Solnyshkov}}, \bibinfo {author} {\bibfnamefont {G.}~\bibnamefont {Malpuech}}, \bibinfo {author} {\bibfnamefont {J.}~\bibnamefont {Yao}},\ and\ \bibinfo {author} {\bibfnamefont {H.}~\bibnamefont {Fu}},\ }\bibfield  {title} {\bibinfo {title} {Experimental measurement of the divergent quantum metric of an exceptional point},\ }\href {https://doi.org/10.1103/PhysRevLett.127.107402} {\bibfield  {journal} {\bibinfo  {journal} {Phys. Rev. Lett.}\ }\textbf {\bibinfo {volume} {127}},\ \bibinfo {pages} {107402} (\bibinfo {year} {2021})}\BibitemShut {NoStop}%
\bibitem [{\citenamefont {Cuerda}\ \emph {et~al.}(2024{\natexlab{b}})\citenamefont {Cuerda}, \citenamefont {Taskinen}, \citenamefont {K\"allman}, \citenamefont {Grabitz},\ and\ \citenamefont {T\"orm\"a}}]{Cuerda2024}%
  \BibitemOpen
  \bibfield  {author} {\bibinfo {author} {\bibfnamefont {J.}~\bibnamefont {Cuerda}}, \bibinfo {author} {\bibfnamefont {J.~M.}\ \bibnamefont {Taskinen}}, \bibinfo {author} {\bibfnamefont {N.}~\bibnamefont {K\"allman}}, \bibinfo {author} {\bibfnamefont {L.}~\bibnamefont {Grabitz}},\ and\ \bibinfo {author} {\bibfnamefont {P.}~\bibnamefont {T\"orm\"a}},\ }\bibfield  {title} {\bibinfo {title} {Observation of quantum metric and non-{Hermitian} {Berry} curvature in a plasmonic lattice},\ }\href {https://doi.org/10.1103/PhysRevResearch.6.L022020} {\bibfield  {journal} {\bibinfo  {journal} {Phys. Rev. Res.}\ }\textbf {\bibinfo {volume} {6}},\ \bibinfo {pages} {L022020} (\bibinfo {year} {2024}{\natexlab{b}})}\BibitemShut {NoStop}%
\bibitem [{\citenamefont {Hu}\ \emph {et~al.}(2024)\citenamefont {Hu}, \citenamefont {Ostrovskaya},\ and\ \citenamefont {Estrecho}}]{YMRobinHu2024}%
  \BibitemOpen
  \bibfield  {author} {\bibinfo {author} {\bibfnamefont {Y.-M.~R.}\ \bibnamefont {Hu}}, \bibinfo {author} {\bibfnamefont {E.~A.}\ \bibnamefont {Ostrovskaya}},\ and\ \bibinfo {author} {\bibfnamefont {E.}~\bibnamefont {Estrecho}},\ }\bibfield  {title} {\bibinfo {title} {Generalized quantum geometric tensor in a non-{Hermitian} exciton-polariton system [invited]},\ }\href {https://doi.org/10.1364/OME.497010} {\bibfield  {journal} {\bibinfo  {journal} {Opt. Mater. Express}\ }\textbf {\bibinfo {volume} {14}},\ \bibinfo {pages} {664} (\bibinfo {year} {2024})}\BibitemShut {NoStop}%
\bibitem [{\citenamefont {Mostafazadeh}(2002{\natexlab{a}})}]{Mostafazadeh2002}%
  \BibitemOpen
  \bibfield  {author} {\bibinfo {author} {\bibfnamefont {A.}~\bibnamefont {Mostafazadeh}},\ }\bibfield  {title} {\bibinfo {title} {Pseudo-{Hermiticity} versus {PT} symmetry: The necessary condition for the reality of the spectrum of a non-{Hermitian} {Hamiltonian}},\ }\href {https://doi.org/10.1063/1.1418246} {\bibfield  {journal} {\bibinfo  {journal} {J. Math. Phys.}\ }\textbf {\bibinfo {volume} {43}},\ \bibinfo {pages} {205} (\bibinfo {year} {2002}{\natexlab{a}})}\BibitemShut {NoStop}%
\bibitem [{\citenamefont {Mostafazadeh}(2002{\natexlab{b}})}]{Mostafazadeh2002b}%
  \BibitemOpen
  \bibfield  {author} {\bibinfo {author} {\bibfnamefont {A.}~\bibnamefont {Mostafazadeh}},\ }\bibfield  {title} {\bibinfo {title} {Pseudo-{Hermiticity} versus {PT}-symmetry. {II}. a complete characterization of non-{Hermitian} {Hamiltonians} with a real spectrum},\ }\href {https://doi.org/10.1063/1.1461427} {\bibfield  {journal} {\bibinfo  {journal} {\textit{ibid.}}\ }\textbf {\bibinfo {volume} {43}},\ \bibinfo {pages} {2814} (\bibinfo {year} {2002}{\natexlab{b}})}\BibitemShut {NoStop}%
\bibitem [{\citenamefont {Mostafazadeh}(2002{\natexlab{c}})}]{Mostafazadeh2002c}%
  \BibitemOpen
  \bibfield  {author} {\bibinfo {author} {\bibfnamefont {A.}~\bibnamefont {Mostafazadeh}},\ }\bibfield  {title} {\bibinfo {title} {Pseudo-{Hermiticity} versus {PT}-symmetry {III}: Equivalence of pseudo-{Hermiticity} and the presence of antilinear symmetries},\ }\href {https://doi.org/10.1063/1.1489072} {\bibfield  {journal} {\bibinfo  {journal} {\textit{ibid.}}\ }\textbf {\bibinfo {volume} {43}},\ \bibinfo {pages} {3944} (\bibinfo {year} {2002}{\natexlab{c}})}\BibitemShut {NoStop}%
\bibitem [{\citenamefont {Mostafazadeh}(2003)}]{Mostafazadeh2003}%
  \BibitemOpen
  \bibfield  {author} {\bibinfo {author} {\bibfnamefont {A.}~\bibnamefont {Mostafazadeh}},\ }\bibfield  {title} {\bibinfo {title} {Exact {PT}-symmetry is equivalent to {Hermiticity}},\ }\href {https://doi.org/10.1088/0305-4470/36/25/312} {\bibfield  {journal} {\bibinfo  {journal} {J. Phys. A: Math. Gen.}\ }\textbf {\bibinfo {volume} {36}},\ \bibinfo {pages} {7081} (\bibinfo {year} {2003})}\BibitemShut {NoStop}%
\bibitem [{\citenamefont {Pinske}\ \emph {et~al.}(2019)\citenamefont {Pinske}, \citenamefont {Teuber},\ and\ \citenamefont {Scheel}}]{Pinske2019}%
  \BibitemOpen
  \bibfield  {author} {\bibinfo {author} {\bibfnamefont {J.}~\bibnamefont {Pinske}}, \bibinfo {author} {\bibfnamefont {L.}~\bibnamefont {Teuber}},\ and\ \bibinfo {author} {\bibfnamefont {S.}~\bibnamefont {Scheel}},\ }\bibfield  {title} {\bibinfo {title} {Holonomic gates in pseudo-{Hermitian} quantum systems},\ }\href {https://doi.org/10.1103/PhysRevA.100.042316} {\bibfield  {journal} {\bibinfo  {journal} {Phys. Rev. A}\ }\textbf {\bibinfo {volume} {100}},\ \bibinfo {pages} {042316} (\bibinfo {year} {2019})}\BibitemShut {NoStop}%
\bibitem [{\citenamefont {Chu}\ \emph {et~al.}(2020)\citenamefont {Chu}, \citenamefont {Liu}, \citenamefont {Liu},\ and\ \citenamefont {Cai}}]{YMChu2020}%
  \BibitemOpen
  \bibfield  {author} {\bibinfo {author} {\bibfnamefont {Y.}~\bibnamefont {Chu}}, \bibinfo {author} {\bibfnamefont {Y.}~\bibnamefont {Liu}}, \bibinfo {author} {\bibfnamefont {H.}~\bibnamefont {Liu}},\ and\ \bibinfo {author} {\bibfnamefont {J.}~\bibnamefont {Cai}},\ }\bibfield  {title} {\bibinfo {title} {Quantum sensing with a single-qubit pseudo-{Hermitian} system},\ }\href {https://doi.org/10.1103/PhysRevLett.124.020501} {\bibfield  {journal} {\bibinfo  {journal} {Phys. Rev. Lett.}\ }\textbf {\bibinfo {volume} {124}},\ \bibinfo {pages} {020501} (\bibinfo {year} {2020})}\BibitemShut {NoStop}%
\bibitem [{\citenamefont {Rigolin}\ \emph {et~al.}(2008)\citenamefont {Rigolin}, \citenamefont {Ortiz},\ and\ \citenamefont {Ponce}}]{Rigolin2008}%
  \BibitemOpen
  \bibfield  {author} {\bibinfo {author} {\bibfnamefont {G.}~\bibnamefont {Rigolin}}, \bibinfo {author} {\bibfnamefont {G.}~\bibnamefont {Ortiz}},\ and\ \bibinfo {author} {\bibfnamefont {V.~H.}\ \bibnamefont {Ponce}},\ }\bibfield  {title} {\bibinfo {title} {Beyond the quantum adiabatic approximation: Adiabatic perturbation theory},\ }\href {https://doi.org/10.1103/PhysRevA.78.052508} {\bibfield  {journal} {\bibinfo  {journal} {Phys. Rev. A}\ }\textbf {\bibinfo {volume} {78}},\ \bibinfo {pages} {052508} (\bibinfo {year} {2008})}\BibitemShut {NoStop}%
\bibitem [{\citenamefont {De~Grandi}\ and\ \citenamefont {Polkovnikov}(2010)}]{DeGrandi2010b}%
  \BibitemOpen
  \bibfield  {author} {\bibinfo {author} {\bibfnamefont {C.}~\bibnamefont {De~Grandi}}\ and\ \bibinfo {author} {\bibfnamefont {A.}~\bibnamefont {Polkovnikov}},\ }\bibinfo {title} {Adiabatic perturbation theory: From {Landau-Zener} problem to quenching through a quantum critical point},\ in\ \href {https://doi.org/10.1007/978-3-642-11470-0_4} {\emph {\bibinfo {booktitle} {Quantum Quenching, Annealing and Computation}}},\ \bibinfo {editor} {edited by\ \bibinfo {editor} {\bibfnamefont {A.~K.}\ \bibnamefont {Chandra}}, \bibinfo {editor} {\bibfnamefont {A.}~\bibnamefont {Das}},\ and\ \bibinfo {editor} {\bibfnamefont {B.~K.}\ \bibnamefont {Chakrabarti}}}\ (\bibinfo  {publisher} {Springer},\ \bibinfo {address} {Berlin},\ \bibinfo {year} {2010})\ pp.\ \bibinfo {pages} {75--114}\BibitemShut {NoStop}%
\bibitem [{\citenamefont {Nenciu}\ and\ \citenamefont {Rasche}(1992)}]{Nenciu1992}%
  \BibitemOpen
  \bibfield  {author} {\bibinfo {author} {\bibfnamefont {G.}~\bibnamefont {Nenciu}}\ and\ \bibinfo {author} {\bibfnamefont {G.}~\bibnamefont {Rasche}},\ }\bibfield  {title} {\bibinfo {title} {On the adiabatic theorem for nonself-adjoint {Hamiltonians}},\ }\href {https://doi.org/10.1088/0305-4470/25/21/027} {\bibfield  {journal} {\bibinfo  {journal} {J. Phys. A: Math. Gen.}\ }\textbf {\bibinfo {volume} {25}},\ \bibinfo {pages} {5741} (\bibinfo {year} {1992})}\BibitemShut {NoStop}%
\bibitem [{\citenamefont {Zhang}\ and\ \citenamefont {Wu}(2019)}]{QZhang2019}%
  \BibitemOpen
  \bibfield  {author} {\bibinfo {author} {\bibfnamefont {Q.}~\bibnamefont {Zhang}}\ and\ \bibinfo {author} {\bibfnamefont {B.}~\bibnamefont {Wu}},\ }\bibfield  {title} {\bibinfo {title} {Non-{Hermitian} quantum systems and their geometric phases},\ }\href {https://doi.org/10.1103/PhysRevA.99.032121} {\bibfield  {journal} {\bibinfo  {journal} {Phys. Rev. A}\ }\textbf {\bibinfo {volume} {99}},\ \bibinfo {pages} {032121} (\bibinfo {year} {2019})}\BibitemShut {NoStop}%
\bibitem [{\citenamefont {Huang}\ \emph {et~al.}(2023)\citenamefont {Huang}, \citenamefont {He}, \citenamefont {Lang},\ and\ \citenamefont {Zhu}}]{ZHHuang2023}%
  \BibitemOpen
  \bibfield  {author} {\bibinfo {author} {\bibfnamefont {Z.-H.}\ \bibnamefont {Huang}}, \bibinfo {author} {\bibfnamefont {P.}~\bibnamefont {He}}, \bibinfo {author} {\bibfnamefont {L.-J.}\ \bibnamefont {Lang}},\ and\ \bibinfo {author} {\bibfnamefont {S.-L.}\ \bibnamefont {Zhu}},\ }\bibfield  {title} {\bibinfo {title} {Quantum circuit for measuring an operator's generalized expectation values and its applications to non-{Hermitian} winding numbers},\ }\href {https://doi.org/10.1103/PhysRevA.107.052205} {\bibfield  {journal} {\bibinfo  {journal} {Phys. Rev. A}\ }\textbf {\bibinfo {volume} {107}},\ \bibinfo {pages} {052205} (\bibinfo {year} {2023})}\BibitemShut {NoStop}%
\bibitem [{\citenamefont {Wagner}\ \emph {et~al.}(2024)\citenamefont {Wagner}, \citenamefont {Schwartzman-Nowik}, \citenamefont {Paiva}, \citenamefont {Te'eni}, \citenamefont {Ruiz-Molero}, \citenamefont {Barbosa}, \citenamefont {Cohen},\ and\ \citenamefont {Galv\~ao}}]{Wagner2024}%
  \BibitemOpen
  \bibfield  {author} {\bibinfo {author} {\bibfnamefont {R.}~\bibnamefont {Wagner}}, \bibinfo {author} {\bibfnamefont {Z.}~\bibnamefont {Schwartzman-Nowik}}, \bibinfo {author} {\bibfnamefont {I.~L.}\ \bibnamefont {Paiva}}, \bibinfo {author} {\bibfnamefont {A.}~\bibnamefont {Te'eni}}, \bibinfo {author} {\bibfnamefont {A.}~\bibnamefont {Ruiz-Molero}}, \bibinfo {author} {\bibfnamefont {R.~S.}\ \bibnamefont {Barbosa}}, \bibinfo {author} {\bibfnamefont {E.}~\bibnamefont {Cohen}},\ and\ \bibinfo {author} {\bibfnamefont {E.~F.}\ \bibnamefont {Galv\~ao}},\ }\bibfield  {title} {\bibinfo {title} {Quantum circuits for measuring weak values, {Kirkwood-Dirac} quasiprobability distributions, and state spectra},\ }\href {https://doi.org/10.1088/2058-9565/ad124c} {\bibfield  {journal} {\bibinfo  {journal} {Quantum Sci. Technol.}\ }\textbf {\bibinfo {volume} {9}},\ \bibinfo {pages} {015030} (\bibinfo {year} {2024})}\BibitemShut {NoStop}%
\bibitem [{\citenamefont {Chiribella}\ \emph {et~al.}(2024)\citenamefont {Chiribella}, \citenamefont {Simonov},\ and\ \citenamefont {Zhao}}]{Chiribella2024}%
  \BibitemOpen
  \bibfield  {author} {\bibinfo {author} {\bibfnamefont {G.}~\bibnamefont {Chiribella}}, \bibinfo {author} {\bibfnamefont {K.}~\bibnamefont {Simonov}},\ and\ \bibinfo {author} {\bibfnamefont {X.}~\bibnamefont {Zhao}},\ }\bibfield  {title} {\bibinfo {title} {Dimension-independent weak value estimation via controlled {SWAP} operations},\ }\href {https://doi.org/10.1103/PhysRevResearch.6.043043} {\bibfield  {journal} {\bibinfo  {journal} {Phys. Rev. Res.}\ }\textbf {\bibinfo {volume} {6}},\ \bibinfo {pages} {043043} (\bibinfo {year} {2024})}\BibitemShut {NoStop}%
\bibitem [{\citenamefont {Dressel}\ \emph {et~al.}(2014)\citenamefont {Dressel}, \citenamefont {Malik}, \citenamefont {Miatto}, \citenamefont {Jordan},\ and\ \citenamefont {Boyd}}]{Dressel2014}%
  \BibitemOpen
  \bibfield  {author} {\bibinfo {author} {\bibfnamefont {J.}~\bibnamefont {Dressel}}, \bibinfo {author} {\bibfnamefont {M.}~\bibnamefont {Malik}}, \bibinfo {author} {\bibfnamefont {F.~M.}\ \bibnamefont {Miatto}}, \bibinfo {author} {\bibfnamefont {A.~N.}\ \bibnamefont {Jordan}},\ and\ \bibinfo {author} {\bibfnamefont {R.~W.}\ \bibnamefont {Boyd}},\ }\bibfield  {title} {\bibinfo {title} {Colloquium: Understanding quantum weak values: Basics and applications},\ }\href {https://doi.org/10.1103/RevModPhys.86.307} {\bibfield  {journal} {\bibinfo  {journal} {Rev. Mod. Phys.}\ }\textbf {\bibinfo {volume} {86}},\ \bibinfo {pages} {307} (\bibinfo {year} {2014})}\BibitemShut {NoStop}%
\bibitem [{\citenamefont {Hofmann}(2012)}]{Hofmann2012}%
  \BibitemOpen
  \bibfield  {author} {\bibinfo {author} {\bibfnamefont {H.~F.}\ \bibnamefont {Hofmann}},\ }\bibfield  {title} {\bibinfo {title} {How weak values emerge in joint measurements on cloned quantum systems},\ }\href {https://doi.org/10.1103/PhysRevLett.109.020408} {\bibfield  {journal} {\bibinfo  {journal} {Phys. Rev. Lett.}\ }\textbf {\bibinfo {volume} {109}},\ \bibinfo {pages} {020408} (\bibinfo {year} {2012})}\BibitemShut {NoStop}%
\bibitem [{\citenamefont {Wen}\ \emph {et~al.}(2023)\citenamefont {Wen}, \citenamefont {Wang}, \citenamefont {Tian}, \citenamefont {Zhang}, \citenamefont {Li}, \citenamefont {Du}, \citenamefont {Yan},\ and\ \citenamefont {Zhu}}]{YLWen2023}%
  \BibitemOpen
  \bibfield  {author} {\bibinfo {author} {\bibfnamefont {Y.-L.}\ \bibnamefont {Wen}}, \bibinfo {author} {\bibfnamefont {Y.}~\bibnamefont {Wang}}, \bibinfo {author} {\bibfnamefont {L.-M.}\ \bibnamefont {Tian}}, \bibinfo {author} {\bibfnamefont {S.}~\bibnamefont {Zhang}}, \bibinfo {author} {\bibfnamefont {J.}~\bibnamefont {Li}}, \bibinfo {author} {\bibfnamefont {J.-S.}\ \bibnamefont {Du}}, \bibinfo {author} {\bibfnamefont {H.}~\bibnamefont {Yan}},\ and\ \bibinfo {author} {\bibfnamefont {S.-L.}\ \bibnamefont {Zhu}},\ }\bibfield  {title} {\bibinfo {title} {Demonstration of the quantum principle of least action with single photons},\ }\href {https://doi.org/10.1038/s41566-023-01212-1} {\bibfield  {journal} {\bibinfo  {journal} {Nat. Photonics}\ }\textbf {\bibinfo {volume} {17}},\ \bibinfo {pages} {717} (\bibinfo {year} {2023})}\BibitemShut {NoStop}%
\bibitem [{\citenamefont {Brody}(2013)}]{Brody2014}%
  \BibitemOpen
  \bibfield  {author} {\bibinfo {author} {\bibfnamefont {D.~C.}\ \bibnamefont {Brody}},\ }\bibfield  {title} {\bibinfo {title} {Biorthogonal quantum mechanics},\ }\href {https://doi.org/10.1088/1751-8113/47/3/035305} {\bibfield  {journal} {\bibinfo  {journal} {J. Phys. A: Math. Theor.}\ }\textbf {\bibinfo {volume} {47}},\ \bibinfo {pages} {035305} (\bibinfo {year} {2013})}\BibitemShut {NoStop}%
\bibitem [{\citenamefont {Gritsev}\ and\ \citenamefont {Polkovnikov}(2012)}]{Gritsev2012}%
  \BibitemOpen
  \bibfield  {author} {\bibinfo {author} {\bibfnamefont {V.}~\bibnamefont {Gritsev}}\ and\ \bibinfo {author} {\bibfnamefont {A.}~\bibnamefont {Polkovnikov}},\ }\bibfield  {title} {\bibinfo {title} {Dynamical quantum {Hall} effect in the parameter space},\ }\href {https://doi.org/10.1073/pnas.1116693109} {\bibfield  {journal} {\bibinfo  {journal} {Proc. Natl. Acad. Sci. U.S.A.}\ }\textbf {\bibinfo {volume} {109}},\ \bibinfo {pages} {6457} (\bibinfo {year} {2012})}\BibitemShut {NoStop}%
\bibitem [{\citenamefont {Long}\ \emph {et~al.}(2022)\citenamefont {Long}, \citenamefont {Xue},\ and\ \citenamefont {Zhang}}]{YLong2022}%
  \BibitemOpen
  \bibfield  {author} {\bibinfo {author} {\bibfnamefont {Y.}~\bibnamefont {Long}}, \bibinfo {author} {\bibfnamefont {H.}~\bibnamefont {Xue}},\ and\ \bibinfo {author} {\bibfnamefont {B.}~\bibnamefont {Zhang}},\ }\bibfield  {title} {\bibinfo {title} {Non-{Hermitian} topological systems with eigenvalues that are always real},\ }\href {https://doi.org/10.1103/PhysRevB.105.L100102} {\bibfield  {journal} {\bibinfo  {journal} {Phys. Rev. B}\ }\textbf {\bibinfo {volume} {105}},\ \bibinfo {pages} {L100102} (\bibinfo {year} {2022})}\BibitemShut {NoStop}%
\bibitem [{\citenamefont {Wu}\ \emph {et~al.}(2019)\citenamefont {Wu}, \citenamefont {Liu}, \citenamefont {Geng}, \citenamefont {Song}, \citenamefont {Ye}, \citenamefont {Duan}, \citenamefont {Rong},\ and\ \citenamefont {Du}}]{YWu2019}%
  \BibitemOpen
  \bibfield  {author} {\bibinfo {author} {\bibfnamefont {Y.}~\bibnamefont {Wu}}, \bibinfo {author} {\bibfnamefont {W.}~\bibnamefont {Liu}}, \bibinfo {author} {\bibfnamefont {J.}~\bibnamefont {Geng}}, \bibinfo {author} {\bibfnamefont {X.}~\bibnamefont {Song}}, \bibinfo {author} {\bibfnamefont {X.}~\bibnamefont {Ye}}, \bibinfo {author} {\bibfnamefont {C.-K.}\ \bibnamefont {Duan}}, \bibinfo {author} {\bibfnamefont {X.}~\bibnamefont {Rong}},\ and\ \bibinfo {author} {\bibfnamefont {J.}~\bibnamefont {Du}},\ }\bibfield  {title} {\bibinfo {title} {Observation of parity-time symmetry breaking in a single-spin system},\ }\href {https://doi.org/10.1126/science.aaw8205} {\bibfield  {journal} {\bibinfo  {journal} {Science}\ }\textbf {\bibinfo {volume} {364}},\ \bibinfo {pages} {878} (\bibinfo {year} {2019})}\BibitemShut {NoStop}%
\bibitem [{\citenamefont {Liu}\ \emph {et~al.}(2021)\citenamefont {Liu}, \citenamefont {Wu}, \citenamefont {Duan}, \citenamefont {Rong},\ and\ \citenamefont {Du}}]{WQLiu2020}%
  \BibitemOpen
  \bibfield  {author} {\bibinfo {author} {\bibfnamefont {W.}~\bibnamefont {Liu}}, \bibinfo {author} {\bibfnamefont {Y.}~\bibnamefont {Wu}}, \bibinfo {author} {\bibfnamefont {C.-K.}\ \bibnamefont {Duan}}, \bibinfo {author} {\bibfnamefont {X.}~\bibnamefont {Rong}},\ and\ \bibinfo {author} {\bibfnamefont {J.}~\bibnamefont {Du}},\ }\bibfield  {title} {\bibinfo {title} {Dynamically encircling an exceptional point in a real quantum system},\ }\href {https://doi.org/10.1103/PhysRevLett.126.170506} {\bibfield  {journal} {\bibinfo  {journal} {Phys. Rev. Lett.}\ }\textbf {\bibinfo {volume} {126}},\ \bibinfo {pages} {170506} (\bibinfo {year} {2021})}\BibitemShut {NoStop}%
\bibitem [{\citenamefont {Wu}\ \emph {et~al.}(2024)\citenamefont {Wu}, \citenamefont {Wang}, \citenamefont {Ye}, \citenamefont {Liu}, \citenamefont {Niu}, \citenamefont {Duan}, \citenamefont {Wang}, \citenamefont {Rong},\ and\ \citenamefont {Du}}]{YWu2024}%
  \BibitemOpen
  \bibfield  {author} {\bibinfo {author} {\bibfnamefont {Y.}~\bibnamefont {Wu}}, \bibinfo {author} {\bibfnamefont {Y.}~\bibnamefont {Wang}}, \bibinfo {author} {\bibfnamefont {X.}~\bibnamefont {Ye}}, \bibinfo {author} {\bibfnamefont {W.}~\bibnamefont {Liu}}, \bibinfo {author} {\bibfnamefont {Z.}~\bibnamefont {Niu}}, \bibinfo {author} {\bibfnamefont {C.-K.}\ \bibnamefont {Duan}}, \bibinfo {author} {\bibfnamefont {Y.}~\bibnamefont {Wang}}, \bibinfo {author} {\bibfnamefont {X.}~\bibnamefont {Rong}},\ and\ \bibinfo {author} {\bibfnamefont {J.}~\bibnamefont {Du}},\ }\bibfield  {title} {\bibinfo {title} {Third-order exceptional line in a nitrogen-vacancy spin system},\ }\href {https://doi.org/10.1038/s41565-023-01583-0} {\bibfield  {journal} {\bibinfo  {journal} {Nat. Nanotechnol.}\ }\textbf {\bibinfo {volume} {19}},\ \bibinfo {pages} {160} (\bibinfo {year} {2024})}\BibitemShut {NoStop}%
\bibitem [{\citenamefont {Wu}\ \emph {et~al.}(2025)\citenamefont {Wu}, \citenamefont {Zhu}, \citenamefont {Wang}, \citenamefont {Rong},\ and\ \citenamefont {Du}}]{YWu2025}%
  \BibitemOpen
  \bibfield  {author} {\bibinfo {author} {\bibfnamefont {Y.}~\bibnamefont {Wu}}, \bibinfo {author} {\bibfnamefont {D.}~\bibnamefont {Zhu}}, \bibinfo {author} {\bibfnamefont {Y.}~\bibnamefont {Wang}}, \bibinfo {author} {\bibfnamefont {X.}~\bibnamefont {Rong}},\ and\ \bibinfo {author} {\bibfnamefont {J.}~\bibnamefont {Du}},\ }\bibfield  {title} {\bibinfo {title} {Experimental observation of {Dirac} exceptional points},\ }\href {https://doi.org/10.1103/PhysRevLett.134.153601} {\bibfield  {journal} {\bibinfo  {journal} {Phys. Rev. Lett.}\ }\textbf {\bibinfo {volume} {134}},\ \bibinfo {pages} {153601} (\bibinfo {year} {2025})}\BibitemShut {NoStop}%
\bibitem [{\citenamefont {Dogra}\ \emph {et~al.}(2021)\citenamefont {Dogra}, \citenamefont {Melnikov},\ and\ \citenamefont {Paraoanu}}]{Dogra2021}%
  \BibitemOpen
  \bibfield  {author} {\bibinfo {author} {\bibfnamefont {S.}~\bibnamefont {Dogra}}, \bibinfo {author} {\bibfnamefont {A.~A.}\ \bibnamefont {Melnikov}},\ and\ \bibinfo {author} {\bibfnamefont {G.~S.}\ \bibnamefont {Paraoanu}},\ }\bibfield  {title} {\bibinfo {title} {Quantum simulation of parity-time symmetry breaking with a superconducting quantum processor},\ }\href {https://doi.org/10.1038/s42005-021-00534-2} {\bibfield  {journal} {\bibinfo  {journal} {Commun. Phys.}\ }\textbf {\bibinfo {volume} {4}},\ \bibinfo {pages} {26} (\bibinfo {year} {2021})}\BibitemShut {NoStop}%
\bibitem [{\citenamefont {G\"unther}\ and\ \citenamefont {Samsonov}(2008)}]{Gunther2008}%
  \BibitemOpen
  \bibfield  {author} {\bibinfo {author} {\bibfnamefont {U.}~\bibnamefont {G\"unther}}\ and\ \bibinfo {author} {\bibfnamefont {B.~F.}\ \bibnamefont {Samsonov}},\ }\bibfield  {title} {\bibinfo {title} {{Naimark}-dilated $\mathcal{P}\mathcal{T}$-symmetric brachistochrone},\ }\href {https://doi.org/10.1103/PhysRevLett.101.230404} {\bibfield  {journal} {\bibinfo  {journal} {Phys. Rev. Lett.}\ }\textbf {\bibinfo {volume} {101}},\ \bibinfo {pages} {230404} (\bibinfo {year} {2008})}\BibitemShut {NoStop}%
\bibitem [{\citenamefont {Jin}\ \emph {et~al.}(2025)\citenamefont {Jin}, \citenamefont {Jiang}, \citenamefont {Zhu}, \citenamefont {Bao}, \citenamefont {Shen}, \citenamefont {Wang}, \citenamefont {Zhu}, \citenamefont {Xu}, \citenamefont {Song}, \citenamefont {Chen}, \citenamefont {Tan}, \citenamefont {Wu}, \citenamefont {Zhang}, \citenamefont {Gao}, \citenamefont {Wang}, \citenamefont {Zou}, \citenamefont {Zhang}, \citenamefont {Li}, \citenamefont {Zhong}, \citenamefont {Cui}, \citenamefont {Han}, \citenamefont {He}, \citenamefont {Wang}, \citenamefont {Yang}, \citenamefont {Wang}, \citenamefont {Shen}, \citenamefont {Liu}, \citenamefont {Deng}, \citenamefont {Dong}, \citenamefont {Zhang}, \citenamefont {Li}, \citenamefont {Yuan}, \citenamefont {Lu}, \citenamefont {Sun}, \citenamefont {Li}, \citenamefont {Zhang}, \citenamefont {Song}, \citenamefont {Wang}, \citenamefont {Guo}, \citenamefont {Machado}, \citenamefont {Kemp}, \citenamefont {Iadecola}, \citenamefont {Yao}, \citenamefont {Wang},\ and\ \citenamefont {Deng}}]{FTJin2025}%
  \BibitemOpen
  \bibfield  {author} {\bibinfo {author} {\bibfnamefont {F.}~\bibnamefont {Jin}}, \bibinfo {author} {\bibfnamefont {S.}~\bibnamefont {Jiang}}, \bibinfo {author} {\bibfnamefont {X.}~\bibnamefont {Zhu}}, \bibinfo {author} {\bibfnamefont {Z.}~\bibnamefont {Bao}}, \bibinfo {author} {\bibfnamefont {F.}~\bibnamefont {Shen}}, \bibinfo {author} {\bibfnamefont {K.}~\bibnamefont {Wang}}, \bibinfo {author} {\bibfnamefont {Z.}~\bibnamefont {Zhu}}, \bibinfo {author} {\bibfnamefont {S.}~\bibnamefont {Xu}}, \bibinfo {author} {\bibfnamefont {Z.}~\bibnamefont {Song}}, \bibinfo {author} {\bibfnamefont {J.}~\bibnamefont {Chen}}, \bibinfo {author} {\bibfnamefont {Z.}~\bibnamefont {Tan}}, \bibinfo {author} {\bibfnamefont {Y.}~\bibnamefont {Wu}}, \bibinfo {author} {\bibfnamefont {C.}~\bibnamefont {Zhang}}, \bibinfo {author} {\bibfnamefont {Y.}~\bibnamefont {Gao}}, \bibinfo {author} {\bibfnamefont {N.}~\bibnamefont {Wang}}, \bibinfo {author} {\bibfnamefont {Y.}~\bibnamefont {Zou}}, \bibinfo {author} {\bibfnamefont {A.}~\bibnamefont {Zhang}}, \bibinfo {author} {\bibfnamefont {T.}~\bibnamefont {Li}}, \bibinfo {author} {\bibfnamefont {J.}~\bibnamefont {Zhong}}, \bibinfo {author} {\bibfnamefont {Z.}~\bibnamefont {Cui}}, \bibinfo {author} {\bibfnamefont {Y.}~\bibnamefont {Han}}, \bibinfo {author} {\bibfnamefont {Y.}~\bibnamefont {He}}, \bibinfo {author} {\bibfnamefont {H.}~\bibnamefont {Wang}}, \bibinfo {author} {\bibfnamefont {J.-N.}\ \bibnamefont {Yang}}, \bibinfo {author} {\bibfnamefont {Y.}~\bibnamefont {Wang}}, \bibinfo {author} {\bibfnamefont {J.}~\bibnamefont {Shen}}, \bibinfo {author} {\bibfnamefont {G.}~\bibnamefont {Liu}}, \bibinfo {author} {\bibfnamefont {J.}~\bibnamefont {Deng}}, \bibinfo {author} {\bibfnamefont {H.}~\bibnamefont {Dong}}, \bibinfo {author} {\bibfnamefont {P.}~\bibnamefont {Zhang}}, \bibinfo {author} {\bibfnamefont {W.}~\bibnamefont {Li}}, \bibinfo {author} {\bibfnamefont {D.}~\bibnamefont {Yuan}}, \bibinfo {author} {\bibfnamefont {Z.}~\bibnamefont {Lu}}, \bibinfo {author} {\bibfnamefont {Z.-Z.}\ \bibnamefont {Sun}}, \bibinfo {author} {\bibfnamefont {H.}~\bibnamefont {Li}}, \bibinfo {author} {\bibfnamefont {J.}~\bibnamefont {Zhang}}, \bibinfo {author} {\bibfnamefont {C.}~\bibnamefont {Song}}, \bibinfo {author} {\bibfnamefont {Z.}~\bibnamefont {Wang}}, \bibinfo {author} {\bibfnamefont {Q.}~\bibnamefont {Guo}}, \bibinfo {author} {\bibfnamefont {F.}~\bibnamefont {Machado}}, \bibinfo {author} {\bibfnamefont {J.}~\bibnamefont {Kemp}}, \bibinfo {author} {\bibfnamefont {T.}~\bibnamefont {Iadecola}}, \bibinfo {author} {\bibfnamefont {N.~Y.}\ \bibnamefont {Yao}}, \bibinfo {author} {\bibfnamefont {H.}~\bibnamefont {Wang}},\ and\ \bibinfo {author} {\bibfnamefont {D.-L.}\ \bibnamefont {Deng}},\ }\bibfield  {title} {\bibinfo {title} {Topological prethermal strong zero modes on superconducting processors},\ }\href {https://doi.org/10.1038/s41586-025-09476-z} {\bibfield  {journal} {\bibinfo  {journal} {Nature (London)}\ }\textbf {\bibinfo {volume} {645}},\ \bibinfo {pages} {626} (\bibinfo {year} {2025})}\BibitemShut {NoStop}%
\bibitem [{\citenamefont {Chau}\ and\ \citenamefont {Wilczek}(1995)}]{HFChau1995}%
  \BibitemOpen
  \bibfield  {author} {\bibinfo {author} {\bibfnamefont {H.~F.}\ \bibnamefont {Chau}}\ and\ \bibinfo {author} {\bibfnamefont {F.}~\bibnamefont {Wilczek}},\ }\bibfield  {title} {\bibinfo {title} {Simple realization of the {Fredkin} gate using a series of two-body operators},\ }\href {https://doi.org/10.1103/PhysRevLett.75.748} {\bibfield  {journal} {\bibinfo  {journal} {Phys. Rev. Lett.}\ }\textbf {\bibinfo {volume} {75}},\ \bibinfo {pages} {748} (\bibinfo {year} {1995})}\BibitemShut {NoStop}%
\bibitem [{\citenamefont {Smolin}\ and\ \citenamefont {DiVincenzo}(1996)}]{Smolin1996}%
  \BibitemOpen
  \bibfield  {author} {\bibinfo {author} {\bibfnamefont {J.~A.}\ \bibnamefont {Smolin}}\ and\ \bibinfo {author} {\bibfnamefont {D.~P.}\ \bibnamefont {DiVincenzo}},\ }\bibfield  {title} {\bibinfo {title} {Five two-bit quantum gates are sufficient to implement the quantum {Fredkin} gate},\ }\href {https://doi.org/10.1103/PhysRevA.53.2855} {\bibfield  {journal} {\bibinfo  {journal} {Phys. Rev. A}\ }\textbf {\bibinfo {volume} {53}},\ \bibinfo {pages} {2855} (\bibinfo {year} {1996})}\BibitemShut {NoStop}%
\bibitem [{\citenamefont {Yu}\ and\ \citenamefont {Ying}(2015)}]{NKYu2015}%
  \BibitemOpen
  \bibfield  {author} {\bibinfo {author} {\bibfnamefont {N.}~\bibnamefont {Yu}}\ and\ \bibinfo {author} {\bibfnamefont {M.}~\bibnamefont {Ying}},\ }\bibfield  {title} {\bibinfo {title} {Optimal simulation of {Deutsch} gates and the {Fredkin} gate},\ }\href {https://doi.org/10.1103/PhysRevA.91.032302} {\bibfield  {journal} {\bibinfo  {journal} {Phys. Rev. A}\ }\textbf {\bibinfo {volume} {91}},\ \bibinfo {pages} {032302} (\bibinfo {year} {2015})}\BibitemShut {NoStop}%
\bibitem [{\citenamefont {{Google Quantum AI and Collaborators}}(2025)}]{Google2025}%
  \BibitemOpen
  \bibfield  {author} {\bibinfo {author} {\bibnamefont {{Google Quantum AI and Collaborators}}},\ }\bibfield  {title} {\bibinfo {title} {Observation of constructive interference at the edge of quantum ergodicity},\ }\href {https://doi.org/10.1038/s41586-025-09526-6} {\bibfield  {journal} {\bibinfo  {journal} {Nature (London)}\ }\textbf {\bibinfo {volume} {646}},\ \bibinfo {pages} {825} (\bibinfo {year} {2025})}\BibitemShut {NoStop}%
\bibitem [{\citenamefont {AbuGhanem}(2025)}]{IBM2025}%
  \BibitemOpen
  \bibfield  {author} {\bibinfo {author} {\bibfnamefont {M.}~\bibnamefont {AbuGhanem}},\ }\bibfield  {title} {\bibinfo {title} {{IBM} quantum computers: evolution, performance, and future directions},\ }\href {https://doi.org/10.1007/s11227-025-07047-7} {\bibfield  {journal} {\bibinfo  {journal} {J.Supercomputing}\ }\textbf {\bibinfo {volume} {81}},\ \bibinfo {pages} {687} (\bibinfo {year} {2025})}\BibitemShut {NoStop}%
\bibitem [{\citenamefont {Kolodrubetz}\ \emph {et~al.}(2017)\citenamefont {Kolodrubetz}, \citenamefont {Sels}, \citenamefont {Mehta},\ and\ \citenamefont {Polkovnikov}}]{Kolodrubetz2017}%
  \BibitemOpen
  \bibfield  {author} {\bibinfo {author} {\bibfnamefont {M.}~\bibnamefont {Kolodrubetz}}, \bibinfo {author} {\bibfnamefont {D.}~\bibnamefont {Sels}}, \bibinfo {author} {\bibfnamefont {P.}~\bibnamefont {Mehta}},\ and\ \bibinfo {author} {\bibfnamefont {A.}~\bibnamefont {Polkovnikov}},\ }\bibfield  {title} {\bibinfo {title} {Geometry and non-adiabatic response in quantum and classical systems},\ }\href {https://doi.org/https://doi.org/10.1016/j.physrep.2017.07.001} {\bibfield  {journal} {\bibinfo  {journal} {Phys. Rep.}\ }\textbf {\bibinfo {volume} {697}},\ \bibinfo {pages} {1} (\bibinfo {year} {2017})}\BibitemShut {NoStop}%
\bibitem [{\citenamefont {Huang}()}]{dataset}%
  \BibitemOpen
  \bibfield  {author} {\bibinfo {author} {\bibfnamefont {Z.-H.}\ \bibnamefont {Huang}},\ }\bibfield  {title} {\bibinfo {title} {Data for "direct measurement of quantum geometric tensor in pseudo-{Hermitian} systems"},\ }\href {https://doi.org/10.5281/zenodo.17639628} {10.5281/zenodo.17639628}\BibitemShut {NoStop}%
\bibitem [{\citenamefont {Haldane}(1988)}]{Haldane1988}%
  \BibitemOpen
  \bibfield  {author} {\bibinfo {author} {\bibfnamefont {F.~D.~M.}\ \bibnamefont {Haldane}},\ }\bibfield  {title} {\bibinfo {title} {Model for a quantum {Hall} effect without {Landau} levels: Condensed-matter realization of the "parity anomaly"},\ }\href {https://doi.org/10.1103/PhysRevLett.61.2015} {\bibfield  {journal} {\bibinfo  {journal} {Phys. Rev. Lett.}\ }\textbf {\bibinfo {volume} {61}},\ \bibinfo {pages} {2015} (\bibinfo {year} {1988})}\BibitemShut {NoStop}%
\bibitem [{\citenamefont {Li}\ \emph {et~al.}(2025)\citenamefont {Li}, \citenamefont {Wang}, \citenamefont {Kong}, \citenamefont {Lv}, \citenamefont {Jia}, \citenamefont {Tao}, \citenamefont {Li},\ and\ \citenamefont {Liu}}]{RJLi2025}%
  \BibitemOpen
  \bibfield  {author} {\bibinfo {author} {\bibfnamefont {R.}~\bibnamefont {Li}}, \bibinfo {author} {\bibfnamefont {W.}~\bibnamefont {Wang}}, \bibinfo {author} {\bibfnamefont {X.}~\bibnamefont {Kong}}, \bibinfo {author} {\bibfnamefont {B.}~\bibnamefont {Lv}}, \bibinfo {author} {\bibfnamefont {Y.}~\bibnamefont {Jia}}, \bibinfo {author} {\bibfnamefont {H.}~\bibnamefont {Tao}}, \bibinfo {author} {\bibfnamefont {P.}~\bibnamefont {Li}},\ and\ \bibinfo {author} {\bibfnamefont {Y.}~\bibnamefont {Liu}},\ }\bibfield  {title} {\bibinfo {title} {Realization of a non-{Hermitian} {Haldane} model in circuits},\ }\href {https://doi.org/https://doi.org/10.15302/frontphys.2025.044204} {\bibfield  {journal} {\bibinfo  {journal} {Front. Phys.}\ }\textbf {\bibinfo {volume} {20}},\ \bibinfo {pages} {044204} (\bibinfo {year} {2025})}\BibitemShut {NoStop}%
\end{thebibliography}

%

\end{document}